\documentclass[10pt]{article}

\newcommand{\beq}{\begin{equation}}
\newcommand{\eeq}{\end{equation}}
\newcommand{\beqa}{\begin{eqnarray}}
\newcommand{\eeqa}{\end{eqnarray}}

\usepackage{feynmf}

\usepackage{adjustbox}
\usepackage{standalone}
\usepackage{amsmath}

\usepackage{xcolor}
\usepackage{epsfig}
\usepackage{cancel}
\usepackage{caption}
\usepackage{feynmf} 
\usepackage{tikz-feynman}
\usepackage{amssymb}
\usepackage{amsfonts}
\usepackage{epsf}
\usepackage{rotating}

\usepackage{amsmath}
\usepackage{fancyhdr}

\usepackage{pstricks}
\usepackage{color}
\usepackage{frontespizio}
\usepackage{hyperref}
\hypersetup{
	colorlinks,
	citecolor=green,
	filecolor=black,
	linkcolor=blue,
	urlcolor=black
}
\usepackage{type1ec}
\usepackage[T1]{fontenc}
\usepackage{lettrine}
\usepackage{bbold}
\usepackage{calligra}
\usepackage{tikz}

\usepackage{mathrsfs}
\usepackage{curve2e}
\usepackage{setspace}
\usepackage{indentfirst}
\usepackage{emptypage}
\usepackage{cite}
\usepackage[font=small,labelfont=bf,labelsep=quad]{caption} 
\usepackage{graphicx} 
\usepackage{listings} 
\usetikzlibrary{patterns}
\usepackage{relsize}
\usetikzlibrary{intersections,positioning}
\usetikzlibrary{decorations.pathmorphing,decorations.markings,arrows,positioning}
\usepackage{braket}
\usepackage{mathrsfs}
\usepackage{stackengine}
\usepackage{calc}
\newlength\shlength
\newcommand\xshlongvec[2][0]{\setlength\shlength{#1pt}%
	\stackengine{-5.6pt}{$#2$}{\smash{$\kern\shlength%
			\stackengine{7.55pt}{$\mathchar"017E$}%
			{\rule{\widthof{$#2$}}{.57pt}\kern.4pt}{O}{r}{F}{F}{L}\kern-\shlength$}}%
	{O}{c}{F}{T}{S}}
\newcommand{\pd}{\partial}

\newcommand{\de}{\delta}
\newcommand{\s}[1]{\sqrt{#1}}

\newcommand{\dfun}[2]{ \frac{\delta #1}{\delta #2}}

\def\lt{\left(}
\def\rt{\right)}

\newcommand{\sdfrac}[2]{\mbox{\small$\displaystyle\frac{#1}{#2}$}}

\IfFileExists{dsfont.sty}
{\usepackage{dsfont}
	\let\mathbb=\mathds
	\newcommand{\id}{\mathds{1}}}
{\typeout{Package dsfont.sty was not found, using alternative macros.}
	\let\mathds=\mathbb
	\newcommand{\id}{\mbox{1 \kern-.59em {\rm l}}}}

\usepackage{slashed}
\usepackage{units}
\usepackage{setspace}
\newcommand{\nn}{\nonumber}
\let\a=\alpha   \let\b=\beta   \let\g=\gamma   \let\d=\delta
\let\e=\epsilon         
    \let\k=\kappa  \let\l=\lambda  \let\m=\mu
\let\n=\nu           \let\p=\pi      \let\r=\rho
\let\s=\sigma

\let\d=\delta
\let\s=\sigma
\newcommand{\sm}{\mathcal{S}}
\newcommand{\figref}[1]{Fig.~\ref{#1}}			
\newcommand{\secref}[1]{Section~\ref{#1}}		
\newcommand{\appref}[1]{Appendix~\ref{#1}}		
\renewcommand{\a}{\alpha}

\newcommand{\vare}{\varepsilon}
\newcommand{\ka}{\kappa}

\renewcommand{\r}{\rho}

\def\nbox#1#2{\vcenter{\hrule \hbox{\vrule height#2in
			\kern#1in \vrule} \hrule}}
\def\sq{\,\raise.5pt\hbox{$\nbox{.09}{.09}$}\,}
\def\sqb{\,\raise.5pt\hbox{$\overline{\nbox{.09}{.09}}$}\,}

\newcommand{\bea}{\begin{eqnarray}}
\newcommand{\eea}{\end{eqnarray}}
\newcommand{\be}{\begin{equation}}
\newcommand{\ee}{\end{equation}}

\newcommand{\bes}{\begin{subequations}}
	\newcommand{\ees}{\end{subequations}}

\def\nn{\nonumber\\}

\def\Box{\sq}

\numberwithin{equation}{section}

\usepackage{accents}

\makeatletter
\newcommand{\xLine}[2][]{\ext@arrow 0359\Rightarrowfill@{#1}{#2}}
\makeatother
\xdefinecolor{darkgreen}{RGB}{102, 204, 70}
\xdefinecolor{darkblue}{RGB}{0, 0, 153}
\usepackage{amssymb}
\usepackage{pifont}
\newcommand{\cmark}{\ding{51}}%
\newcommand{\xmark}{\ding{55}}%

\begin{document}
\begin{center}
\vspace{1.5cm}
{\Large \bf The Conformal Anomaly Action  to Fourth Order  (4T) in $d=4$ \\}
\vspace{0.1cm}
{\Large \bf in Momentum Space\\} 
\vspace{0.3cm}
\vspace{1cm}
{\bf $^{(1)}$Claudio Corian\`o, $^{(1,2)}$Matteo Maria Maglio and $^{(1)}$Dimosthenis Theofilopoulos\\}
\vspace{1cm}
{\it  $^{(1)}$Dipartimento di Matematica e Fisica, Universit\`{a} del Salento \\
and INFN Sezione di Lecce, Via Arnesano 73100 Lecce, Italy\\}
\vspace{0.5cm}
{\it  $^{(2)}$Galileo Galilei Institute for Theoretical Physics, \\
	Largo Enrico Fermi 2, I-50125 Firenze, Italy}

\begin{abstract}
We elaborate on the structure of the conformal anomaly effective action up to 4-th order, in an expansion in the gravitational fluctuations $(h)$ of the background metric, in the flat spacetime limit. For this purpose we discuss the renormalization of 4-point functions containing insertions of stress-energy tensors (4T), in conformal field theories in four spacetime dimensions with the goal of identifying the structure of  the anomaly action. We focus on a separation of the correlator into its transverse/traceless and longitudinal components, applied to the trace and conservation Ward identities (WI) in momentum space. These are sufficient  to identify, from their hierarchical structure, the anomaly contribution, without the need to proceed with a complete determination of all of its independent form factors. Renormalization induces sequential bilinear graviton-scalar mixings on single, double and multiple trace terms, corresponding to $R\square^{-1}$ interactions of the scalar curvature, with intermediate virtual massless exchanges. These dilaton-like terms couple to the conformal anomaly, as for the chiral anomalous WIs. We show that at 4T level a new traceless component appears after renormalization. We comment on future extensions of this result to more general backgrounds, with possible applications to non local cosmologies. \end{abstract}

\end{center}

\newpage
\tableofcontents
\section{Introduction} 
The analysis of multi-point correlation functions in conformal field theory (CFT) in higher dimensions $(D>2)$ plays an important role in the attempt to understand such symmetry at quantum level. In this class of theories, lower-point correlators are fixed up to 3-point functions, except for few constants. Higher point functions, on the other end, are expected to be determined in terms of their "conformal data", and constrained by their associated conformal Ward identities (CWI's) \cite{Coriano:2020ees}.\\
 In the more general case of $n$-point functions, the CWI's predict a specific structure of the correlators, except for some arbitrary functions of the conformal invariant ratios, whose number depends on $n$. \\
Most of such investigations, in the past, have  been performed in coordinate space, but more recently there has been considerable interest in extending these approaches to momentum space. This has spurred a significant 
activity and has allowed to proceed with the characterization of their salient features 
\cite{Isono:2018rrb,Isono:2019wex,Chen:2019gka, Gillioz:2018mto,Gillioz:2018kwh,Loebbert:2020hxk,Loebbert:2016cdm,Jain:2021wyn,Jain:2020rmw}
. \\
Examples are the classification of the minimal number of form factors present in their expansions in the external momenta, the identification of the arbitrary functions which appear in the solution of the corresponding CWIs for $n> 3$,  
\cite{Bzowski:2013sza,Bzowski:2015yxv,Bzowski:2020kfw,Bzowski:2015pba}, or the search for exact solutions in the presence of dual conformal symmetry \cite{Maglio:2019grh}.  \\
Among these correlators, an important role is played by those involving the stress energy tensor, due to the appearance of a conformal (trace) anomaly in even spacetime dimensions (see \cite{Bastianelli:2006rx} for a general discussion). Examples which have been investigated are those involving multiple $T$ insertions ($nT$), or combined with scalars ($O$) and/or conserved vectors currents $(J)$. In particular, $nT$  correlators play a special role and represent the essential part of the generating functional which characterize the coupling of a conformal field theory to gravity. \\   
Indeed, in any CFT a crucial role is played by the stress energy tensor $T_{\mu\nu}$ and by its trace, both in the definition of the CFT and in the description of its breaking, due to the presence of a conformal anomaly in even spacetime dimensions \cite{Bastianelli:2006rx}. 
$nT$ correlators characterize the functional expansion of the anomaly effective action, which collects such contributions for any integer $n$, and is significantly constrained by the underlying CWI's.  \\
We believe that these analysis are necessary in order to nail down the structure of the anomaly action using a direct approach, independently of those entertained so far, based on the variational solutions of the anomaly functional, which take either to a non local Riegert action \cite{Riegert:1987kt} or, alternatively, to local ones, such those derived by the Noether (leaking) method (see for instance \cite{Coriano:2012dg,Coriano:2013nja}), with the inclusion of an asymptotic dilaton. The analysis of the conformal constraints and the way they impact the structure of the anomaly action is the main goal of our work. \\
 For this purpose, we recall that variational solutions of the anomaly equations are naturally limited by the mathematical procedure of integration of  the underlying action, since they differ by arbitrary Weyl-invariant contributions. \\
As shown in several previous analysis, such a breaking is characterised by the appearance of massless poles \cite{Giannotti:2008cv,2009PhLB..682..322A,Armillis:2010qk,Armillis:2009pq,Coriano:2014gja} in specific form factors of correlation functions involving one or more stress energy tensors, in the form of bilinear mixings. Such contributions are ubiquitous in all the (chiral, conformal, superconformal) correlation functions investigated so far, indicating that their appearance is directly linked to 
an explicitly broken phase induced by the anomaly, due to renormalization. Our approach defines a way to isolate such contributions to the anomaly action, working up to $n=4$, extending previous analysis \cite{Coriano:2017mux,Coriano:2018zdo,Coriano:2018bsy,Coriano:2020ees}.\\
\subsection{Bilinear mixings}
Bilinear mixings are, even in ordinary field theory, the signature that the functional expansion of the effective action is taking place in a nontrivial vacuum, as for the Higgs mechanism. \\
 In an ordinary gauge theory such mixings are removed by a suitable gauge choice, such as the `t Hooft or the unitary gauge. In the conformal case, a massless pole represents a virtual (nonlocal) interaction, which is directly coupled to the anomaly and defines a skeleton expansion which can be taken as a possible definition the anomaly action.  \\
 As we are going to elaborate in a following section, such terms correspond to an expansion of the same action - in coordinate space and in the flat limit- in the dimensionless variable $R\,\square^{-1}$. Here, $R$ is the scalar curvature, which has very often appeared in the analysis of nonlocal cosmologies, such as in 
 $f(R\,\square^{-1})$ models \cite{Deser:2007jk}. In this case, the goal has been that of explaining the late-time dark energy dominance of our universe. \\
These analysis have been limited in the past to 3-point functions \cite{Bzowski:2020kfw,Coriano:2019nkw,Maglio:2019grh}. Four-point functions have received attention only more recently \cite{Coriano:2019nkw,Serino:2020pyu}.
For 3-point functions, the analysis of conformal and non-conformal correlators in realistic 
theories such as QED and QCD, at one-loop, has covered the $TJJ$ and the $TTT$ (3T)\cite{Giannotti:2008cv,Coriano:2018bsy,Coriano:2018bbe,Armillis:2009pq}.\\
 In this work we are going to show that this feature is generic. No complete solution of the CWIs at quartic level ($O(h^4)$ in flat space, with $ h$ the metric fluctuation) is needed in order to extract such behaviour, at least in flat space, once the contribution of the anomaly is correctly taken into account.  \\
Notice that the advantage of solving explicitly the CWIs, as in the case of conformal 3-point functions,  beside its indisputable value,  is that it gives the opportunity to apply and verify the renormalization procedure on the solution, using the known gravitational counterterms, as given in Eq. \eqref{counter} below. 
However, for the rest, the method becomes increasingly prohibitive if the goal is to infer the 
 all-order behaviour of the anomaly action.\\
  The purpose of our paper is to show that a certain amount of information is hidden in the structure of the WIs, if we assume that they can be renormalized using the standard (known) counterterms typical of the anomaly functional. We work our way from this assumption backward, in order to characterize the structure of the CWIs and identify the implications for the structure of the anomaly action to a given order in $h^{\mu\nu}$ $(h)$, the fluctuation of the metric around its flat spacetime limit. \\
Our approach is, in a way, quite direct, and can be extended to n-point functions and to any spacetime dimension, following the same strategy. More details on these further developments will be presented in forthcoming work.
Here we intend to characterize the method in its simplest formulation, by working in $d=4$ spacetime dimensions, and relying on an expansion of the anomaly functional around a flat background in Dimensional Regularization (DR).
\subsection{Nonlocal interactions in the 3T}
\begin{figure}[t]
	\centering
	{\includegraphics[scale=0.6]{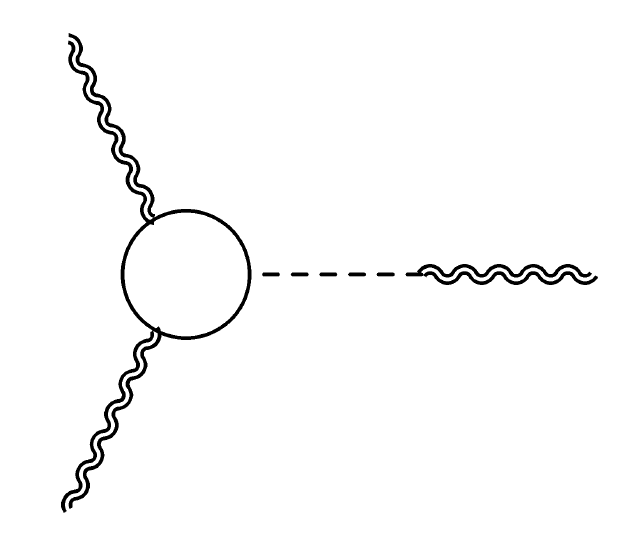}} \hspace{0.2cm}
	{\includegraphics[scale=0.6]{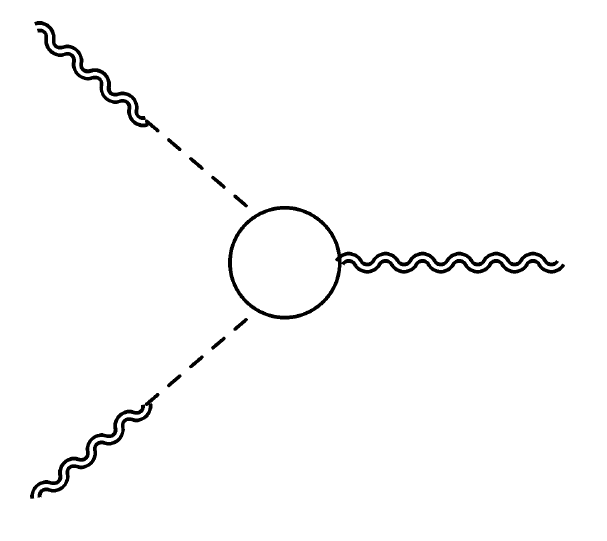}}\hspace{0.2cm}
	{\includegraphics[scale=0.6]{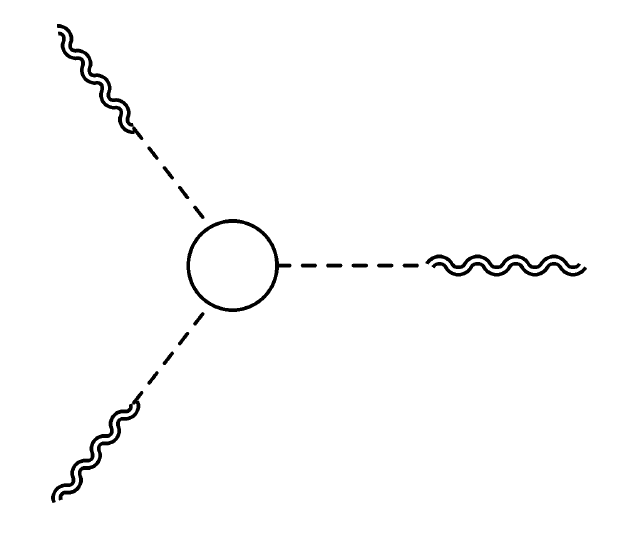}}
	\caption{Expansion of the anomaly contributions to the renormalized vertex for the 3T.\label{FeynTTTX}}
\end{figure}
In the case of the 3T all the contributions to the trace anomaly of the renormalized vertex  can be summarised by the three diagrams shown in \figref{FeynTTTX} and will be discussed below. \\ 
In this figure, the dashed lines indicate the inclusion of an operator
\begin{equation}
\hat{\pi}^{\mu\nu} =g^{\mu\nu}p^2  - p^\mu p^\nu
\end{equation}
on the external lines, which projects on the subspace transverse to a given momentum $p$. The projector is accompanied by a pole. Together,  $\hat{\pi}$ and the pole define the ordinary transverse projector 
\begin{equation}
\label{pole}
\pi^{\mu\nu}=\hat{\pi}^{\mu\nu}\frac{1}{p^2} 
\end{equation}
inserted on all the external lines in one, two and three copies. 
Poles in separate variables $1/p_i^2$,  $1/(p_i^2 p_j^2) (i\neq j)$ and $1/(p_1^2 p_2^2 p_3^2)$, are connected to separate external graviton lines, and each momentum invariant appears as a single pole. \\
This result had been obtained in \cite{Coriano:2018bsy} by performing a complete perturbative analysis of the same vertex using free field theory realizations. In this case, the inclusion of 3 sectors, a scalar, a  fermion and a spin 1, allows to obtain the most general expression of this correlator, and its renormalization, performed by the addition of the general counterterm \eqref{counter}, has been verified.   \\
Our goal in the next few sections will be to show that the Weyl-variant contribution to $\sm_A$  can be identified directly from the CWIs, under the assumption that  $\sm_{ct}$, as defined in \eqref{counter}, is all that is needed in order to proceed with the renormalization of a 3- or a 4-point function of correlators of stress energy tensors in $d=4$, in the flat spacetime limit.\\
If we move to coordinate space from momentum space 
and attach the metric fluctuations $h^{\mu\nu}$ on the external lines, it is easy to show that the inclusion of a projector of the form \eqref{pole} induces a bilinear mixing between the graviton and a virtual scalar propagating in the intermediate virtual (dashed) line. The interaction takes the form  
\begin{equation}
\label{coupling}
\frac{1}{p^2} \hat\pi^{\mu\nu} \leftrightarrow R^{(1)}\frac{1}{\square}
\end{equation}
In the 3T case, as one can immediately figure out from \figref{FeynTTTX}, such expressions can be easily traced back to nonlocal terms in the anomaly action $\sm_A$ of the form
\begin{equation}
\mathcal{S}_{A}\sim \int d^4 x\, d^4 y R^{(1)}(x)  \left(\frac{1}{\square}\right)(x,y) \left( b'\, E_4^{(2)}(y) + b\, (C^2)^{(2)}(y)\right),
\end{equation}
where the labels $(1)$, $(2)$ refer to the first and second variations of the invariants $R$, the scalar curvature, $C^2$ and $E_4$.
\begin{figure}[t]
	\centering
	{\includegraphics[scale=0.6]{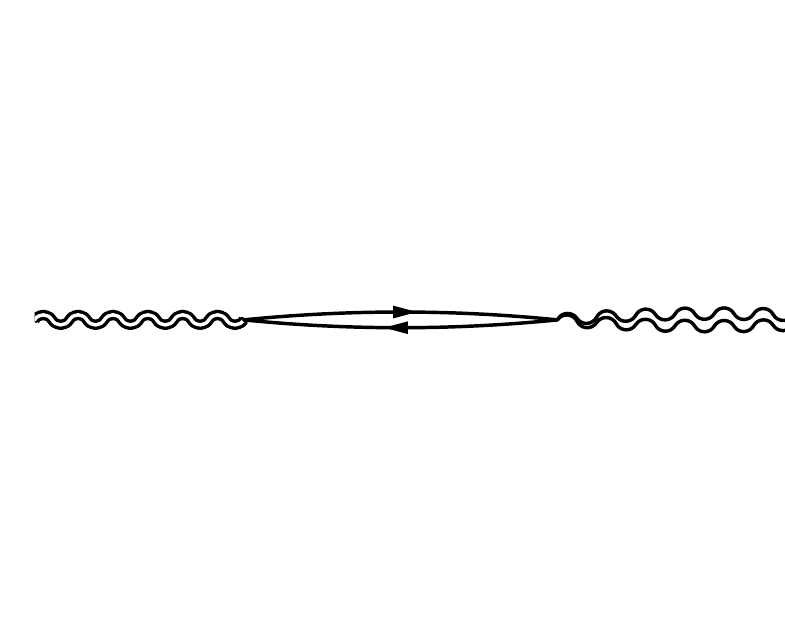}} \hspace{1cm}
	\raisebox{-1.5ex}{\includegraphics[scale=0.6]{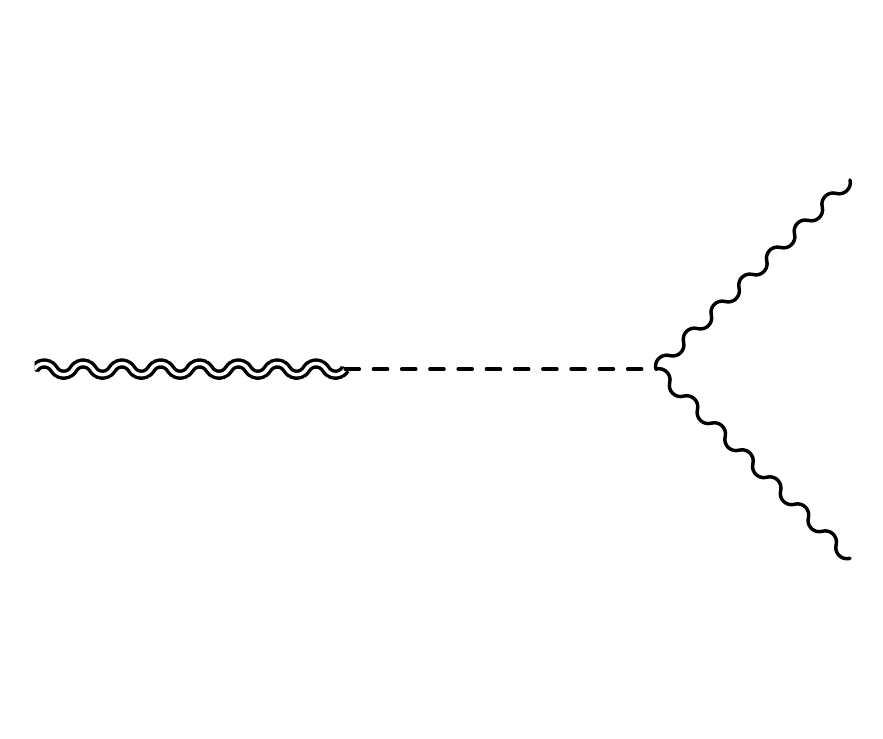}} 
	\caption{Expansion of the anomaly contributions to the renormalized vertex for the TJJ. Left: The collinear region in momentum space responsible for the origin of the pole. Right: Its interpretation as a scalar exchange. \label{FeynTJJ}}
\end{figure}
Analogously, in the case of the TJJ correlator in QED, the anomaly contribution, extracted by a complete perturbative analysis,  is shown in \figref{FeynTJJ} and takes the form
\begin{equation}
\mathcal{S}_{A}\sim \beta(e) \int d^4 x\, d^4 y \,R^{(1)}(x)  \left(\frac{1}{\square}\right)(x,y) F^{\mu\nu} F_{\mu\nu}(y),
\end{equation}
with $F^{\mu\nu}$ being the QED field strength and $\beta(e)$ the corresponding beta-function of the gauge coupling. Notice that in both examples the intermediate propagating virtual state is directly coupled to the anomaly. \\
This characterization of the anomaly action $\sm_A$ at cubic order, in each of the two cases, has been obtained by a complete analysis of the conformal constraints, using the explicit expression of their solutions, which for 3 point functions depends only on three constant for the 3T and on two for the TJJ. In both cases, the solutions, which are given by hypergeometric functions $F_4$, can be mapped to  general free field theory realizations, characterized by the inclusion at 1-loop level of an arbitrary number of scalars, fermions and spin-1 fields in the Feynman expansion. The map between the two approaches is an exact one in $d=3,5$.\\
In \cite{Coriano:2017mux} it has been shown that such nonzero trace contributions are automatically generated by a nonlocal conformal anomaly action of Riegert type \cite{Riegert:1984kt} expanded to first order in $\delta g$ and to second order in the external gauge field $A_{\mu}$, respectively. \\
As we move from 3- to 4-point functions, the solutions of the CWIs are affected by the appearance of arbitrary functions, which is a general feature of CFT,  and this approach is not of immediate help, in the sense that even if such bilinear mixings would be found in the perturbative description of a certain correlator, they would not find an immediate counterpart in the general solution of the conformal constraints. \\
Obviously, we would like to provide a proof of such behaviour with no reference to free field theory. 
We recall that the CWIs' impose, at any $n$, hierarchical relations connecting $n$ to $n-1$ point functions, which we are going to investigate in great detail in momentum space. While these relations are not sufficient, for a generic CFT, to reconstruct the expression of $\sm(g)$ beyond $O(\delta g^3)$ (i.e. $n=3$), they nevertheless  constrain significantly its structure after renormalization.\\
 As we are going to show, the CWIs predict a well-defined structure for the anomaly contribution to $\sm_A$ at the level of the 4T, once we assume that their renormalization proceeds via the counterterm action \eqref{counter}. As already mentioned, this analysis does not require a complete classification of all the form factors which appear in the correlator, but just a careful study of the structure of the CWIs satisfied by it and by the 3T.\\
For this reason, we are going to illustrate how this simplified procedure works first in the case of the 3T, before moving to the 4T in this new approach, showing how the structures of $\sm_A$ is constrained by the renormalized CWIs to assume a form very similar to the 3T case. \\
 As in previous studies in momentum space, we will rely on decompositions of the 3T and 4T  into a transverse traceless, a longitudinal, and a trace part, extending the approach formulated in \cite{Bzowski:2013sza} for 3-point functions to 4-point functions.

\subsection{Content of our work}
Our work is organised as follows. We will define our conventions and characterise the main features of the anomaly action in section 2, highlighting all the simplifications that are present when we consider the flat spacetime limit in the definition of the correlation functions. Conveniently, the discussion, in this section, is carried out in full generality in regard to the choice of the background metric, and provides a wider perspective on the structure of the effective action $(\mathcal{S})$ for arbitrary metrics, and on the increased complexity of the CWIs around such backgrounds. \\
It is well known that in more general backgrounds, for instance for Weyl flat/conformally flat metrics, the structure of the quantum average of the energy momentum tensor (its one-point function), is affected by tadpole contributions which are fixed by the anomaly, and need to be included in the analysis, and call for a generalization of our method. \\
Then we will review the structure of the CWIs in the longitudinal/transverse separation introduced in \cite{Bzowski:2013sza} and developed further in \cite{Bzowski:2018fql,Bzowski:2015pba,Bzowski:2015yxv}, followed by the analysis of the anomalous CWI's. In \secref{heres2} a new derivation, performed directly in $d=4$, of the special CWI's is presented for the $4T$. The approach  extends the one developed in \cite{Coriano:2017mux} for the 3T to the new case.\\
The renormalization of the 4T and its longitudinal/transverse decomposition is worked out in a following section, where, just for completeness, we also classify the singular form factors which are affected by renormalization.  \\
The structure of the anomaly action is slowly built starting from the 2-point function, that we review. The vanishing of the anomaly contribution to the anomaly action $\mathcal{S}$  
at $O(h^2)$ is illustrated in great detail. Then we derive the structure of the local (i.e. $t_{loc}$ or longitudinal stress energy tensor) contributions to the CWI in the renormalization procedure, and separate the anomaly contributions from the unknown but finite, renormalized  parts. This is the content of \secref{fourp}. \\
Even if the finite local terms are not explicitly given, we show that they are not necessary in order to identify the structure of the anomaly action at the level of the 4T. \\
The result is obtained  by working directly with the longitudinal components $t_{loc}$ of the counterterms to the same 4T correlator, combined with their trace WIs. We show that a structure with bilinear mixings appears quite naturally from the decomposition, together with an extra, trace-free contribution. 
This extra term, absent in the 3-point function, appears for the first time at the level of the 
4T. A comment on our result follows next, before our Conclusions.     
   
\section{The anomaly action }
\label{two}
In general, the effective action for a given Lagrangian CFT, $\mathcal{S}(g)$ (for former discussions of $\mathcal{S}(g)$ see 
\cite{Cappelli:1988vw,Cappelli:2001pz,Erdmenger:1998xv}) which is a functional of the external metric background $g$, can be generated, for instance, by integrating out some conformal matter in the path integral, leaving the external metric arbitrary. The simplest realization is provided, for instance, by a scalar free field theory coupled to gravity, as discussed in several perturbative studies \cite{Coriano:2018bbe,Coriano:2018bsy,Serino:2020pyu}. \\
Compared to Sakharov's induced gravity, where integration over ordinary matter, for a generic metric $g$, is expected to generate terms of the form  \cite{Visser:2002ew}
\begin{equation} 
\sm_{eff}\sim \int d^4 x \sqrt{g}\left( \Lambda + c_1(g) R + c_2``R^2" \right),
\end{equation}
corresponding to a cosmological constant, the Einstein-Hilbert action and to generic $``R^2"$ terms, the integration over conformal matter should not introduce any scale, if the result of the integration turns out to be finite.\\
This turns out to be the case, in $d$ spacetime dimensions, at least for a scalar theory, as far as we stay away from even dimensions, but a renormalization procedure is required in the limit $d\to 4$, with the inclusion of a renormalization scale $\mu$. \\
This appears as a balancing factor $\mu^{-\epsilon}$ - with $\epsilon=d-4$ - in the structure of the counterterms, which are expressed in terms of Weyl-invariant operators in $d=4$. \\
The renormalized action then acquires a $\log k^2/\mu^2$ dependence, where $k$ is a generic momentum, determining the breaking of dilatation invariance. The trace  
anomaly part is simply associated to a well-defined pole structure, on which we are going to elaborate below. This is associated to the emergence of the $R\square^{-1}$ dimensionless operator attached to the external legs of the correlator. We will come back to comment on this result in a final section.

For a general CFT, when no Lagrangian is present, the constraints induced by the CWI's are identical to those derived from the functional integral (Lagrangian) approach. In this case  $\mathcal{S}(g)$ can be simply defined to be the functional which collects all the correlation functions with single and multiple T- insertions, and it is clearly not related to a path integral formulation. Therefore, the CWIs remain identical both for Lagrangian and non-Lagrangian CFTs.\\
In the latter case, however, one could resort to the operatorial formulation of the theory and, for primary fields such as the stress energy tensor, derive the CWIs in a completely independent manner (see for instance 
\cite{Osborn:1993cr,Erdmenger:1996yc} and \cite{Coriano:2018bbe} for coordinate and momentum space derivations of tensorial correlators).\\
We recall that in an ordinary field theory the relation between the partition function and the functional of all the connected correlators is obviously given by the functional relation

\begin{equation}
\label{defg}
e^{-\mathcal{S}(g)}=Z(g) \leftrightarrow \mathcal{S}(g)=-\log Z(g).
\end{equation}
 As mentioned, $Z(g)$ can be thought of as related to a functional integral in which we integrate the action of a generic CFT  over a field $(\phi)$, or in general, a collection of fields,  in a background metric $g_{\mu\nu}$, 
\begin{equation} 
\label{induced}
Z(g)=N \int D\phi e^{-S_0(g,\phi)},   \qquad Z(\eta)=1.
\end{equation}
Its logarithm, $\mathcal{S}(g)$, is our definition of the effective action. $S_0(g,\phi)$ is the classical action. As usual $Z(g)$, in the Feynman diagrammatic expansion, will contain both connected and disconnected graphs, while $\mathcal{S}(g)$ collects only connected graphs. It is easy to verify that this 
collection corresponds also to 1PI (1 particle irreducible) graphs only in the case of free field theories embedded in external (classical) gravity.  \\
The emergence of bilinear mixing on the external graviton lines, as we are going to realize at the end of our analysis, should then be interpreted as a dynamical response 
of the theory, induced by the process of renormalization, with the generation of an intermediate dynamical degree of freedom propagating with the $1/\square$ operator. For this reason, the presence of such terms does not invalidate the 1PI nature of this functional.  

As a reference for our discussion, as already mentioned, we may assume that $S_0(\phi,g)$ describes, for instance, a free scalar field $\phi$ in a generic background. The action, in  this case, is given by 
\bea
\label{phi}
S_0(g,\phi)&=&\frac{1}{2}\int\, d^dx\,\sqrt{-g}\left[g^{\mu\nu}\nabla_\mu\phi\nabla_\nu\phi-\chi\, R\,\phi^2\right]
\eea
where we have included a conformal coupling $\chi(d)=\frac{1}{4}\frac{(d-2)}{(d-1)}$, and $R$ is the scalar curvature.
 This choice of $\chi(d)$ guarantees the conformal invariance of this action in $d$ dimensions as well as introduces a term of improvement for the stress-energy tensor in the flat limit, which becomes symmetric and traceless. A general perturbative analysis of this term and its role in the renormalization of the stress energy tensor of the Standard Model, which requires a conformally coupled Higgs sector, can be found in 
 \cite{Coriano:2011zk,Coriano:2011ti}.\\
Dimensional counting on this action can be performed, as usual, in two ways, either by considering the canonical dimension of the fields - as determined by their kinetic term in the Lagrangian - 
or by their response to Weyl  transformations. In the first case we have 
\begin{equation}
x_\mu \to \lambda x_\mu \qquad  g_{\mu\nu} \to  g_{\mu\nu} \qquad  \phi\to \lambda^{(1-d/2)} \phi.
\end{equation}
The dimensional counting for the Weyl scaling proceeds differently. The scaling  of a length scale $ds^2$ ($ds^2\to \lambda^2 ds^2$), is accounted for only by the metric 
\begin{equation}
\label{WW}
g_{\mu\nu}\to\lambda^2 g_{\mu\nu}, 
\end{equation}
while keeping the same scale dimensions of the scalar field, and one obtains 
\beq
  x_\mu\to x_\mu \qquad  \sqrt{g}\to \lambda^d\sqrt{g} \qquad  \phi\to \lambda^{1-d/2} \phi.
 \eeq
 The locality of the transformation is ensured with $\lambda=e^{\sigma(x)}$, for an arbitrary function $\sigma(x)$.\\
 $\mathcal{S}(g)$, in this example, is the functional that collects correlation functions with multiple insertions of the stress energy tensor
  \bea
 \label{defT}
T^{\mu\nu}_{scalar}
&\equiv&\frac{2}{\sqrt{g}}\frac{\delta S_0}{\delta g_{\mu\nu}}\nonumber \\
&=&\nabla^\mu \phi \, \nabla^\nu\phi - \frac{1}{2} \, g^{\mu\nu}\,g^{\alpha\beta}\,\nabla_\alpha \phi \, \nabla_\beta \phi
+ \chi \bigg[g^{\mu\nu} \Box - \nabla^\mu\,\nabla^\nu + \frac{1}{2}\,g^{\mu\nu}\,R - R^{\mu\nu} \bigg]\, \phi^2 .
\eea
the free field theory realization stays conformal at quantum level, modulo the appearance  of an anomaly in even dimensions. In the interacting case, things turn out to be different. Anomalous dimensions appear as soon as we switch-on a marginal interaction in \eqref{phi}.\\ 
For instance,  a $\lambda_0 \phi^4/4!$ potential will induce anomalous dimensions and a $\beta$ 
function which in flat space $(d=4-2 \epsilon)$ take the form
 \begin{eqnarray}
\Delta_\phi &=&\frac{d-2}{2} +\frac{\lambda_0^2}{12 (4 \pi)^4} +O(\lambda_0^3)\nonumber \\
\beta(\lambda_0)&=&- 2 \epsilon \lambda_0 +3 \frac{\lambda_0^2}{(4 \pi)^2} +O(\lambda_0^3),
\end{eqnarray}
causing a breaking of the classical conformal symmetry. For this reason we stick to the $\lambda_0=0$ case. \\
$\mathcal{S}[g]$ can be computed in a generic background, but its structure and the conditions that it has to satisfy vary considerably with the choice of $g$.\\
If we consider a generic metric background, there are significant changes on the structure of the CWIs. The flat spacetime limit corresponds to the simplest nontrivial case. Results obtained in this case, using the gravitational formulation of $\mathcal{S}[g]$, 
are equivalent to those obtained for ordinary conformal field theory in flat space, which can be naturally defined without any reference to gravity. \\
It is also clear, from this correspondence,  that the renormalization of these theories, in the flat limit, should involve only counterterms of mass dimension four.\\
In the most general case, one can derive CWI's in backgrounds of the form ($\overline{g}_{\mu\nu}$), with $g_{\mu\nu}=\overline{g}_{\mu\nu} +\delta g_{\mu\nu}$ ($\delta g^{\mu\nu}= h^{\mu\nu}$), and the formalism that we are going to discuss needs to be extended in order to encompass also this scenario. \\
In general, the variation \eqref{WW} on the functional $\sm(g)$ takes to the relation 
\begin{equation}
\label{anomx}
\frac{\delta \sm}{\delta \sigma(x)}=\sqrt{g} \,g_{\mu\nu}\,\langle T^{\mu\nu}\rangle, 
\qquad \langle T^{\mu\nu}\rangle = \frac{2}{\sqrt{g}}\frac{\delta \sm}{\delta g_{\mu\nu}}
\end{equation} 
and its invariance under Weyl  \begin{equation}
\delta_\sigma g_{\mu\nu}= 2 \sigma g_{\mu\nu} 
\end{equation}
and diffeomorphisms   
\begin{equation} 
\delta_\epsilon g_{\mu\nu}=-\nabla_{\mu}\epsilon_{\nu}- \nabla_{\nu}\epsilon_{\mu}, 
\end{equation}
summarised by the relations 
\begin{equation} 
\label{eww}
\delta_\sigma \sm=0 \qquad \delta_\epsilon \sm=0,
\end{equation}
take to trace and conservation conditions of the quantum averages of $T^{\mu\nu}$
 \begin{equation}
\label{comby}
\langle T^\mu_\mu\rangle=0 \qquad \qquad \nabla_\mu\langle T^{\mu\nu}\rangle=0.
\end{equation}
Trace and conformal WI's can be derived from the equations above by functional differentiations of $\sm(g)$ with respect to the metric background. However, the relations above are modified in the presence of an anomaly.\\
 The anomalous trace WIs can be derived by allowing for an anomaly contribution on the rhs of the $\sigma$ variation in \eqref{eww}
\begin{equation}
\label{plus}
\delta_\sigma \sm=\int d^4 x\sqrt{g}\,\sigma\,\bar{\mathcal{A}}(x) \qquad \langle T^\mu_\mu\rangle =\bar{\mathcal{A}}(x),
\end{equation}
which violates Weyl invariance. Here, $\sqrt{g}\, \bar{\mathcal{A}}(x)$ is the anomaly. Functional differentiations of this relation take to the hierarchy of trace WIs that we will be defining below.

The identification of the special CWIs in the formalism of the effective action, in the presence of an anomaly, is less straightforward compared to space, and can be addressed in this formulation by defining currents which are associated to symmetries of $\sm$. An example is the current 
\begin{equation} 
\label{confcur}
\langle J^\mu\rangle =\epsilon_\nu^{(K)} \langle T^{\mu\nu}\rangle, 
\end{equation}
where $\epsilon_\mu^{(K)}(x)$ is a Killing vector field of the metric $g$, which is conserved. The proof follows closely the classical geometric derivation of the conservation of such current. For this purpose, we recall that $\epsilon_\mu^{(K)}(x)$ characterizes the isometries of 
$g$
\begin{equation}
 (d s')^2=(d s)^2 \qquad \leftrightarrow  \qquad \nabla_\mu\epsilon_{\nu}^{(K)} + \nabla_\nu\epsilon_{\mu}^{(K)}=0.
\end{equation}
Then, the requirement of diffeomorphism invariance of the effective action $\sm(g)$, summarised by the second condition in \eqref{comby},
take to the conservation equation
\begin{equation}
\label{iso}
\nabla\cdot \langle J\rangle =0. 
\end{equation}
Such equation can be re-obtained  by requiring the invariance of $\sm$ under a specific variation respect to Killing vectors (KVs)  $\epsilon_\mu^{(K)}$ of the form
\begin{equation}
\delta_{KV}\sm=0.  
\end{equation}
If we require that the metric background allows conformal Killing vectors (CKVs) and the effective action is invariant under such transformations, then we have the possibility to take into account the anomaly contribution to the equations.  \\
We recall that the CKVs are solution of the equation
\begin{equation}
(ds')^2 = e^{2 \sigma(x)}(ds)^2 \qquad \leftrightarrow\qquad  \nabla_\mu \epsilon_\nu^{(K)} + \nabla_\nu \epsilon_\mu^{(K)} = 2 \sigma \delta_{\mu\nu}\qquad \sigma=\frac{1}{4}\nabla\cdot\epsilon^{(K)}.
\end{equation}
Notice that if we introduce a conformal current $J_c$, defined as in \eqref{confcur} but now using the CKVs of the background metric, if conditions \eqref{comby} are respected by $\sm$, then $J_c$ is conserved as in the isometric case \eqref{iso}. \\
Things are slightly different if we allow for a Weyl-variant term in $\sm$ as in \eqref{plus}, which takes place in $d=4$, after renormalization. \\
In this case the anomaly induces a nonzero trace, and modifies the semiclassical condition \eqref{iso} into the new form
\begin{equation}
\nabla\cdot \langle J_c\rangle =\frac{1}{4}\nabla \cdot \epsilon^{(K)} \langle T^\mu_\mu\rangle  +\epsilon^{(K)}_\nu \nabla_{\mu} \langle T^{\mu\nu}\rangle.
\end{equation}
Notice that this relation can still be imposed as a conservation equation on correlators of the form $J_c T$ in $d$ dimensions, since the anomaly contribution is induced only at $d=4$ while the stress energy tensor is always conserved. This will be the approach exploited by us and described in \secref{heres} for the derivation of the anomalous CWIs.\\
Notice that $\sigma(x)$ is, at the beginning, a generic scalar function, which in a Taylor expansion around a given point $x^\mu$ is characterized by an infinite and arbitrary number of constants. Their number gets drastically reduced if we require that the spacetime manifold with metric $g$ allows a tangent space at each of its points, endowed with a flat conformal symmetry. \\
Indeed, in flat space, the conformal Killing equation takes to CKVs $\epsilon^\mu$ which are at most quadratic in $x$, expressed in terms of the 15 parameters $(a^\mu,\omega^{\mu\nu}, \lambda_s, b^\mu)$ of the conformal group, indicated as $K^\mu(x)$ 
\begin{equation}
\label{Kil}
\epsilon^{\mu}(x)\,\big|_{flat}\to K^\mu(x)= a^\alpha + \omega^{\mu\nu} x^\nu +\lambda_s x^\mu + b^\mu x^2 -2 x^\mu b\cdot x.
\end{equation}
Using such CKVs, the derivation of the special CWIs, following the approach of \cite{Coriano:2017mux}, can be performed directly in $d=4$, and takes to anomalous special CWIs. 

Coming to the definition of our correlators, in our conventions, n-point correlation functions will be defined as
\begin{equation}
\label{exps1}
\langle T^{\mu_1\nu_1}(x_1)\ldots T^{\mu_n\nu_n}(x_n)\rangle \equiv\frac{2}{\sqrt{g_1}}\ldots \frac{2}{\sqrt{g_n}}\frac{\delta^n \sm(g)}{\delta g_{\mu_1\nu_1}(x_1)\delta g_{\mu_2\nu_2}(x_2)\ldots \delta g_{\mu_n\nu_n}(x_n)} 
\end{equation}
with $\sqrt{g_1}\equiv \sqrt{|\textrm{det} \, g_{{\mu_1 \nu_1}}(x_1)} $ and so on. \\
$\sm(g)$ in \eqref{defg} collects all the connected contributions of the correlation functions in the expansion respect to the metric fluctuations, and may as well be expressed in a covariant expansion as 
\begin{equation}
\label{exps2}
\sm(g)=\sm(\bar{g})+\sum_{n=1}^\infty \frac{1}{2^n n!} \int d^d x_1\ldots d^d x_n \sqrt{g_1}\ldots \sqrt{g_n}\,\langle T^{\mu_1\nu_1}\ldots \,T^{\mu_n\nu_n}\rangle_{\bar{g}}\delta g_{\mu_1\nu_1}(x_1)\ldots \delta g_{\mu_n\nu_n}(x_n).
\end{equation}
Diagrammatically, for a scalar theory in a flat background, it takes the form
\begin{align}
\label{figg}
\sm(g)=& \sum_n \quad\raisebox{-8.5ex}{{\includegraphics[width=0.19\linewidth]{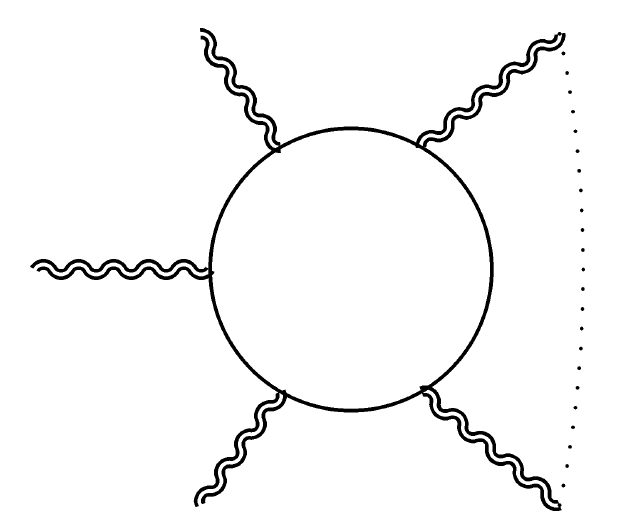}}} \,\scriptstyle \text{(n-point)}
\end{align}
where the external weavy lines represent gravitational fluctuations, in terms of contributions that are classified as tadpoles, 2-point, 3-point and n-point correlation functions of stress energy tensors. Tadpoles are removed in DR in flat space, and the sum in \eqref{figg} starts from 2-point functions. The anomaly contributions to the trace of the correlators start from 3-point functions.\\ 
Expression \eqref{figg} needs to be renormalized by suitable counterterms. In a free-field theory realization the expansion covers only one-loop diagrams and the origin of the anomaly lies entirely in the renormalization of such  contributions. The structure of the counterterms is crucial in order to identify its renormalized expression.\\
 \subsection{The expansion in a general background}
The correlation functions collected into $\sm(g)$  are obviously affected by contact terms, which can be clearly identified. These are generated from the definition of the nT's  correlators - due to the inclusion of the $\sqrt{g}/2$ prefactor in \eqref{defT} - but also from the quantum averages of multiple functional derivatives of the stress energy tensor, that we are going to classify.\\ We generalize the 
example discussed above, by turning to the expansion of a generic functional of the metric, still denoted as $\sm(g)$, now for a generic CFT.\\
 $\sm(g)$, as already mentioned, collects all the nT's multi-point functions, which are hierarchically constrained by the symmetry of the corresponding CFT. The expansion, similarly to the free-field theory example in \eqref{figg}, also in this case can be 
conveniently thought as generated by the integration of a conformal theory with a fundamental action $S$.  
  
In order to simplify the index structure of the equations, we will introduce some notations. For instance we define  
\begin{equation} 
\frac{\delta^n Z(g)}{\delta g_1\ldots  \delta g_n}\equiv\frac{\delta^n Z(g)}{\delta g_{ \mu_1\nu_1}(x_1) \ldots  \delta g_{ \mu_n\nu_n}(x_n) },
\end{equation}
to indicate the functional derivatives of the partition functions $Z$, as well for the effective action $\sm(g)$. 
As mentioned, since the conformal matter $\phi$ in \eqref{defg} and \eqref{induced} is integrated out, it is quite obvious that no scale should appear in all the quantum averages. If a specific scale pops out, it will be automatically related to the renormalization  of $\sm$, inducing a violation of the Weyl symmetry.  \\
For the moment we keep our background generic and investigate the structure of the correlation functions and of their contact terms.

The usual relation between $Z$ and $\sm$ in \eqref{defg} gives, in the case of the 3-point correlator (3T) the expansion
\begin{align}
\label{gamma3}
\frac{\delta^3 \sm }{\delta g_1\delta g_2\delta g_3 }=&2 \left\langle \frac{{\delta\,S_0}}{\delta g_1}\right\rangle  \left\langle \frac{\text{$\delta	S_0$}}{\delta g_2}\right\rangle  \left\langle \frac{{\delta\,S_0}}{\text{$\delta	$g}_3}\right\rangle
+\left\langle \frac{\delta ^3 S_0}{\delta g_1\delta g_2\text{$\delta
		$g}_3}\right\rangle 
-\left\langle \frac{{\delta\,S_0}}{\delta g_3}\right\rangle  \left\langle
\frac{{\delta\,S_0}}{\delta g_1} \frac{{\delta\,S_0}}{\text{$\delta
		$g}_2}\right\rangle +\left\langle \frac{{\delta\,S_0}}{\delta g_1} \frac{\text{$\delta
		S_0$}}{\delta g_2} \frac{{\delta\,S_0}}{\delta g_3}\right\rangle
\nonumber \\
&  +\left\langle
\frac{{\delta\,S_0}}{\delta g_3}\right\rangle  \left\langle \frac{\delta
	^2 S_0}{\delta g_1\delta g_2}\right\rangle -\left\langle \frac{\text{$\delta
		S_0$}}{\delta g_3} \frac{\delta ^2 S_0}{\delta g_1\delta g_2}\right\rangle + \text{sym}.
\end{align}

In the case of the 4-point correlator, which will be relevant for our discussion, we have a similar expansion, which in terms of the  averages of the classical action and its derivatives takes the form

\begin{eqnarray}
\frac{\delta^4 \sm}{\delta g_1\delta g_2\delta g_3\delta g_4}&=&6 \left\langle \frac{{\delta\,S_0}}{\delta g_1}\right\rangle 
\left\langle \frac{\text{$\delta
		S_0$}}{\delta g_2}\right\rangle 
\left\langle \frac{\text{$\delta
		S_0$}}{\delta g_3}\right\rangle 
\left\langle \frac{\text{$\delta
		S_0$}}{\delta g_4}\right\rangle
-2 \left\langle \frac{\text{$\delta
		S_0$}}{\delta g_3}\right\rangle
\left\langle \frac{\text{$\delta
		S_0$}}{\delta g_4}\right\rangle 
\left\langle \frac{\text{$\delta
		S_0$}}{\delta g_1}
\frac{\text{$\delta
		S_0$}}{\delta g_2}\right\rangle 
+\left\langle \frac{\text{$\delta
		S_0$}}{\delta g_4}\right\rangle 
\left\langle \frac{\text{$\delta
		S_0$}}{\delta g_1}
\frac{\text{$\delta
		S_0$}}{\delta g_2}
\frac{\text{$\delta
		S_0$}}{\delta g_3}\right\rangle \nonumber \\
&&    +\left\langle \frac{\text{$\delta
		S_0$}}{\delta g_1}
\frac{\text{$\delta
		S_0$}}{\delta g_4}\right\rangle 
\left\langle \frac{\text{$\delta
		S_0$}}{\delta g_2}
\frac{\text{$\delta
		S_0$}}{\delta g_3}\right\rangle
-\left\langle \frac{\text{$\delta
		S_0$}}{\delta g_1}
\frac{\text{$\delta
		S_0$}}{\delta g_2}
\frac{\text{$\delta
		S_0$}}{\delta g_3}
\frac{\text{$\delta
		S_0$}}{\delta g_4}\right\rangle
+2 \left\langle \frac{\text{$\delta
		S_0$}}{\delta g_3}\right\rangle 
\left\langle \frac{\text{$\delta
		S_0$}}{\delta g_4}\right\rangle 
\left\langle \frac{\delta
	^2S_0}{\text{$\delta
		$g}_1\delta g_2}\right\rangle \nonumber \\
&&  -\left\langle \frac{\text{$\delta
		S_0$}}{\delta g_4}\right\rangle 
\left\langle \frac{\text{$\delta
		S_0$}}{\delta g_3}
\frac{\delta ^2 S_0}{\text{$\delta
		$g}_1\delta g_2}\right\rangle
-\left\langle \frac{\delta
	^2 S_0}{\text{$\delta
		$g}_1\delta g_2}\right\rangle 
\left\langle \frac{\text{$\delta
		S_0$}}{\delta g_3}
\frac{\text{$\delta
		S_0$}}{\delta g_4}\right\rangle
+\left\langle \frac{\text{$\delta
		S_0$}}{\delta g_3}
\frac{\text{$\delta
		S_0$}}{\delta g_4}
\frac{\delta ^2 S_0}{\text{$\delta
		$g}_1\delta g_2}\right\rangle \nonumber \\
&& +\left\langle \frac{\delta
	^2 S_0}{\text{$\delta
		$g}_1\delta g_4}\right\rangle 
\left\langle \frac{\delta
	^2 S_0}{\text{$\delta
		$g}_2\delta g_3}\right\rangle
-\left\langle \frac{\delta
	^2S}{\text{$\delta
		$g}_1\delta g_4}
\frac{\delta ^2 S_0}{\text{$\delta
		$g}_2\delta g_3}\right\rangle
+\left\langle \frac{\text{$\delta
		S_0$}}{\delta g_4}\right\rangle 
\left\langle \frac{\delta
	^3 S_0}{\text{$\delta
		$g}_1\text{$\delta
		$g}_2\delta g_3}\right\rangle
\nonumber \\
&& -\left\langle \frac{\text{$\delta
		S_0$}}{\delta g_4}
\frac{\delta ^3 S_0}{\text{$\delta
		$g}_1\text{$\delta
		$g}_2\delta g_3}\right\rangle +\left\langle \frac{\delta
	^4 S_0}{\text{$\delta
		$g}_1\text{$\delta
		$g}_2\text{$\delta
		$g}_3\delta g_4}\right\rangle  + \text{sym},\nonumber \\
\end{eqnarray}

 where the symmetrization is performed only on those terms which are not explicitly symmetric. For instance we have
 \begin{equation}
   \left\langle \frac{{\delta
   S_0}}{{\delta g}_3}\right\rangle 
   \left\langle \frac{{\delta
   S_0}}{{\delta g}_4}\right\rangle 
   \left\langle \frac{{\delta
   S_0}}{{\delta g}_1}
   \frac{{\delta
   S_0}}{{\delta g}_2}\right\rangle \equiv \langle 3,4,12\rangle \to  \langle 3,4,12\rangle + \langle 1,2,34\rangle  + 
    \langle 2,4,13\rangle +  \langle 2,3,14\rangle  + \langle 1,4,23\rangle + \langle 1,3,24\rangle
   \end{equation}
   and so on. They can be re-expressed in terms of correlators of lower orders (2- and 3- point functions) and of contact interactions. 
Diagrammatically, the expansion of the 4T correlator of $\sm$ takes the form 

\begin{align}
\label{darker}
\left(\frac{\delta^4 \sm }{{\delta g_1\delta g_2}\delta g_3\delta g_4 }\right)&=6\,\raisebox{-3ex}{{\includegraphics[width=0.19\linewidth]{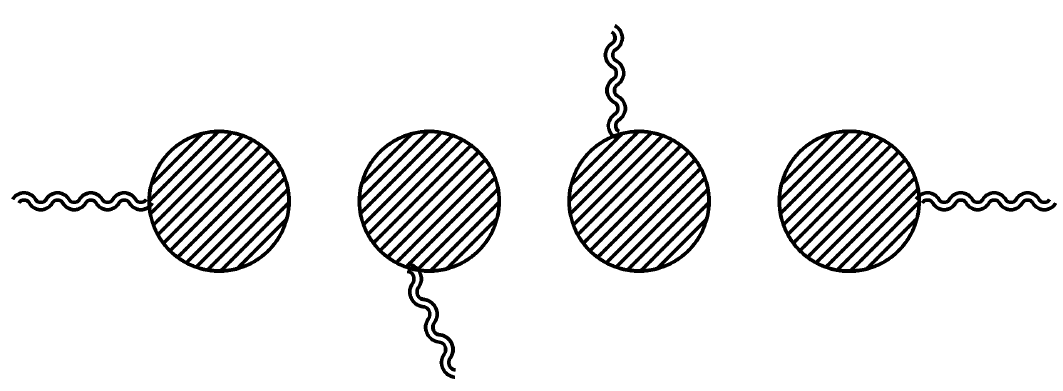}}}-2\,\raisebox{-2.5ex}{{\includegraphics[width=0.17\linewidth]{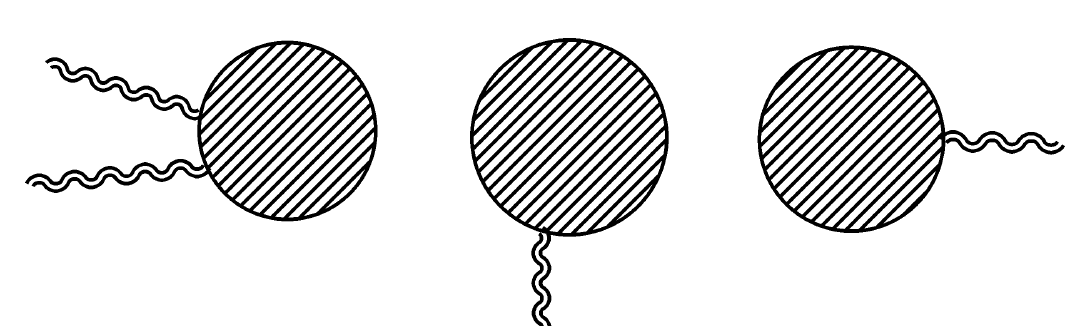}}}+\raisebox{-2.5ex}{{\includegraphics[width=0.12\linewidth]{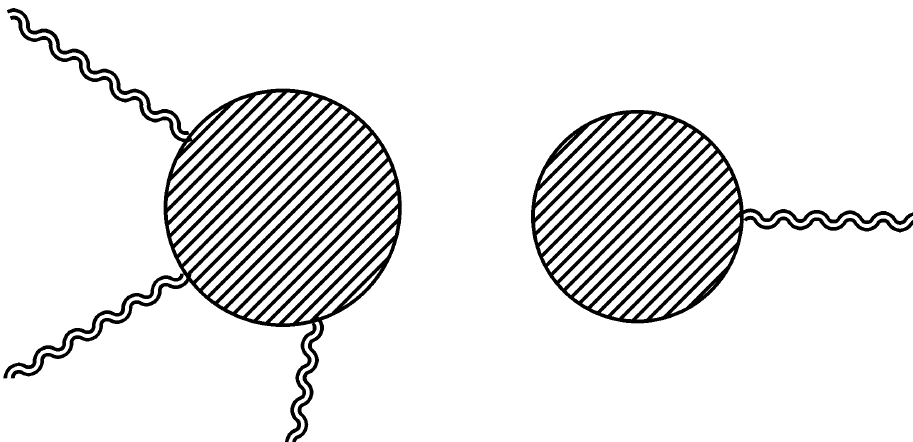}}}\notag\\
&+\raisebox{-1.5ex}{{\includegraphics[width=0.10\linewidth]{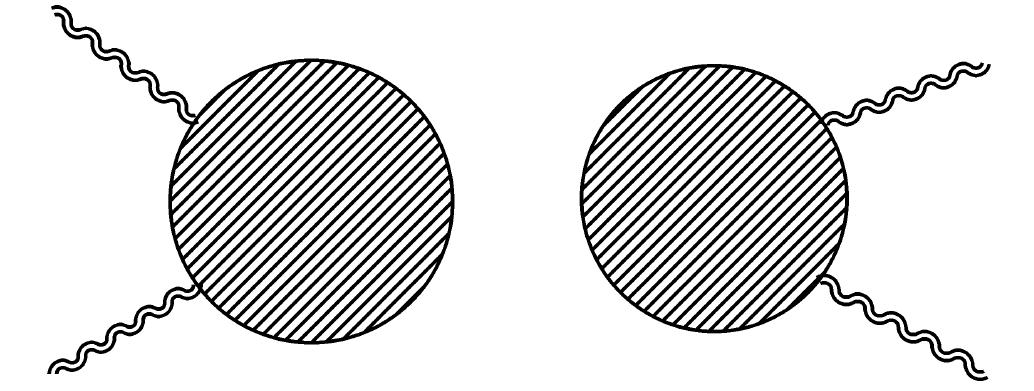}}}-\raisebox{-2.2ex}{{\includegraphics[width=0.09\linewidth]{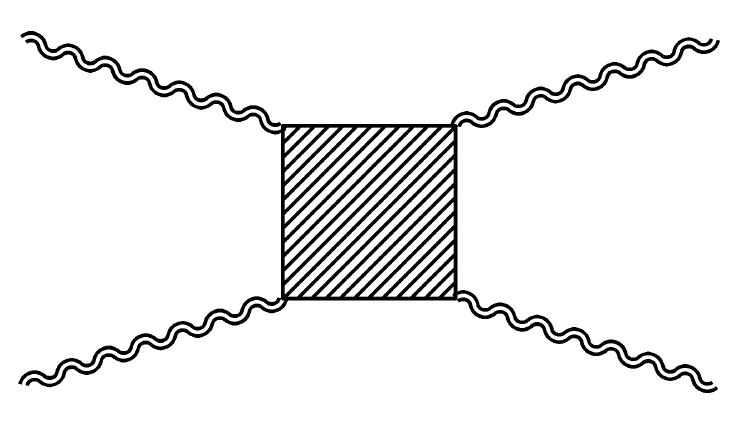}}}+2\raisebox{-1.4ex}{{\includegraphics[width=0.14\linewidth]{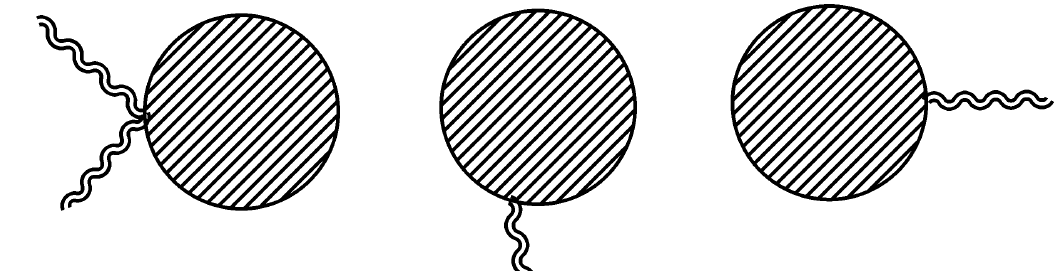}}}-\raisebox{-2.4ex}{{\includegraphics[width=0.14\linewidth]{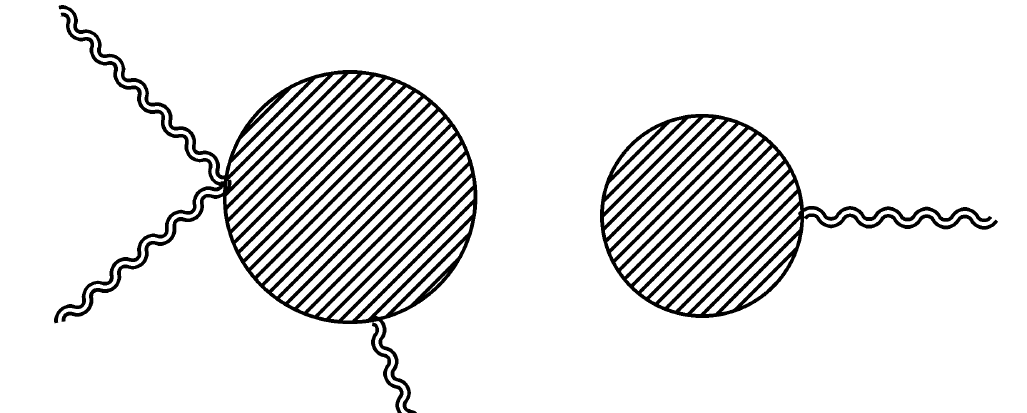}}}\notag\\
&-\raisebox{-1ex}{{\includegraphics[width=0.1\linewidth]{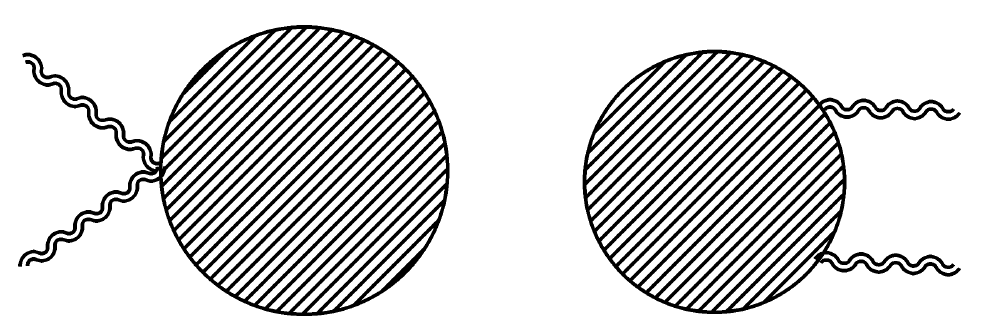}}}+\raisebox{-2ex}{{\includegraphics[width=0.06\linewidth]{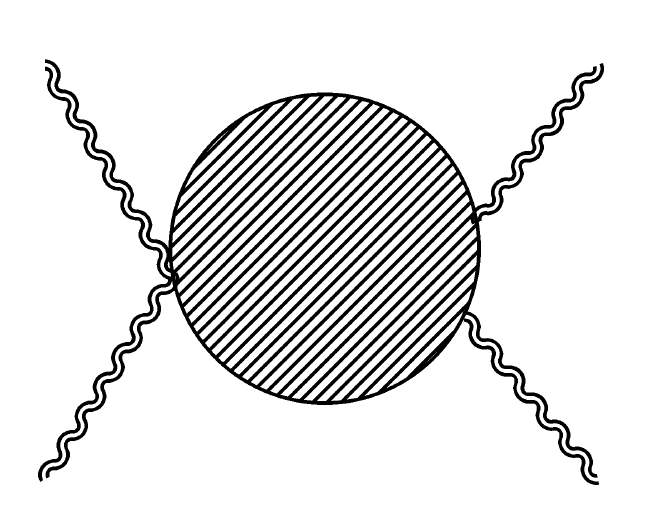}}}+\raisebox{-1.4ex}{{\includegraphics[width=0.10\linewidth]{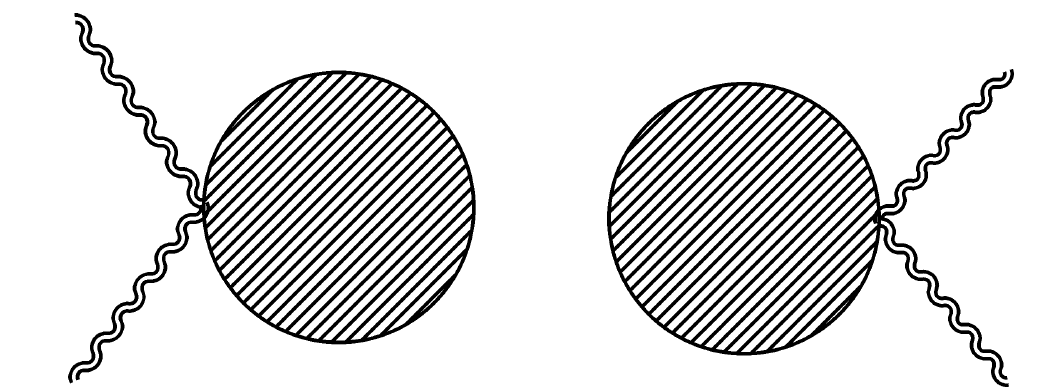}}}-\raisebox{-2.ex}{{\includegraphics[width=0.06\linewidth]{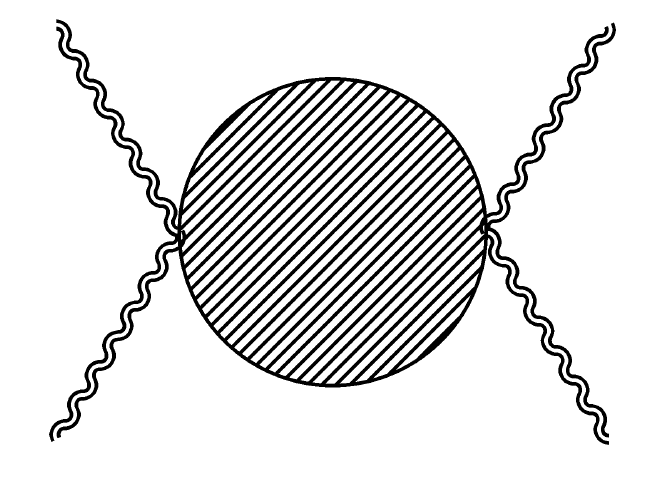}}}
+\raisebox{-1.3ex}{{\includegraphics[width=0.1\linewidth]{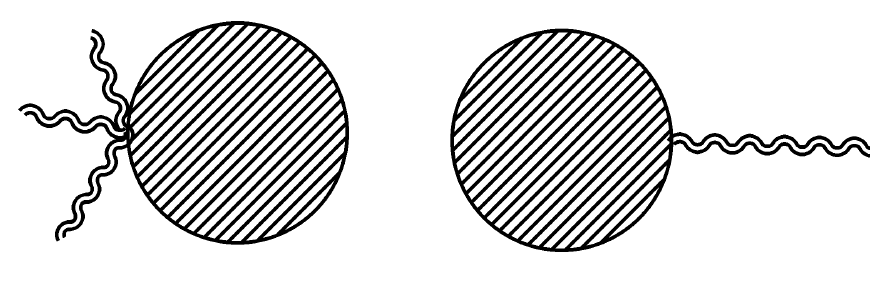}}}\notag\\&
-\raisebox{-1.8ex}{{\includegraphics[width=0.08\linewidth]{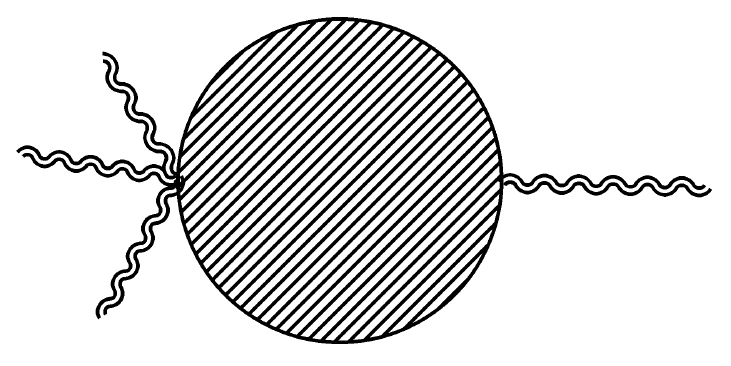}}}+\raisebox{-2ex}{{\includegraphics[width=0.07\linewidth]{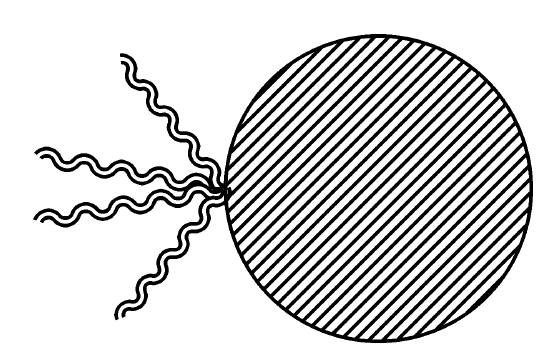}}}+
\textrm{sym}
\end{align}
where the averages correspond to direct insertions of ${\delta S}/{\delta g}$ operators for each separate external graviton vertex. Notice that the "${S}$" correlators, which contain insertions of single and multiple derivatives of the fundamental action, do not correspond, according to the definition \eqref{exps1}, to insertions of stress energy tensors, but differ from the latter by a certain number of contact terms, that we intend to identify.\\
It is convenient, for this purpose, to introduce some condensed notation, in order to avoid, at least in part, the proliferation of indices in the analysis below. For instance, the relation
\begin{equation}
\frac{\delta\sqrt{ g(x_a)}}{\delta g_{\mu_b\nu_b}(x_b)}=\frac{1}{2}\sqrt{g(x_a)}\,g^{\mu_b\nu_b}(x_a)\delta^d(x_a - x_b), 
\end{equation}
will be summarized below in the expression 
\begin{equation}
\frac{\delta \sqrt{dg(a)}}{\delta g(b)}=\frac{1}{2}\sqrt{dg(a)}\,g(a)\delta(a,b), 
\end{equation}
where $dg(a)\equiv g(x_a)$ denotes the metric determinant at point $x_a$, $(a=1,2,3,4)$ and 
$\delta(a,b)\equiv \delta(x_a-x_b)$.\\
 Differentiation is performed, in our conventions, with respect to the metric with covariant indices $(g_{\mu_a\nu_a})$ on the left-hand side of the equations, but the result on the right-hand side has all the indices contravariant, e.g. 
$g(a)\equiv g^{\mu_a\nu_a}$, $T(a)\equiv T^{\mu_a\nu_a}$. \\
Contact terms are generated either by differentiation of the metric, as above, or by multiple differentiation of the action $S$. Using the notation above, the stress energy tensor $T$ is related to $S$ in the form
\begin{equation}
\label{TS}
\left\langle \frac{\delta S_0}{\delta g(a)}\right\rangle =\frac{1}{2} \sqrt{\text{dg}(a)} \langle
   T(a)\rangle.
\end{equation}
and by differentiating with respect to the metric this relation, we obtain the identity
\begin{equation}
\label{2S}
\left\langle\frac{\delta^2 S_0 }{\delta g(b)\delta g(a) }\right\rangle=\frac{1}{4} \sqrt{\text{dg}(b)} g(b) \delta (b,a) \langle T(a)\rangle +\frac{1}{2}
\sqrt{\text{dg}(a)} \left\langle \frac{\text{$\delta $T}(a)}{\text{$\delta
		$g}(b)}\right\rangle, 
\end{equation}
 where the differentiation of the stress energy tensor $\left\langle
   \frac{\text{$\delta $T}(b)}{\text{$\delta
   $g}(a)}\right\rangle\sim c(g)\delta(a,b)$ introduces one contact term. Contact terms of higher orders are present in the other derivatives, for example
\begin{align}
&\left\langle\frac{\delta^3 S_0 }{\delta g(c)\delta g(b)\delta g(a) }\right\rangle=\frac{1}{4} \sqrt{\text{dg}(b)} g(b) \delta (b,a) \left\langle \frac{\text{$\delta
		$T}(a)}{\delta g(c)}\right\rangle +\frac{1}{4} \sqrt{\text{dg}(c)} g(c) \delta
(c,a) \left\langle \frac{\text{$\delta $T}(a)}{\delta g(b)}\right\rangle
\nonumber \\[1.5ex]
&+\frac{1}{2} \sqrt{\text{dg}(a)} \left\langle \frac{\delta ^2 T(a)}{\delta g(b)
	\delta g(c)}\right\rangle+\frac{1}{8} g(b) \sqrt{\text{dg}(c)} g(c) \delta (b,a) \delta (c,b) \langle T(a)\rangle
+\frac{1}{4} \sqrt{\text{dg}(b)} \delta (b,a) \delta (c,b) s(c,b) \langle T(a)\rangle,
\end{align}
where 
\begin{equation} \label{S come somma di g}
s(a,b)\equiv S^{\mu_a\nu_a\mu_b\nu_b} (x) \equiv
-\frac{1}{2} \Big(g^{\mu_a\mu_b} (x) g^{\nu_a\nu_b}(x)+g^{\mu_a\nu_b} (x) g^{\nu_a\mu_b}  (x) \Big),
\end{equation}
and
\begin{align}\label{S derivative}
 s(c,a,b)&\equiv\dfun{ S^{\mu_a \nu_a \mu_b \nu_b} (x)}{g_{\alpha_c \beta_c} (y)} \notag\\
 &= -\frac{1}{2}\lt 
\dfun{g^{\mu_a \mu_b}(x) }{g_{\alpha_c \beta_c}(y)} g^{\nu_a \nu_b}(x) +
\dfun{g^{\nu_a \nu_b}(x) }{g_{\alpha_c \beta_c}(y)} g^{\mu_a \mu_b}(x) +
\dfun{g^{\mu_a \nu_b}(x) }{g_{\alpha_c \beta_c}(y)} g^{\nu_a \mu_b}(x) +\dfun{g^{\nu_a \mu_b}(x) }{g_{\alpha_c \beta_c}(y)} g^{\mu_a \nu_b}(x) \rt \notag\\
&=-\frac{1}{2} \de^d(x-y) \big(
S^{\alpha \beta \mu_1 \mu_2} g^{\nu_1 \nu_2}+
S^{\alpha \beta \nu_1 \nu_2} g^{\mu_1 \mu_2}+
S^{\alpha \beta \mu_1 \nu_2} g^{\nu_1 \mu_2}+
S^{\alpha \beta \nu_1 \mu_2} g^{\mu_1 \nu_2}\big). 
\end{align}
Contact terms generated by higher derivatives of the fundamental action can be found in \appref{contact}.\\
Correlators with multiple $T$'s and those with ${S}$-insertions are then related by contact terms involving expectation values of a single $T$ and of their derivatives. For instance, for 2-point functions, the relation 
between the ${S}$ and T correlators, the latter defined by \eqref{exps1}, takes the form    
\begin{align}
&\left\langle \frac{\delta S_0}{\delta g_{\mu_1\nu_1}(x_1)} \frac{\delta S_0}{\delta g_{\mu_2\nu_2}(x_2)}\right\rangle =\frac{\sqrt{g (x_1)}}{2}\frac{\sqrt{g(x_2)}}{2}\langle T^{\mu_1\nu_1}(x_1)T^{\mu_2\nu_2}(x_2)\rangle +\frac{1}{2}  \sqrt{g(x_1)}\left\langle \frac{\delta T^{\mu_1\nu_1}(x_1)}{\delta g_{\mu_2\nu_2}(x_2)}\right\rangle\notag\\[1.2ex]
&\qquad+ \frac{\sqrt{g(x_1)}}{2}\frac{\sqrt{g(x_2)}}{2}\langle T^{\mu_1\nu_1}(x_1)\rangle 
\langle T^{\mu_2\nu_2}(x_2)\rangle + \frac{1}{4}\sqrt{g(x_1)}g^{\mu_2\nu_2}(x_2)\delta(x_1-x_2) \langle T^{\mu_1\nu_1}(x_1)\rangle,
\end{align}
where we have used the tadpole relation \eqref{2S}. The expression above shows that the 2-point function of stress energy tensors - defined as $\langle TT\rangle $ -  and the 
result obtained by direct S-insertions, agree if we are allowed to remove all the tadpoles from the expansion of the functional. \\
Since we will be defining the hierarchy of the CWI's in flat space, this is possible if we adopt a regularization scheme in which the tadpoles vanish. This requirement, in the context of a conformal theory, has a deep physical meaning and is 
regularization dependent, for being associated to a diverging quartic contribution, if the computation is performed covariantly with the inclusion of a UV cutoff $(\Lambda^4)$. 
Such cutoff dependence epitomizes the hierarchy problem for the cosmological constant, which is clearly hidden in DR, since massless tadpoles, in this regularization scheme, are set to zero.\\
 A recent analysis \cite{Donoghue:2020hoh} has concluded that this issue, even if the computation is performed covariantly and by a cutoff, can be ameliorated by the inclusion of extra singular interactions coming from the path integral measure, not considered before. \\
Since our analysis is performed in flat space, we will proceed  eliminating such contributions,  by adopting a DR scheme.  \\
There is a large list of such contributions which are removed from our analysis as $g\to \delta$, for a flat background. In this case we obtain
\begin{align}
\label{wick}
&\Bigg(\frac{\delta^4 \sm}{\delta g_1\delta g_2\delta g_3\delta g_4}\Bigg)=\notag\\
&=\left\langle \frac{\delta S_0}{\delta g_1}\frac{\delta S_0}{\delta g_4}\right\rangle \left\langle \frac{\delta S_0}{\delta g_2}
\frac{\delta S_0}{\delta g_3}\right\rangle-\left\langle \frac{\delta S_0}{\delta g_1}\frac{\delta S_0}{\delta g_2}
\frac{\delta S_0}{\delta g_3}\frac{\delta S_0}{\delta g_4}\right\rangle-\left\langle \frac{\delta^2S_0}{\delta g_1\delta g_2}\right\rangle \left\langle \frac{\delta S_0}{\delta g_3}\frac{\delta S_0}{\delta g_4}\right\rangle
+\left\langle \frac{\delta S_0}{\delta g_3}\frac{\delta S_0}{\delta g_4}\frac{\delta ^2 S_0}{\delta g_1\delta g_2}\right\rangle \nonumber \\
&\quad+\left\langle \frac{\delta^2 S_0}{\delta g_1\delta g_4}\right\rangle  \left\langle \frac{\delta^2 S_0}{\delta g_2\delta g_3}\right\rangle-\left\langle \frac{\delta^2 S_0}{\delta g_1\delta g_4}
\frac{\delta ^2 S_0}{\delta g_2\delta g_3}\right\rangle
-\left\langle \frac{\delta S_0}{\delta g_4}
\frac{\delta ^3 S_0}{\delta
		g_1\,\delta g_2\delta g_3}\right\rangle +\left\langle \frac{\delta^4 S_0}{\delta g_1\,\delta g_2\,\delta g_3\delta g_4}\right\rangle + \text{sym}.
\end{align}
 where all the correlators are computed in Euclidean space. The expression of the contact terms are traced back to the contributions
\begin{align}
\left\langle \frac{\delta S_0}{\delta g(a)}\right\rangle &=\frac{1}{2} \left\langle
T(a)\right\rangle=0,\\
\left\langle \frac{\delta^2 S_0}{\delta g(a)\delta g(b)}\right\rangle &=\frac{1}{2}
\left\langle
\frac{\delta T(b)}{\delta g(a)}\right\rangle,\\
\langle \frac{\delta^3 S_0}{\delta g(a)\delta g(b)\delta g(c)} \rangle&= \frac{1}{4} \delta ^c \delta (c,a) \left\langle \frac{\delta T(a)}{\delta g(b)}\right\rangle +\frac{1}{4} \delta ^b \delta (b,a) \left\langle \frac{\delta T(a)}{\delta g(c)}\right\rangle +\frac{1}{2} \left\langle \frac{\delta ^2
	T(a)}{\delta g(b) \delta g(c)}\right\rangle,
\end{align}
where $\delta^a\equiv \delta^{\mu_a\nu_a}$, obtained by the flat limit $g(a)\to \delta^a$, 
and in a general background these contact terms would be related to additional conformal data, not present in our analysis.\\
 Notice that \eqref{wick}, in a diagrammatic expansion, defines a connected 
functional if we resort to free field theory and allow the ordinary Wick contractions of the corresponding operators on its rhs to take place. \\
Indeed, in the case of the scalar free field theory presented above, the topological contributions coming from the 4-T  are just summarized by the vertices

\begin{align}
\label{lighter}
\sm_4[g]=\quad\raisebox{-4.9ex}{{\includegraphics[width=0.20\linewidth]{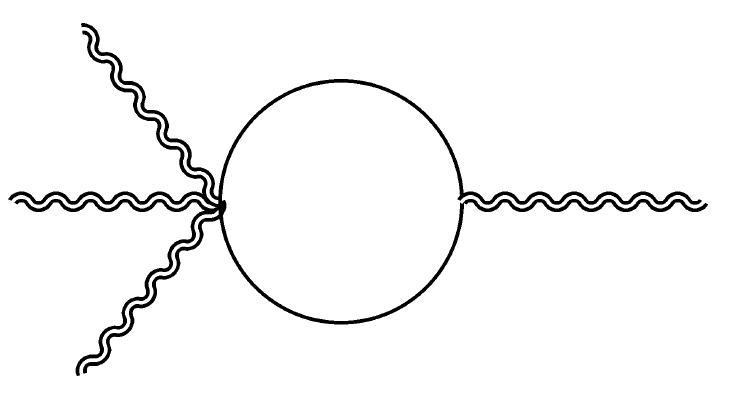}}}+\quad\raisebox{-4.5ex}{{\includegraphics[width=0.15\linewidth]{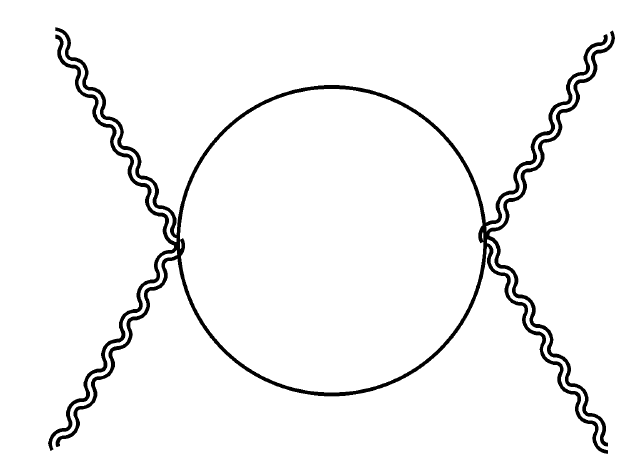}}}+\quad\raisebox{-3.5ex}{{\includegraphics[width=0.15\linewidth]{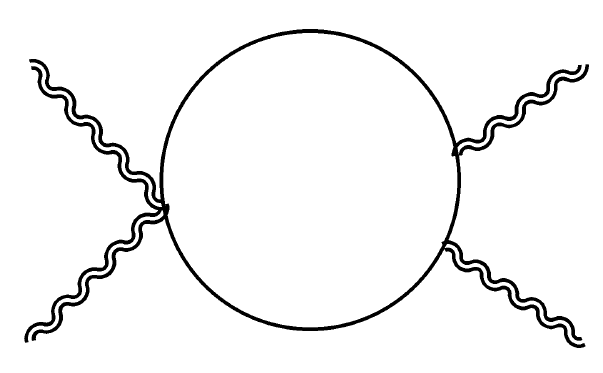}}}+\quad\raisebox{-6ex}{{\includegraphics[width=0.2\linewidth]{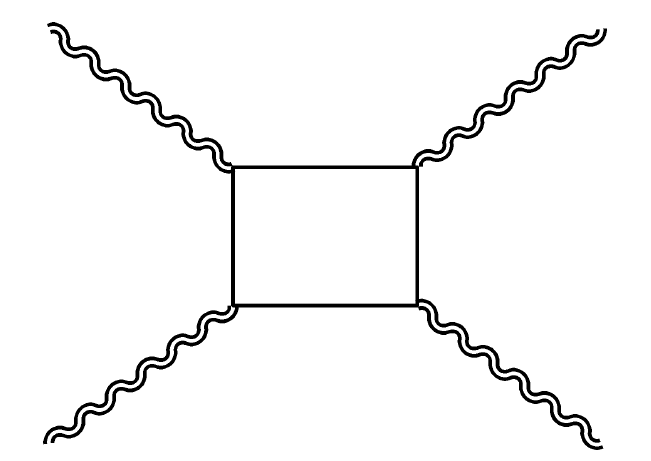}}}\nonumber \\
\end{align}
which can be directly computed in perturbation theory \cite{Serino:2020pyu}. If we indicate by $\langle\ldots \rangle_c$ the connected contributions to this expansion, then the expression of the 
4T in the flat limit, using DR, takes the form   
\begin{align}
\label{gamma4}
\left\langle T^{\mu_1\nu_1} T^{\mu_2\nu_2} T^{\mu_3\nu_3} T^{\mu_4\nu_4}\right\rangle=&
16\Bigg(-\left\langle \frac{\delta	S_0}{\delta g_1}\frac{\delta S_0}{\delta g_2}\frac{\delta S_0}{\delta g_3}\frac{\delta S_0}{\delta g_4}\right\rangle_c+\left\langle \frac{\delta S_0}{\delta g_3}\frac{\delta S_0}{\delta g_4}\frac{\delta ^2 S_0}{\delta g_1\delta g_2}\right\rangle_c \nonumber \\
&  -\left\langle \frac{\delta^2 S_0}{\delta g_1\delta g_4}\frac{\delta ^2 S_0}{\delta g_2\delta g_3}\right\rangle_c-\left\langle \frac{\delta S_0}{\delta g_4}\frac{\delta ^3 S_0}{\delta g_1\delta g_2\delta g_3}\right\rangle_c +\left\langle \frac{\delta^4 S_0}{\delta g_1\delta g_2\delta g_3\delta g_4}\right\rangle_c + \text{sym}\Bigg).
\end{align}
A similar result holds for the 3-point function 3T, from \eqref{gamma3} 
\begin{equation}
\label{gamma3p}
\langle T^{\mu_1\nu_1} T^{\mu_2\nu_2} T^{\mu_3\nu_3}\rangle =
8\left( \left\langle \frac{\delta\,S_0}{\delta g_1} \frac{\delta S_0}{\delta g_2} \frac{\delta\,S_0}{\delta g_3}\right\rangle_c -\left\langle \frac{\delta S_0}{\delta g_3} \frac{\delta ^2 S_0}{\delta g_1\delta g_2}\right\rangle_c + \text{sym}\right).
\end{equation}
Notice that \eqref{gamma4} and \eqref{gamma3p} are derived from 
a path integral realization of a certain CFT, in the flat spacetime limit. The condition of integrating out the conformal matter in the path integral  forces us to consider only one-loop terms generated by the ordinary Wick  contractions, accompanied by an arbitrary number of external graviton lines. This is equivalent to going from a general expansion
 \eqref{darker} to the simplified one \eqref{lighter}. \\
It is also clear that this result is specific for the flat background case. For instance, in the conformally flat (Weyl-flat) case where 
\begin{equation}
g_{\mu\nu}=e^{2 \sigma(x)}\eta_{\mu\nu}
\end{equation}
this expansion requires the inclusion of massless tadpoles. 
In particular, the expansion of the dominant contribution to $\sm$ would start with a single insertion of stress energy tensor contracted with the gravitational fluctuation $\sim \langle T\cdot h\rangle$, and would be entirely defined by the anomaly functional. The tadpoles would contribute to all orders in an expansion $h$. 
This is related to the fact that the entire stress energy tensor is proportional to the anomaly coefficient of the Gauss-Bonnet  term, as discussed in \cite{Brown:1977sj,
Herzog:2013ed}.   

\section{The counterterms}
If we expand around a flat metric background and rely on a  mass-independent regularization scheme, then the structure of the counterterms is simply polynomial in momentum space. \\
Indeed, in general, the breaking of Weyl invariance takes to the anomalous variation
\begin{equation}
 \delta_\sigma \sm=\frac{1}{(4 \pi)^2}\int d^4 x \sqrt{g}\,\sigma\,\left(c_1 R_{\mu\nu\rho\sigma}R^{\mu\nu\rho\sigma} + c_2 R_{\mu\nu}R^{\mu\nu} +c_3 R^2 + c_4\square R\right) 
\end{equation}
which is constrained by the Wess-Zumino consistency condition 
\begin{equation}
\label{WZ}
\left[\delta_{\sigma_1},\delta_{\sigma_2}\right]\sm=0
\end{equation}
to take the form 
\beq
\label{anom1}
\delta_\sigma \sm=\frac{1}{(4\pi)^2}\int d^4 x\sqrt{g}\,\sigma\left( b_1 C^{(4)}_{\mu\nu\rho\sigma}C^{(4)\mu\nu\rho\sigma} + b_2 E_{4} +b_3 \square R\right),
\eeq
given in terms of the dimension-4 curvature invariants
\begin{align}
\label{fourd}
E_4&\equiv R_{\mu\nu\alpha\beta}R^{\mu\nu\alpha\beta}-4R_{\mu\nu}R^{\mu\nu}+R^2  \\
( C^{(4)})^2&\equiv R_{\mu\nu\alpha\beta}R^{\mu\nu\alpha\beta}-2R_{\mu\nu}R^{\mu\nu}+\frac{1}{3}R^2, \label{fourd2}
\end{align}
which are the Euler-Gauss-Bonnet (GB) invariant and the square of the Weyl conformal tensor, respectively, in $d=4$. \\
We pause for few comments.
From now on, when referring to $C^2$ without any subscript, we will be indicating the $(C^{(d)})^2$ expression of this invariant, with a parametric dependence on $d$

\beq\label{Geometry1}
C^{(d) \alpha\beta\gamma\delta}C^{(d)}_{\alpha\beta\gamma\delta}
=
R^{\alpha\beta\gamma\delta}R_{\alpha\beta\gamma\delta} -\frac{4}{d-2}R^{\alpha\beta}R_{\alpha\beta}+\frac{2}{(d-2)(d-1)}R^2
\eeq
where
\beq
C^{(d)}_{\alpha\beta\gamma\delta} = R_{\alpha\beta\gamma\delta} -
\frac{2}{d-2}( g_{\alpha\gamma} \, R_{\delta\beta} + g_{\alpha\delta} \, R_{\gamma\beta}
- g_{\beta\gamma} \, R_{\delta\alpha} - g_{\beta\delta} \, R_{\gamma\alpha} ) +
\frac{2}{(d-1)(d-2)} \, ( g_{\alpha\gamma} \, g_{\delta\beta} + g_{\alpha\delta} \, g_{\gamma\beta}) R\, .
\eeq
The choice of $(C^{(d)})^2\equiv C^2$ instead of $(C^{(4)})^2$ in the counterterm action that we will define below, takes to variations which are deprived of total derivative ($\square R$) term in \eqref{anom1}. 
With such a choice of $(C^{(d)})^2$, one derives the relation
\begin{equation}
\frac{\delta}{(d-4)\delta \sigma (x) }\int d^d x \sqrt{-g} (C^{(d)})^2 =\sqrt{-g} (C^{(d)})^2,
\label{oneq}
\end{equation}
which differs from an analogous one
\begin{equation}
\frac{\delta}{(d-4)\delta \sigma (x) }\int d^d x \sqrt{-g} (C^{(4-\epsilon)})^2 =\sqrt{-g}\Bigg( (C^{(4)})^2
-\frac{2}{3}\square R\Bigg) 
\label{twoq}
\end{equation}
obtained by the replacement of $(C^{(d)})^2\to (C^{(4-\epsilon)})^2$ in the integrand, followed by an expansion of the parametric dependence on $\epsilon$, which induces a finite renormalization of the effective action. Explicitly
\beq
C^{(d)\alpha\beta\gamma\de}C_{(d)\alpha\beta\gamma\de}=R^{\alpha\beta\gamma\de}R_{\alpha\beta\gamma\de}-2R^{\alpha\beta}R_{\alpha\beta}+\frac{1}{3}R^2+\epsilon\lt -R^{\alpha\beta}R_{\alpha\beta}+\frac{5}{18}R^2 \rt,
\eeq
that is
\begin{equation}
\label{interm}
(C^{(4-\epsilon)})^2=(C^{(4)})^2 +\epsilon\left( - (R_{\mu\nu})^2 + \frac{5}{18}R^2\right).
\end{equation}
Using the property that the integration measure scales in $d$ dimensions as 
\begin{equation} 
g=\textrm{det} g_{\alpha\beta} =\epsilon^{\mu_1\ldots \mu_d}g_{0 \mu_1}\ldots g_{0\mu_d}\to e^{2 d \sigma} g 
\end{equation}
and the fact that both $C^{(d)}$ and $C^{(4)}$ carry the same Weyl scaling 
\begin{equation}  
C{}_{\lambda\mu\nu\rho}=e^{2 \sigma} \bar{C}_{\lambda\mu\nu\rho}\qquad C^{\lambda\mu\nu\rho}=e^{-6  \sigma} \bar{C}^{ \lambda\mu\nu\rho}
\end{equation}
 one obtains 
\begin{equation} 
\frac{\delta}{\delta \sigma(x)}\int d^d x \sqrt{-g}(C^{4)})^2(x)=\epsilon \sqrt{g}(C^{(4)})^2,
\end{equation}
as well as 
\begin{equation} 
\frac{\delta}{\delta \sigma(x)}\int d^d x \sqrt{-g}(C^{d)})^2(x)=\epsilon \sqrt{g}(C^{(d)})^2.
\end{equation}
Using
\begin{equation} 
\frac{\delta}{\delta \sigma(x)}\int d^d x \sqrt{-g}\left( - R_{\mu\nu}^2 + \frac{5}{18}R^2\right)=
-\frac{2}{3}\epsilon \sqrt{g} \Box R
\end{equation}
one finally derives the relation \eqref{twoq}, with the inclusion of a local (scheme dependent) term $\Box R$. \\The result shows that the expansion in the parametric dependence of $C^{(d)}$ on $d$, around $d=4$, and the Weyl variation do not commute. \\
The number of such invariants depends on the dimension. For instance, the case of $d=6$ is discussed in\cite{Bastianelli:2000hi}.\\
One can derive an anomaly action of the Wess-Zumino form starting from these invariants using the Weyl gauging approach \cite{Iorio:1996ad,Codello:2012sn}. An example, in the case of $d=6$, can be found in \cite{Coriano:2013nja}. 
However, the procedure followed in any approach based on the implementation of a Weyl-gauging, is quite different from the one formulated in this work, in which the anomaly action is directly identified from the anomalous CWIs in an expansion with respect to the fluctuations around a certain external background.\\
 Indeed, all the approaches based on Weyl gauging require the addition of one extra degree of freedom in the spectrum (a dilaton) and correspond to variational solutions of the anomaly equation \eqref{plus}.\\
 The result of that variational method is a local action, with extra interactions induced by the 
 dilaton field, whose structure depends on the spacetime dimensions. \\  
All of this is avoided in our approach, since we retain into $\sm(g)$ only the 
genuine constraints coming from  the anomalous CWI's, with no inclusion of any extra, intermediate  compensator field. \\
In generic (even) spacetime dimensions, the structure of the countertem Lagrangian, is modified accordingly, with the Euler/ Gauss-Bonnet  density, given by
\bea
\label{Ed}
E_d = \frac{ 1}{2^{d/2}}
\delta_{\mu_1 \cdots \mu_d}^{\nu_1 \cdots \nu_d}
{R^{\mu_1 \mu_2}}_{\nu_1 \nu_2} \cdots
{R^{\mu_{d-1} \mu_d}}_{\nu_{d-1} \nu_d} \ ,
\eea
which, for $d=4$, is quadratic in the curvatures, and is indeed given by \eqref{fourd}. 
 It is the latter, together with the other invariant $C^2$, the only extra, additional operator present in $d=4$, that has to be varied around $d=4$ in order to generate the counterterms needed for the renormalization of $\sm(g)$. We recall that the Gauss-Bonnet term satisfies the relation  

\begin{equation}
\label{form}
\frac{\delta}{(d-4)\delta \sigma (x) }\int d^d x \sqrt{-g} E_4 =\sqrt{-g} E_4.
\end{equation}
and it is topological only in $d=4$.\\
  From \eqref{oneq} and \eqref{form} we will be deriving a series of important functional constraints, as we are going to see next. 
In the expressions above, all the traces are performed in $d$ spacetime dimensions, as in ordinary DR, and the renormalization of the entire functional is then obtained by the addition of the counterterm action. Unless explicitly stated, in all the equations that follow we will be referring to $(C^{(d)})^2$ simply as $C^2$. Specifically we will choose
\begin{align}
\label{counter}
\sm_{ct}&=-\frac{\mu^{-\varepsilon}}{\varepsilon}\,\int\,d^dx\,\sqrt{-g}\left(b\,C^2+b'\,E\right),
\end{align}
as in \eqref{fourd} and \eqref{fourd2}. 
 The two Weyl-covariant terms introduced by the renormalization procedure that we will be using in our analysis below, can be separately defined in the form 
\begin{align}
V_{C^2}\equiv & \mu^{-\varepsilon}\int\,d^dx\,\sqrt{-g}\, C^2 \notag \\
V_{E}\equiv &\mu^{-\varepsilon} \int\,d^dx\,\sqrt{-g}\,E  
\end{align}
where $\mu$ is a renormalization scale, while $\varepsilon=d-4$ and the counterterm vertices will be simply obtained by multiple differentiations of the two expressions above. \\
 As we are going to show, by performing an expansion of $V_E$ with respect to 
$d$ and using a dimensional reduction of the result at $d=4$, one can derive in a simple way that all the functional derivatives (with open indices) of $V_E$ vanish.\\
 In the explicit empirical check of this result, one may need to re-express the reduced Kronecker deltas 
$\delta^{(d)}_{\mu\nu}$ as 4-dimensional $\delta^{(4)}_{\mu\nu}$, written in terms of the external momenta $p_i$  and of their orthogonal component $n^\mu$ (the $n$-$p$ basis that will be introduced below). This allows to account for the evanescent terms which are present in the expansion of higher point functions. \\
The anomaly contribution, in this formulation, is simply generated by the derivative respect to $d$ of $V_E$, once we perform a Weyl variation (i.e. a $g_{\mu\nu}\delta/\delta g_{\mu\nu}$ operation) on this functional. These points will be addressed in a section below.

\subsection{The finite renormalization induced by $V_E$} 
It is quite obvious that the inclusion of $V_E$ induces a finite renormalization of the effective action in $d=4$. This can be simply shown by noticing that both $V_E$ and $V_{C^2}$ manifest an explicit dependence on $\varepsilon$, i.e. 
\begin{equation}
V_{E/C^2}\equiv V_{E/C^2}(d),
\end{equation} 
and the counterterm contributions can be expanded around $d=4$. 
For this purpose, given a scalar functional  $f(d)$, it will be convenient to denote its Taylor expansion around $d=4$ in the form 
\begin{equation} 
f(d)=\left[f\right] + \varepsilon \left[f'\right] \qquad \left[f\right]\equiv f(4), \left[f'\right] \equiv 
f'(4) 
\end{equation}
with coefficients which are square bracketed once they are computed at $d=4$. The expansion of $V_{E/C^2}(d)$, using these notations, takes the form 

\begin{equation} 
\label{expand}
\frac{1}{\varepsilon}V_{E/C^2}(d)=\frac{\mu^{-\varepsilon}}{\varepsilon}\left( \left[V_{E/C^2}\right] + \varepsilon 
\left[V_{E/C^2}'\right] +O(\varepsilon^2) \right),
\end{equation}
where the first correction to the residue at the pole in $\varepsilon$ comes from the derivative respect to the dimension $d$. 
Notice that $\left[V_E\right]$ is a topological term and it is therefore metric-independent. Its value is related to the global topology of the spacetime and it is therefore a pure number. 
Then, it is clear that the $1/\varepsilon$ term will not contribute to the renormalization of the bare 4-T vertex, since each counterterm vertex  is generated by functional differentiation of \eqref{expand} with respect to the background metric. \\
The only contribution of the $V_E$ as $\varepsilon\to 0$  is related to $\left[V'_E\right]$, and it is indeed finite, as a 0/0 contribution in $\varepsilon$. Therefore, the inclusion of $V_E$, will induce only a finite renormalization of the bare vertex, and henceforth of the entire effective action, since this result remains valid to all orders.\\
Finally, we can relate $\left[V'_E\right]$ to the anomaly by the equation 
\begin{equation}
\sqrt{g} E(x)=2 g_{\alpha\beta}\left[ V'_E\right]^{\alpha\beta}
\end{equation} 
where the indices of $\left[ V'_E\right]^{\alpha\beta}$ run in four dimensions. 
\subsubsection{Open indices in $V_E$}
We can investigate this point in more detail.\\
 The extraction of a counterterm vertex, as mentioned above, requires a functional differentiation of \eqref{expand}. 
It will appear in the $\varepsilon\to 0$ limit, in the form
\begin{align}
\label{ff}
\left[ \frac{\delta}{\delta g_{\mu_1\nu_1} (x_1)} \frac{\delta}{\delta g_{\mu_2\nu_2} (x_2)}\ldots  \frac{\delta}{\delta g_{\mu_n\nu_n} (x_n)}\frac{\mu^{-\varepsilon}}{\varepsilon}V_{E}(d)\right]= &
\frac{\delta}{\delta g_{\mu_2\nu_2} (x_2)}\ldots  \frac{\delta}{\delta g_{\mu_n\nu_n} (x_n)}\left[ V_{E}'\right]\notag \\
&\equiv \left[ V_{E}'\right]^{\mu_1\nu_1\ldots  \mu_n\nu_n}(x_1,\ldots x_n).\notag \\
\end{align} 
The rhs of the this expression is clearly regular as $\varepsilon\to 0$, and defines a finite renormalization of the corresponding $n$-graviton vertex. Obviously, the same is not true for the $V_{C^2}$ counterterm since $\left[V_{C^2}\right]$ will be metric-dependent. In this second case the analogous of \eqref{ff} is 
\begin{align}
\label{open}
\left[ \frac{\delta}{\delta g_{\mu_1\nu_1} (x_1)} \frac{\delta}{\delta g_{\mu_2\nu_2} (x_2)}\ldots  \frac{\delta}{\delta g_{\mu_n\nu_n} (x_n)}\frac{\mu^{-\varepsilon}}{\varepsilon}V_{C^2}(d)\right]= &
\frac{1}{\varepsilon} \left[V_{C^2}\right]^{\mu_1\nu_1\ldots  \mu_n\nu_n}(x_1,\ldots x_n)+ \left[V_{C^2}'\right]^{\mu_1\nu_1\ldots  \mu_n\nu_n}(x_1,\ldots x_n).\notag \\
\end{align} 
The second, finite term of this expression  ($V'_{C^2}$), will appear in all the renormalized 
anomalous WI that we will present below, when we will perform the  $\varepsilon\to 0$ limit of those eqs. in a flat background. \\
One important comment concerns possible ambiguities arising from the differentiation of 
terms such as $\left[V'_E\right]$ with respect to the metric $g_{\mu\nu}$. We mention that such ambiguities arise only in the presence of contractions with a metric, and not otherwise. We will be dealing with this second case, in the subsection below.\\
As in any practical application of DR, once the renormalization of a vertex is completed, a tensor structure with open indices, which is generated by the procedure, is automatically dimensionally reduced to the $d=4$ subspace, giving the final, finite expression of such vertex. \\
Indeed, expanding around $d=4$, \eqref{expand} can be rewritten in the form 
\begin{align}
 \frac{\delta}{\delta g_{\mu_1\nu_1} (x_1)} \frac{\delta}{\delta g_{\mu_2\nu_2} (x_2)}\ldots  \frac{\delta}{\delta g_{\mu_n\nu_n} (x_n)}V_{E}(d)= &
\left[V_{E}\right]^{\mu_1\nu_1\ldots  \mu_n\nu_n}(x_1,\ldots x_n)+ \varepsilon\left[V_{E}'\right]^{\mu_1\nu_1\ldots  \mu_n\nu_n}(x_1,\ldots x_n),\notag \\
\end{align}  
which in the $d\to 4$ limit vanishes, since the two terms on the rhs vanish separately. 
Notice that the first term $\left[V_{E}\right]^{\mu_1\nu_1\ldots  \mu_n\nu_n}$ is topological and is simply obtained by the differentiation of $V_E$ in $d=4$. As far as we leave all the indices open, then the following relation holds 
\begin{align}
\label{frr}
\lim_{d\to 4} \left(\frac{\delta}{\delta g_{\mu_1\nu_1} (x_1)} \frac{\delta}{\delta g_{\mu_2\nu_2} (x_2)}\ldots  \frac{\delta}{\delta g_{\mu_n\nu_n} (x_n)}V_{E}(d)\right)=&
 \frac{\delta}{\delta g_{\mu_1\nu_1} (x_1)} \frac{\delta}{\delta g_{\mu_2\nu_2} (x_2)}\ldots  \frac{\delta}{\delta g_{\mu_n\nu_n} (x_n)} \lim_{d\to 4} V_{E}(d) \notag \\
=& 0
\end{align} 
 In other words, the operation of functional differentiation with open indices and limit to $d=4$ commute. The topological nature of this part of the 3-point vertex counterterm was originally noted in \cite{Osborn:1993cr} (see also \cite{Zwiebach:1985uq}).
The explicit check of this topological relation requires the dimensional reduction of the $\delta^{d}_{\mu\nu}$ to $\delta^{(4)}_{\mu\nu}$, using the $n$-$p$ basis (discussed in Sec. \ref{nnp}) to take into account momentum degeneracies in specific dimensions), and indeed it has been verified for $n=3$ and $n=4$ \cite{Bzowski:2015pba,Serino:2020pyu}, i.e. for three- and four-point functions. We stress once again that \eqref{frr} holds as far as we do not perform any contraction of the tensor indices with the external metric. \\ 
In the presence of metric contractions, these results can be re-addressed using a Weyl variation. We are going to illustrate this point in some detail below, defining a straightforward procedure in order to handle this second case correctly. \\
Contributions proportional to $V'_E$ will be present once we contract the WIs with the $\Sigma$ projector defined below in \eqref{Lproj}, using the longitudinal-trace/ transverse-traceless decomposition of the correlators. 
One finds, by a direct computation, the emergence of virtual scalar exchanges (or mixing terms) in the decomposed WI, which signal the breaking of the conformal symmetry induced by the conformal anomaly. Such terms are easy to derive simply due to the presence of a tcombined $\pi^{\mu\nu}\delta^{\alpha\beta}$ projector in $\Sigma$, in each of the external legs of the gravitational vertex. This defines a coupling of a scalar pole to the anomaly functional. 
\subsubsection{Closed indices}
The analysis of relations involving the topological counterterm $V_E$ in the presence of contraction with the metric cannot be performed as above, but, as we have mentioned, 
can be addressed correctly by relating the contraction to a Weyl variation. 
For this purpose consider the Weyl scaling relations 
\begin{equation}
\sqrt{g}=e^{d \sigma.}\sqrt{\bar{g}}\qquad E=e^{-4 \sigma}\bar{E},
\end{equation}
from which we derive the constraints 
 \begin{equation}
 \label{third}
 \frac{\delta}{\delta \sigma(x_3)} \frac{\delta}{\delta \sigma(x_2)} \frac{\delta}{\delta \sigma(x_1)}V_E=\varepsilon^3 \sqrt{g}E \delta^d(x_1-x_2)\delta^d(x_2-x_3)
\end{equation}
which can be generalized to any multiple derivative 
\begin{equation} 
\label{general}
\frac{\delta}{\delta \sigma(x_n)} \ldots\frac{\delta}{\delta \sigma(x_2)} \frac{\delta}{\delta \sigma(x_1)}V_E=\varepsilon^n \sqrt{g}E \delta^d(x_1-x_2)\delta^d(x_2-x_3)\ldots \delta^d(x_{n-1}-x_n).
\end{equation}
If we use 
\begin{equation}
\label{var}
2 g_{\alpha\beta} \frac{\delta}{\delta g_{\alpha\beta}}=\frac{\delta}{\delta \sigma(x)}
\end{equation} 
and
\begin{equation}
\frac{\delta}{\delta \sigma(x_2)}g^{\mu\nu}(x_1)=-2 g^{\mu\nu}\delta^4(x_2-x_1) 
\end{equation}
we derive several relations. For instance, expanding the left-hand side of \eqref{general} for $n=2$ we obtain  
\begin{align}
\frac{\delta}{\delta\sigma(x_2)}\frac{\delta}{\delta\sigma(x_1)}V_E=& 4 g_{\mu\nu} V_E^{\mu\nu}(x_1)
\delta^d(x_1-x_2) + 4 g_{\alpha\beta}(x_2)g_{\mu\nu}(x_1) V_E^{\alpha\beta\mu\nu}(x_2,x_1)\notag \\
=&\varepsilon^2{\sqrt{g(x_1)}}E(x_1)\delta^d(x_1-x_2).
\end{align}
The expression above can be evaluated in the flat limit both in 4 and $d$ dimensions. Consider, for instance, the flat limit with $d$ generic. We use the fact that $E$ and $V_E^{\mu\nu}$ vanish in the flat limit to obtain the constraint
\begin{align}
\label{double}
\delta^{(d)}_{\alpha\beta}\delta^{(d)}_{\mu\nu}V_E^{\alpha\beta\mu\nu}=0, 
\end{align}
which remains valid for $d=4$ since the equation is analytic in $d$, and in particular in $d=4$. 
Similarly, one can derive relations for the traces of higher point functions, which are discussed in the \appref{TraceRelations}. 
\subsubsection{Open and closed indices}
We can generalize the method and perform combined variations respect to $\sigma$ and 
to the metric in order to derive some additional relations. 
For this purpose we consider the expression 

\begin{equation}
J^{\mu_2\nu_2\mu_3\nu_3\mu_4\nu_4}\equiv
\frac{\delta}{\delta \sigma(x_1)}\frac{\delta}{\delta g_{\mu_2\nu_2}(x_2)}\frac{\delta}{\delta g_{\mu_3\nu_3}(x_3)}\frac{\delta}{\delta g_{\mu_4\nu_4}(x_4)}V_{E}.
\end{equation}
We can use the commutation relation
\begin{equation}
\left[ \frac{\delta }{\delta \sigma(x_1)},\frac{\delta}{\delta g_{\mu_2\nu_2}(x_2)}\right] V_E=
-2 V_E^{\mu_2\nu_2}(x_2)\delta^d(x_1-x_2)
\end{equation}
 to rearrange the differentiations in the form 
\begin{align}
J^{\mu_2\nu_2\mu_3\nu_3\mu_4\nu_4}=&
\frac{\delta}{\delta g_{\mu_2\nu_2}(x_2)}\frac{\delta}{\delta g_{\mu_3\nu_3}(x_3)}\frac{\delta}{\delta g_{\mu_4\nu_4}(x_4)}\frac{\delta}{\delta \sigma(x_1)}V_{E} \notag \\- 
& 2 V_E^{\mu_2\nu_2\mu_3\nu_3\mu_4\nu_4}(x_2,x_3,x_4) \left(\delta^d(x_1-x_2) +
\delta^d(x_1-x_3) +\delta^d(x_1-x_4)\right)\notag \\
\end{align} 
and use \eqref{form} to derive the constraint
\begin{align}
J^{\mu_2\nu_2\mu_3\nu_3\mu_4\nu_4}=& (d-4)\left(\sqrt{g} E\right)^{\mu_2\nu_2\mu_3\nu_3\mu_4\nu_4}(x_2,x_3,x_4) \notag \\- 
& 2 V_E^{\mu_2\nu_2\mu_3\nu_3\mu_4\nu_4}(x_2,x_3,x_4) \left(\delta^d(x_1-x_2) +
\delta^d(x_1-x_3)+ \delta^d(x_1-x_4)\right),\notag \\
\end{align}
 which is \eqref{mom} in coordinate space. The second term in the expression above, related to the pinched contributions $\delta (x_1-x_i)\, (i=2,3,4)$ is also d-dimensional and needs to be expanded around $d=4$. The expansion is performed as in \eqref{expand}  
 \begin{equation} 
V_E^{\mu_2\nu_2\mu_3\nu_3\mu_4\nu_4}= \left[V_E^{\mu_2\nu_2\mu_3\nu_3\mu_4\nu_4}\right] + \varepsilon \left[{V'_E}^{\mu_2\nu_2\mu_3\nu_3\mu_4\nu_4}\right]. 
 \end{equation}
Notice that the first contribution vanishes for being purely topological, while the second, transformed to momentum space, will appear in all the renormalized anomalous CWIs 
that we will discuss in the sections below 
\begin{equation}
(d-4)\left[{V'_E}^{\mu_2\nu_2\mu_3\nu_3\mu_4\nu_4}\right]\to
(d-4)\left[{V'}_{E}^{\m_2\n_2\dots\mu_4\nu_4}(p_1+p_2,p_3,p_4)\right]. 
\end{equation} 
It accounts for a finite $0/0$ contribution in the expansion of the $1/\varepsilon \, V_E(d)$ counterterm in the variable $\varepsilon$.  

 \subsubsection{Simplifications in the $n$-$p$ basis} 
 \label{nnp}
Simplifications in the structure of the renormalized 4-point function are possible once we re-express all the contributions of the previous section in terms of the tensorial basis formed by the $n^\mu$ and $p_1,p_2$ and $p_3$ four-vectors
 \begin{equation}
n^{\mu}=\epsilon^{\m \a \b\g}p_{i,\a}p_{j,\b}p_{k,\g}, \quad i\neq j \neq k =1,2,3,4,
\end{equation}
with $(n^\mu, p_i,p_j, p_k)$ forming a tetrad that can be used as a basis of expansion in Minkowski space. The $n$-$p$ parametrization is also discussed in \appref{np}.
 We just recall that in the computation of the residue of the $\sim V_E$ counterterm, we 
need to parameterize the Kronecker $\delta_{\mu\nu}$ in this basis in the form 

\begin{equation}
\d^{(4)}_{\m \n}=\sum_{i,j}^4 p_i^{\m} p_j^{\n} (Z^{-1})_{j i},
\end{equation}
where $(Z^{-1})_{j i}$ is the inverse of the Gramm matrix, defined as $Z=[p_i\cdot p_j]_{i,j=1}^d$.\\
Using the expression of $\delta^{\mu\nu}$ in the $n$-$p$ basis, denoted as $\delta^{(4)}_{\mu\nu}$ 

\begin{align}\label{deltaepsilon}
\d^{(4)}_{\m \n}=&\frac{1}{n^2}\Bigg(2p_1^{(\mu } p_2^{\nu) } \left(p_3^2
   p_1\cdot p_2-p_1\cdot p_3 p_2\cdot p_3\right)+2p_1^{(\nu } p_3^{\mu )}
   \left(p_2^2 p_1\cdot p_3-p_1\cdot p_2 p_2\cdot p_3\right) \notag \\ 
 & \qquad \qquad  +2p_2^{(\nu } p_3^{\mu) }
   \left(p_1^2 p_2\cdot p_3-p_1\cdot p_2 p_1\cdot p_3\right)\notag +p_1^{\mu }
   p_1^{\nu } \left((p_2\cdot p_3)^2-p_2^2 p_3^2\right)\notag \\
   & \qquad \qquad +p_2^{\mu } p_2^{\nu }\left((p_1\cdot p_3)^2-p_1^2 p_3^2\right) +p_3^{\mu } p_3^{\nu }\left((p_1\cdot p_2)^2-p_1^2 p_2^2\right)+n^{\mu } n^{\nu } \Bigg)
   \end{align}
one derives several relations in $d=4$. We easily derive the constraint
\begin{align}\label{projtwon}
\Pi^{\m_i \n_i}_{\a_i \b_i}(p_i)n^{\a_i } n^{\b_i }=& -\Pi^{\m_i \n_i}_{\a_i \b_i}(p_i)\Bigg(2p_j^{(\b_i  } p_k^{\a_i) }
   \left(p_i^2 p_j\cdot p_k-p_i\cdot p_k p_i\cdot p_j\right)\notag \\
   & +p_j^{\a_i } p_j^{\b_i  }\left((p_i\cdot p_k)^2-p_i^2 p_k^2\right) +p_k^{\a_i } p_k^{\b_i  }\left((p_i\cdot p_j)^2-p_i^2 p_j^2\right) \Bigg).
\end{align}
while the relation $\d^{(4)}_{\a_i \b_i}\Pi(p_i)^{(4)\m_i \n_i}_{\a_i \b_i}=0$ is obviously satisfied in the new basis. Using these relations, it is possible to show the vanishing of 
$\left[ V^{\mu_1\nu_1\ldots \mu_n\nu_n}\right] $, which is obviously defined at $d=4$. 
An explicit check has been discussed in \cite{Serino:2020pyu} for $n=4$, using the $n$-$p$ decomposition. As we have elaborated above, this results holds in general, due to 
\eqref{frr}.

\section{Renormalization in momentum space}

For a generic $nT$ correlator, the only counterterm needed for its renormalization, is obtained by the inclusion of a classical gravitational vertex generated by the differentiation of \eqref{counter} $n$ times.\\
The renormalized effective action $\sm_R$ is then defined by the sum of the two terms
\begin{equation}
\sm_R(g)=\sm(g)+ \sm_{ct}(g)
\end{equation}
\begin{align}
\label{cct1}
\sm_{ct}=\raisebox{-0.8ex}{\includegraphics[width=0.15\linewidth]{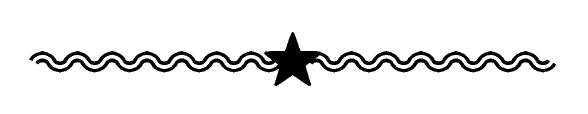}}+\raisebox{-5ex}{\includegraphics[width=0.15\linewidth]{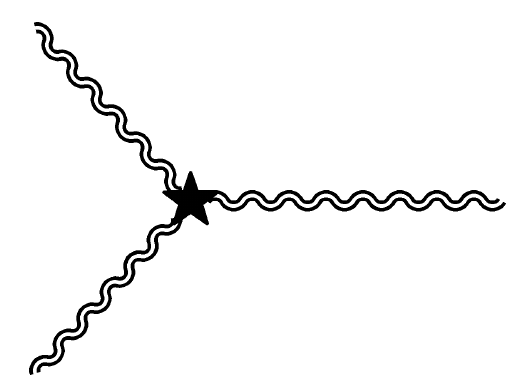}}+\raisebox{-6ex}{\includegraphics[width=0.15\linewidth]{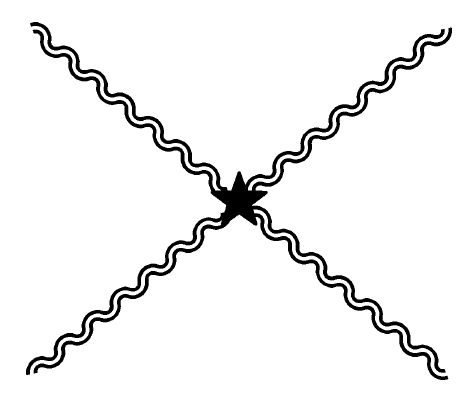}}+...
\end{align}
with $\sm_{ct}$ shown in \eqref{cct1}. Both terms of $\sm_R(g)$ are expanded in the metric fluctuations as in \eqref{cct1}. If we resort to a path integral definition of a certain CFT,  it is clear that any renormalized correlation function appearing in the expansion of $\sm_R(g)$ would be expressed in terms of a bare contribution accompanied by a counterterm vertex.\\
The correlation functions extracted by the renormalized action can be expressed as the sum of a finite $(f)$  correlator and of an anomaly term ($anomaly$) in the form
\begin{align}
\label{cct}
 \braket{T^{\mu_1\nu_1}T^{\mu_2\nu_2}\ldots T^{\mu_n\nu_n}}_{Ren}=
&\bigg[\braket{T^{\mu_1\nu_1}T^{\mu_2\nu_2}\ldots T^{\mu_n\nu_n}}_{bare}+\braket{T^{\mu_1\nu_1}T^{\mu_2\nu_2}\ldots T^{\mu_n\nu_n}}_{count}\bigg]_{d\to4}=\notag\\
&=\braket{T^{\mu_1\nu_1}T^{\mu_2\nu_2}\ldots T^{\mu_n\nu_n}}^{(d=4)}_{f}+\braket{T^{\mu_1\nu_1}T^{\mu_2\nu_2}\ldots T^{\mu_n\nu_n}}^{(d=4)}_{anomaly}
\end{align} 
The renormalized correlator shown above satisfies anomalous CWIs. \\
To characterize the anomaly contribution to each correlation function, we start from the 1-point function. In a generic background $g$, the 1-point function is decomposed as 
\begin{equation}
\label{decomp}
\langle T^{\mu\nu}\rangle_{Ren}=\frac{2}{\sqrt{g}}\frac{\delta \sm_{Ren}}{\delta g_{\mu\nu}} =\langle T^{\mu\nu} \rangle_A  + \langle \overline{T}^{\mu\nu}\rangle_f
\end{equation}
with
\begin{equation}
g^{\mu\nu}\frac{\delta \sm}{\delta g^{\mu\nu}} = g^{\mu\nu}\frac{\delta \sm_A}{\delta g^{\mu\nu}}\equiv \frac{\sqrt{g}}{2} g_{\mu\nu} \langle T^{\mu\nu} \rangle_A \end{equation}
being the trace anomaly equation, and $\langle \overline{T}^{\mu\nu}\rangle_f$ is the Weyl-invariant (traceless) term.
\\
Following the discussion in \eqref{anom1}, these scaling violations may be written for the 1-point function in the form
\begin{equation}
\label{anomeq}
\braket{T^{\mu}_{\ \ \mu}(x)}=\mathcal{A}(x)
\end{equation} 
- having dropped the suffix {\em Ren} from the renormalized stress energy tensor - 
where the finite terms on the right hand side of this equation denote the anomaly contribution
with
\begin{equation}
\label{AF}
\mathcal{A}(x)=\sqrt{-g(x)}\bigg[b\,C^2(x)+b'E(x)\bigg]
\end{equation}
being the anomaly functional. We will be needing several differentiation of this functional, 
evaluated in the flat limit. This procedure generates expressions which are  polynomial in the momenta, that can be found in the \appref{Mvc}.
 In general, one also finds additional dimension-4 local invariants $\mathcal{L}_i$, if there are couplings to other background fields, as for instance in the QED and QCD cases, with coefficients related to the $\beta$ functions of the corresponding gauge couplings. \\
For n-point functions the trace anomaly, as well as all the other CWIs, are far more involved, and take a hierarchical structure. \\
For all the other WIs, in DR the structure of the equations can be analyzed in two different frameworks. \\
In one of them, we are allowed to investigate the correlators directly in $d$ spacetime dimensions, deriving ordinary (anomaly-free) CWI's, which are then modified by the inclusion of the 4-dimensional counterterm as $d\to 4$. In this limit, the conformal constraints become anomalous and the hierarchical equations are modified by the presence of extra terms which are anomaly-related. \\
Alternatively, it is possible to circumvent this limiting procedure by working out the equations directly in $d=4$, with the inclusion of the contributions coming from the anomaly functional, as we are going to show below. This second approach has been formulated in \cite{Coriano:2017mux} and will be extended to the 4-point function in \secref{heres}. \\  
We recall that the counterterm vertex for the nT correlator, in DR, in momentum space takes the form
\begin{equation}
\braket{T^{\mu_1\nu_1}(p_1)\dots T^{\mu_n\nu_n}(\bar{p}_n)}_{count}=-\frac{\mu^{-\varepsilon}}{\varepsilon}\bigg(b\,V_{C^2}^{\mu_1\nu_1\dots\mu_n\nu_n}(p_1,\dots,\bar{p}_n)+b'\,V_{E}^{\mu_1\nu_1\dots\mu_n\nu_n}(p_1,\dots,\bar{p}_n)\bigg)
\label{nTcount},
\end{equation}
 where  
\begin{align}
V_{C^2}^{\m_1\n_1\dots\mu_n\nu_n}(p_1,\dots,\bar{p}_n)
&\equiv 2^n\big[\sqrt{-g}\,C^2\big]^{\m_1\n_1\dots\m_n\n_n}(p_1,\dots,\bar{p}_n)\notag\\
&=2^n\int\,d^dx_1\,\dots\,d^dx_n\,d^dx\,\bigg(\sdfrac{\d^n(\sqrt{-g}C^2)(x)}{\d g_{\m_1\n_1}(x_1)\dots\d g_{\m_n\n_n}(x_n)}\bigg)_{g=\delta}\,e^{-i(p_1\,x_1+\dots+p_nx_n)}\notag\\[2ex]
\end{align}
and
\begin{align}
V_{E}^{\m_1\n_1\dots\mu_n\nu_n}(p_1,\dots,\bar{p}_n)&\equiv 2^n\big[\sqrt{-g}\,E\big]^{\m_1\n_1\dots\m_n\n_n}(p_1,\dots,\bar{p}_n)\notag\\
&=2^n\int\,d^dx_1\,\dots\,d^dx_n\,d^dx\,\bigg(\sdfrac{\d^n(\sqrt{-g}E)(x)}{\d g_{\m_1\n_1}(x_1)\dots\d g_{\mu_n\nu_n}(x_n)}\bigg)_{g=\delta}\,e^{-i(p_1\,x_1+\dots+p_nx_n)}.
\end{align}
are the expressions of the two contributions present in \eqref{nTcount} in momentum space. 
One can also verify the following trace relations 
\begin{align}
\label{expabove}
&\delta_{\mu_1\nu_1}\,V_{C^2}^{\m_1\n_1\dots\mu_n\nu_n}(p_1,\dots,p_n)=2^{n-1}(d-4)\,\left[\sqrt{-g}C^2\right]^{\mu_2\nu_2\dots\mu_n\nu_n}(p_2,\dots,p_n)\notag\\
&-2\bigg[V_{C^2}^{\m_2\n_2\dots\mu_n\nu_n}(p_1+p_2,p_3,\dots,p_n)+V_{C^2}^{\m_2\n_2\dots\mu_n\nu_n}(p_2,p_1+p_3,\dots,p_n)+\dots+V_{C^2}^{\m_2\n_2\dots\mu_n\nu_n}(p_2,p_3,\dots,p_1+p_n)\bigg]
\end{align}
\begin{align}
\label{mom}
&\delta_{\mu_1\nu_1}\,V_{E}^{\m_1\n_1\dots\mu_n\nu_n}(p_1,\dots,p_n)=2^{n-1}(d-4)\,\left[\sqrt{-g}E\right]^{\mu_2\nu_2\dots\mu_n\nu_n}(p_2,\dots,p_n)\notag\\
&-2\bigg[V_{E}^{\m_2\n_2\dots\mu_n\nu_n}(p_1+p_2,p_3,\dots,p_n)+V_{E}^{\m_2\n_2\dots\mu_n\nu_n}(p_2,p_1+p_3,\dots,p_n)+\dots+V_{E}^{\m_2\n_2\dots\mu_n\nu_n}(p_2,p_3,\dots,p_1+p_n)\bigg]
\end{align}
that hold in general $d$ dimensions. \\
Obviously, the effective action that results from the renormalization can be clearly separated in terms of two contributions, as evident from \eqref{decomp}, 
\begin{equation} 
\sm_R(g)=\sm_{A}(g) +{\sm}_f(g),
\end{equation}
corresponding to an anomaly part $\sm_{A}[g]$ and to a finite, Weyl-invariant term, which can be expanded in terms of fluctuations over a background $\bar{g}$ as for the entire effective action $\sm(g)$
\begin{equation}
\label{exps3}
\sm_f(g)=\sm(\bar{g})+\sum_{n=1}^\infty \frac{1}{2^n n!} \int d^d x_1\ldots d^d x_n \sqrt{g_1}\ldots \sqrt{g_n}\,\langle T^{\mu_1\nu_1}\ldots \,T^{\mu_n\nu_n}\rangle_{f }\delta g_{\mu_1\nu_1}(x_1)\ldots \delta g_{\mu_n\nu_n}(x_n).
\end{equation}
This functional collects finite correlators \eqref{cct} in $d=4$.
A similar expansion holds also for $\sm_{A}$, the anomaly part.\\ 
The anomaly effective action $\sm_A$ that results from this analysis in momentum space is a rational function of the external momenta, characterised by well-defined tensor structures, and it is free of logarithmic terms, as shown in direct perturbative studies of the TJJ and 3T \cite{Giannotti:2008cv,Armillis:2009pq,Armillis:2010qk}. \\
\section{Conservation Ward identities}
The anomaly action $\sm_A(g)$ is constrained by a hierarchical set of equations which can be derived by the symmetries of the general effective action $\sm$. We proceed assuming that the correlation functions can be derived by varying the path integral definition of $\sm(g)$ \eqref{defg} \eqref{induced} as in \eqref{exps1}.\\ 
Starting from the covariant definition of the stress-energy tensor, expressed in terms of a fundamental action as in \eqref{defT}, but for the rest, generic
\begin{equation}
\braket{T^{\mu\nu}(x)}_g=\frac{2}{\sqrt{-g(x)}}\frac{\delta S(g)}{\delta g_{\mu\nu}(x)},
\end{equation}
this 1-point function satisfies the fundamental Ward identity of covariant conservation in an arbitrary background $g$
\begin{equation}
\label{vat}
^{(g)}\nabla_\mu\braket{T^{\mu\nu}(x)}_g=0, \qquad i.e. \qquad \delta_\epsilon \sm(g)=0
\end{equation}
as a consequence of the invariance of $\sm(g)$ under diffeomorphisms.
Here $^{(g)}\nabla_\mu$ denotes the covariant derivative in the general background metric $g_{\mu\nu}(x)$. It can be expressed in the form 
\begin{equation}
\partial_\nu\left(\frac{\delta S(g)}{\delta g_{\mu\nu}(x)}\right)+\sm^\mu_{\nu\lambda}\left(\frac{\delta S(g)}{\delta g_{\lambda\nu}(x)}\right)=0\label{conserv},
\end{equation}
where $\Gamma^\mu_{\lambda\nu}$ is the Christoffel connection for the general background metric $g_{\mu\nu(x)}$.\\
Our definitions and conventions are summarised in an \appref{Definitions}.
In order to derive the conservation WIs for higher point correlation functions, one has to consider additional variations with respect to the metric of \eqref{vat} and then move to flat space, obtaining 
\begin{align}
&\partial_{\nu_1}\braket{T^{\mu_1\nu_1}(x_1)T^{\mu_2\nu_2}(x_2)T^{\mu_3\nu_3}(x_3)T^{\mu_4\nu_4}(x_4)}=\notag\\
=&-\left[2\left(\frac{\delta\Gamma^{\mu_1}_{\lambda\nu_1}(x_1)}{\delta  g_{\mu_2\nu_2}(x_2)}\right)_{g=\delta}\braket{T^{\lambda\nu_1}(x_1)T^{\mu_3\nu_3}(x_3)T^{\mu_4\nu_4}(x_4)}+(23)+(24)\right]\notag\\
& -\left[4\left(\frac{\delta^2\Gamma^{\mu_1}_{\lambda\nu_1}(x_1)}{\delta  g_{\mu_2\nu_2}(x_2)\delta  g_{\mu_3\nu_3}(x_3)}\right)_{g=\delta}\braket{T^{\lambda\nu_1}(x_1)T^{\mu_4\nu_4}(x_4)}+(24)+(34)\right],\label{transverseX}
\end{align}
where 
\begin{align}
\left(\frac{\delta\Gamma^{\mu_1}_{\lambda\nu_1}(x_1)}{\delta  g_{\mu_i\nu_i}(x_i)}\right)_{g=\delta}&=\frac{1}{2}\left(\delta^{\mu_1(\mu_i}\delta^{\nu_i)}_{\nu_1}\,\partial_\lambda\delta_{x_1x_i}+\delta^{\mu_1(\mu_i}\delta^{\nu_i)}_{\lambda}\,\partial_{\nu_1}\delta_{x_1x_i}-\delta^{(\mu_i}_\lambda\delta^{\nu_i)}_{\nu_1}\,\partial^{\mu_1}\delta_{x_1x_i}\right)\\[1.5ex]
\left(\frac{\delta^2\Gamma^{\mu_1}_{\lambda\nu_1}(x_1)}{\delta  g_{\mu_i\nu_i}(x_i)\delta  g_{\mu_j\nu_j}(x_j)}\right)_{g=\delta}&=\notag\\
&\hspace{-3cm}=-\frac{\delta_{x_1x_i}}{2}\delta^{\mu_1(\mu_i}\delta^{\nu_i)\epsilon}\left(\delta^{(\mu_j}_\epsilon\delta^{\nu_j)}_{\nu_1}\,\partial_\lambda\delta_{x_1x_j}+\delta^{(\mu_j}_\epsilon\delta^{\nu_j)}_{\lambda}\,\partial_{\nu_1}\delta_{x_1x_j}-\delta^{(\mu_j}_\lambda\delta^{\nu_j)}_{\nu_1}\,\partial_\epsilon\delta_{x_1x_j}\right)+(ij),
\end{align}
are the first and second functional derivatives of the connection, in the flat limit.
We have explicitly indicated the symmetrization with respect to the relevant indices using the permutation $(ij)\equiv (i\leftrightarrow j)$. We have defined $ \delta^{(\mu_i}_\lambda\delta^{\nu_i)}_{\nu_1}\equiv1/2(\delta^{\mu_i}_\lambda\delta^{\nu_i}_{\nu_1}+\delta^{\nu_i}_\lambda\delta^{\mu_i}_{\nu_1})$, and introduced a simplified notation for the Dirac delta $\delta_{x_ix_j}\equiv \delta(x_i-x_j)$. All the derivative (e.g. $\partial_\lambda$) are taken with respect to the coordinate $x_1$  $(e.g. \partial/\partial x_1^\lambda)$. \\
We Fourier transform to momentum space with the convention 
\bea
\braket{T^{\mu_1\nu_1}(p_1)T^{\mu_2\nu_2}(p_2)T^{\mu_3\nu_3}(p_3)T^{\mu_4\nu_4}( \overline{p}_4)}
&=&\int d^4 x_1 d^4 x_2 d^4 x_3 e^{-i( p_1\cdot x_1 +p_2\cdot x_2 + p_3\cdot x_3)}\nonumber \\
&& \times \braket{T^{\mu_1\nu_1}(x_1)T^{\mu_2\nu_2}(x_2)T^{\mu_3\nu_3}(x_3)T^{\mu_4\nu_4}(0)}.
\eea

Here we have used the translational invariance of the correlator in flat space, which allows to use momentum conservation to express one of the momenta (in our convention $p_4$) as combination of the remaining ones $\bar{p}_4=-p_1-p_2-p_3$. Details on the elimination of one of the momenta in the derivation of the CWIs and on the modification of the Leibnitz rule in the differentiation of such correlators in momentum space can be found in \cite{Coriano:2018bbe}. \\
The conservation Ward Identity \eqref{transverseX} in flat spacetime may be Fourier transformed, giving the CWIs in momentum space
\begin{align}
&p_{1\nu_1}\braket{T^{\mu_1\nu_1}(p_1)T^{\mu_2\nu_2}(p_2)T^{\mu_3\nu_3}(p_3)T^{\mu_4\nu_4}(\bar p_4)}=\notag\\
&=\Big[4\, \mathcal{B}^{\mu_1\hspace{0.4cm}\mu_2\nu_2\mu_3\nu_3}_{\hspace{0.3cm}\lambda\nu_1}(p_2,p_3)\braket{T^{\lambda\nu_1}(p_1+p_2+p_3)T^{\mu_4\nu_4}(\bar p_4)}+(34)+(2 4)\Big]\notag\\
&\hspace{0.5cm}+\Big[2 \, \mathcal{C}^{\mu_1\hspace{0.4cm}\mu_2\nu_2}_{\hspace{0.3cm}\lambda\nu_1}(p_2)\braket{T^{\lambda\nu_1}(p_1+p_2)T^{\mu_3\nu_3}(p_3)T^{\mu_4\nu_4}(\bar p_4)}+(2 3)+(2 4)\Big],\label{transverseP}
\end{align}
where we have defined
\begin{align}
\label{BB}
\mathcal{B}^{\mu_1\hspace{0.4cm}\mu_2\nu_2\mu_3\nu_3}_{\hspace{0.3cm}\lambda\nu_1}(p_2,p_3)&\equiv -\frac{1}{2}\delta^{\mu_1(\mu_2}{\delta^{\nu_2)\epsilon}}\left(\delta_\epsilon^{(\mu_3}\delta^{\nu_3)}_{\nu_1}\,p_{3\,\lambda}+\delta_\epsilon^{(\mu_3}\delta^{\nu_3)}_{\lambda}\,p_{3\,\nu_1}-\delta_\lambda^{(\mu_3}\delta^{\nu_3)}_{\nu_1}\,p_{3\,\epsilon}\right)+(23)\\[2ex]
\label{CC}
\mathcal{C}^{\mu_1\hspace{0.4cm}\mu_2\nu_2}_{\hspace{0.3cm}\lambda\nu_1}(p_2)&\equiv \frac{1}{2}\left(\delta^{\mu_1(\mu_2}\delta^{\nu_2)}_{\nu_1}\,p_{2\,\lambda}+\delta^{\mu_1(\mu_2}\delta^{\nu_2)}_{\lambda}p_{2\,\nu_1}-\delta^{(\mu_2}_{\lambda}\delta^{\nu_2)}_{\nu_1}p_2^{\mu_1}\right),
\end{align}
related to the second and first functional derivatives of the Christoffel, connection respectively. 

\subsection{Conservation WI's for the counterterms} 
To illustrate the conservation WI in detail, we turn to the expression of the counterterm action \eqref{counter}, which generates counterterm vertices of the form
\begin{align}
&\braket{T^{\mu_1\nu_1}(p_1)T^{\mu_2\nu_2}(p_2)T^{\mu_3\nu_3}(p_3)T^{\mu_4\nu_4}(\bar{p}_4)}_{count}=\notag\\
&\qquad=-\frac{\mu^{-\varepsilon}}{\varepsilon}\bigg(b\,V_{C^2}^{\mu_1\nu_1\mu_2\nu_2\mu_3\nu_3\mu_4\nu_4}(p_1,p_2,p_3,\bar{p}_4)+b'\,V_{E}^{\mu_1\nu_1\mu_2\nu_2\mu_3\nu_3\mu_4\nu_4}(p_1,p_2,p_3,\bar{p}_4)\bigg),\label{TTTTcount}
\end{align}
where on the rhs of the expression above we have introduced the counterterm vertices (with $P= p_1 +\ldots p_4$)
\begin{align}
&V_{C^2}^{\m_1\n_1\m_2\n_2\m_3\n_3\mu_4\nu_4}(p_1,p_2,p_3,\bar{p}_4)
\,\delta^4(P)\equiv 16\, \delta^4(P)\,\big[\sqrt{-g}\,C^2\big]^{\m_1\n_1\m_2\n_2\m_3\n_3\mu_4\nu_4}(p_1,p_2,p_3,\bar{p}_4)\notag\\
=&16\int\,d^dx_1\,\,d^dx_2\,d^dx_3\,d^dx_4\,d^dx\,\bigg(\sdfrac{\d^4(\sqrt{-g}C^2)(x)}{\d g_{\m_1\n_1}(x_1)\d g_{\m_2\n_2}(x_2)\d g_{\m_3\n_3}(x_3)\d g_{\m_4\n_4}(x_4)}\bigg)_{g=\delta}\,e^{-i(p_1\,x_1+p_2\,x_2+p_3\,x_3+p_4x_4)},\notag\\[2ex]
&V_{E}^{\m_1\n_1\m_2\n_2\m_3\n_3\mu_4\nu_4}(p_1,p_2,p_3,\bar{p}_4)\delta^4(P)\equiv 16\,\,\delta^4(P)\big[\sqrt{-g}\,E\big]^{\m_1\n_1\m_2\n_2\m_3\n_3\m_4\n_4}(p_1,p_2,p_3,\bar{p}_4)\notag\\
=&16\int\,d^dx_1\,\,d^dx_2\,d^dx_3\,d^dx_4\,d^dx\,\bigg(\sdfrac{\d^4(\sqrt{-g}E)(x)}{\d g_{\m_1\n_1}(x_1)\d g_{\m_2\n_2}(x_2)\d g_{\m_3\n_3}(x_3)\d g_{\mu_4\nu_4}(x_4)}\bigg)_{g=\delta}\,e^{-i(p_1\,x_1+p_2\,x_2+p_3\,x_3+p_4x_4)},\label{count}
\end{align}
evaluated in the flat spacetime limit.
These vertices share some properties when contracted with flat metric tensors and the external momenta as we have already seen. In particular, from \eqref{expabove} and \eqref{mom}, when $n=4$ and in $d$ dimensions we have
\begin{align}
\delta_{\mu_1\nu_1}\,V_{C^2}^{\m_1\n_1\m_2\n_2\m_3\n_3\mu_4\nu_4}(p_1,p_2,p_3,\bar{p}_4)&=8(d-4)\,\left[\sqrt{-g}C^2\right]^{\mu_2\nu_2\mu_3\nu_3\mu_4\nu_4}(p_2,p_3,\bar{p}_4)\notag\\
&\hspace{-5.5cm}-2V_{C^2}^{\m_2\n_2\m_3\n_3\mu_4\nu_4}(p_1+p_2,p_3,\bar{p}_4)-2V_{C^2}^{\m_2\n_2\m_3\n_3\mu_4\nu_4}(p_2,p_1+p_3,\bar{p}_4)-2V_{C^2}^{\m_2\n_2\mu_3\nu_3\m_4\n_4}(p_2,p_3,p_1+\bar{p}_4),\\[2ex]
\delta_{\mu_1\nu_1}\,V_{E}^{\m_1\n_1\m_2\n_2\m_3\n_3\mu_4\nu_4}(p_1,p_2,p_3,\bar{p}_4)&=8(d-4)\,\left[\sqrt{-g}E\right]^{\mu_2\nu_2\mu_3\nu_3\mu_4\nu_4}(p_2,p_3,\bar{p}_4)\notag\\
&\hspace{-5.5cm}-2V_{E}^{\m_2\n_2\m_3\n_3\mu_4\nu_4}(p_1+p_2,p_3,\bar{p}_4)-2V_{E}^{\m_2\n_2\m_3\n_3\mu_4\nu_4}(p_2,p_1+p_3,\bar{p}_4)-2V_{E}^{\m_2\n_2\mu_3\nu_3\m_4\n_4}(p_2,p_3,p_1+\bar{p}_4),
\end{align}
which play a key role in the renormalization procedure.
Furthermore, the contraction of these vertices with the external momenta generates conservation WIs in $d$ dimensions, similar to \eqref{transverseP}, 
\begin{align}
&p_{1\,\nu_1}\,V_{C^2}^{\m_1\n_1\m_2\n_2\m_3\n_3\mu_4\nu_4}(p_1,p_2,p_3,\bar{p}_4)=\notag\\
&=\Big[4\, \mathcal{B}^{\mu_1\hspace{0.4cm}\mu_2\nu_2\mu_3\nu_3}_{\hspace{0.3cm}\lambda\nu_1}(p_2,p_3)V_{C^2}^{\lambda\n_1\mu_4\nu_4}(p_1+p_2+p_3,\bar{p}_4)+(34)+ (24)\Big]\notag\\
&\hspace{0.5cm}+\Big[2 \, \mathcal{C}^{\mu_1\hspace{0.4cm}\mu_2\nu_2}_{\hspace{0.3cm}\lambda\nu_1}(p_2)V_{C^2}^{\lambda\n_1\m_3\n_3\mu_4\nu_4}(p_1+p_2,p_3,\bar{p}_4)+(2 3)+(24)\Big]\\[2ex]
&p_{1\,\nu_1}\,V_{E}^{\m_1\n_1\m_2\n_2\m_3\n_3\mu_4\nu_4}(p_1,p_2,p_3,\bar{p}_4)=\notag\\
&=\Big[4\, \mathcal{B}^{\mu_1\hspace{0.4cm}\mu_2\nu_2\mu_3\nu_3}_{\hspace{0.3cm}\lambda\nu_1}(p_2,p_3)V_{E}^{\lambda\n_1\mu_4\nu_4}(p_1+p_2+p_3,\bar{p}_4)+(34)+ (24)\Big]\notag\\
&\hspace{0.5cm}+\Big[2 \, \mathcal{C}^{\mu_1\hspace{0.4cm}\mu_2\nu_2}_{\hspace{0.3cm}\lambda\nu_1}(p_2)V_{E}^{\lambda\n_1\m_3\n_3\mu_4\nu_4}(p_1+p_2,p_3,\bar{p}_4)+(2 3)+(24)\Big],
\end{align} 
where  $\mathcal{C}$ and $\mathcal{B}$ are given in \eqref{BB} and \eqref{CC}.
These equations can be generalized to the case of $n$-point functions. 
\section{Conformal Ward Identities}
Turning to the ordinary (i.e. non anomalous) trace and conformal WIs, these can be obtained directly in flat space using the expression of operators of the dilatation and special conformal transformations.
The dilatation WI's for the 4T can be easily constructed from the condition of Weyl invariance of the effective action $\sm$, or equivalently, in the ordinary operatorial approach (see \cite{Coriano:2018bbe}) 

\begin{equation}
	\left[4d+\sum_{j=1}^4\,x_j^\a\sdfrac{\partial}{\partial x_j^\a}\right]\braket{T^{\mu_1\nu_1}(x_1)T^{\mu_2\nu_2}(x_2)T^{\mu_3\nu_3}(x_3)T^{\mu_4\nu_4}(x_4)}=0,
\end{equation}
where we have used the explicit expression of the scaling dimension of the stress energy tensor $\Delta_T=d$. Analogously, the special conformal WIs, corresponding to special conformal transformations, can be derived in the operatorial approach, applied to an ordinary CFT in flat space, 
relying on the change of $T^{\mu\nu}$ under a special conformal transformation, with a generic parameter  
 $b_\mu$, and  $\sigma=-2 b\cdot x$ 
\beq
\delta T^{\mu\nu}(x)=-(b^\alpha x^2 -2 x^\alpha b\cdot x )\, \partial_\alpha  T^{\mu\nu}(x)   - \Delta_T \sigma T^{\mu\nu}(x)+
2(b_\mu x_\alpha- b_\alpha x_\mu)T^{\alpha\nu} + 2 (b_\nu x_\alpha -b_\alpha x_\nu)\, T^{\mu\alpha}(x).
\eeq
 The action of the special conformal operator $\mathcal{K}^\kappa$ on $T$ in its finite form is obtained differentiating respect to the parameter $b_\kappa$ 
\bea
\mathcal{K}^\kappa T^{\mu\nu}(x)&\equiv &\delta_\kappa T^{\mu\nu}(x) =\frac{\partial}{\partial b^\kappa} (\delta T^{\mu\nu})\nonumber \\
&=& -(x^2 \partial_\kappa - 2 x_\kappa x\cdot \partial) T^{\mu\nu}(x) + 2\Delta_T x_\kappa T^{\mu\nu}(x) +
2(\delta_{\mu\kappa}x_\alpha -\delta_{\alpha \kappa}x_\mu) T^{\alpha\nu}(x) \nonumber \\ 
&& + 2 (\delta_{\kappa\nu} x_{\alpha} -\delta_{\alpha \kappa} x_\nu )T^{\mu\alpha}.
\label{ith}
\eea
By using the Leibniz rule for the variation on correlation functions of multiple T's, it can be distributed over the entire correlator as 
\begin{equation}
\label{com}
\mathcal{K}^\kappa\braket{T^{\mu_1\nu_1}(x_1)T^{\mu_2\nu_2}(x_2)T^{\mu_3\nu_3}(x_3)T^{\mu_4\nu_4}(x_4)}=
\sum_{i=1}^4 \langle T(x_1)\ldots \delta_\kappa T(x_i)\ldots T(x_4) \rangle =0
\end{equation}
which takes the form
\begin{align}
	0=&\sum_{j=1}^n\left(2d\,x_j^k+2x_j^\k\,x_j^\a\sdfrac{\partial}{\partial x_j\a}-x_j^2\sdfrac{\partial}{\partial x_{jk}}\right)\,\braket{T^{\mu_1\nu_1}(x_1)T^{\mu_2\nu_2}(x_2)T^{\mu_3\nu_3}(x_3)T^{\mu_4\nu_4}(x_4)}\notag\\
	&+2\sum_{j=1}^4\left[(x_j)_{\a_{j}}\d^{\k\m_{j}}-x_j^{\m_{j}}\d^\k_{\a_{j}}\right]\,\braket{T^{\mu_1\nu_1}(x_1)\dots T^{\alpha_j\nu_j}(x_j) \dots T^{\mu_4\nu_4}(x_4)}\notag\\
	&+2\sum_{j=1}^4\left[(x_j)_{\a_{j}}\d^{\k\nu_{j}}-x_j^{\nu_{j}}\d^\k_{\a_{j}}\right]\,\braket{T^{\mu_1\nu_1}(x_1)\dots T^{\mu_j\alpha_j}(x_j) \dots T^{\mu_4\nu_4}(x_4)},
\end{align}
where $\k$ is now a free Lorentz index. Notice that in order to be allowed to use an operatorial approach, one needs to rely on correlators defined via direct insertions of T's. Such correlation functions, are, in general,  different from the definition given above in \eqref{exps1} due to possible contact terms and the presence of nonvanishing tadpoles, not contemplated in \eqref{com}.\\
For this reason the CWI's derived by this operatorial method and by the functional method that we will present below, are naive expressions which are perfectly well-defined and equivalent, only in the presence of a suitable regularization scheme and of a flat background. In DR, which is well-defined in a flat spacetime, the vanishing of the 1-point function and the inclusion of vertex counterterms shows that we don't need to worry about such issues.  \\
Notice that these constraints are directly written in momentum space as 
\begin{align}
	0=&D\,\braket{T^{\mu_1\nu_1}(p_1)T^{\mu_2\nu_2}(p_2)T^{\mu_3\nu_3}(p_3)T^{\mu_4\nu_4}(\bar{p}_4)}\notag\\
	=&\left(d-\sum_{j=1}^3\,p_j^\a\sdfrac{\partial}{\partial p_j^\a}\right)\braket{T^{\mu_1\nu_1}(p_1)T^{\mu_2\nu_2}(p_2)T^{\mu_3\nu_3}({p}_3)T^{\mu_4\nu_4}(\bar{p}_4)}\label{Dequ}
\end{align}
and
	
\begin{align}	
	0=&\mathcal{K}^\kappa\braket{T^{\mu_1\nu_1}(p_1)T^{\mu_2\nu_2}(p_2)T^{\mu_3\nu_3}(p_3)T^{\mu_4\nu_4}(\bar{p}_4)}\notag\\
	=&\sum_{j=1}^{3}\left((p_j)^\kappa\frac{\partial}{\partial p_j^\alpha}\frac{\partial}{\partial p_{j\,\alpha}}-2p_j^\alpha\frac{\partial}{\partial p_j^\alpha}\frac{\partial}{\partial p_{j\,\kappa}}\right)\braket{T^{\mu_1\nu_1}(p_1)T^{\mu_2\nu_2}(p_2)T^{\mu_3\nu_3}(p_3)T^{\mu_4\nu_4}(\bar{p}_4)}\notag\\
	&\hspace{1cm}+4\left(\delta^{\kappa(\mu_1}\frac{\partial}{\partial p_1^{\alpha_1}}-\delta^\kappa_{\alpha_1}\d_\l^{(\m_1}\frac{\partial}{\partial p_{1\,\l}}\right)\braket{ T^{\nu_1)\alpha_1}(p_1)T^{\mu_2\nu_2}(p_2)T^{\mu_3\nu_3}(p_3)T^{\mu_4\nu_4}(\bar{p}_4)}\notag\\
	&\hspace{1cm}+4\left(\delta^{\kappa(\m_2}\frac{\partial}{\partial p_2^{\a_2}}-\delta^\kappa_{\a_2}\d^{(\m_2}_\l\frac{\partial}{\partial p_{2\,\l}}\right)\braket{ T^{\nu_2)\a_2}(p_2)T^{\mu_1\nu_1}(p_1)T^{\mu_3\nu_3}(p_3)T^{\mu_4\nu_4}(\bar{p}_4)}\notag\\
	&\hspace{1cm}+4\left(\delta^{\kappa(\m_3}\frac{\partial}{\partial p_3^{\a_3}}-\delta^\kappa_{\a_3}\d^{(\m_3}_\l\frac{\partial}{\partial p_{3\,\l}}\right)\braket{ T^{\nu_3)\a_3}(p_3)T^{\mu_1\nu_1}(p_1)T^{\mu_2\nu_2}(p_2)T^{\mu_4\nu_4}(\bar{p}_4)}\label{Keq},
\end{align}
in terms of a dilatation operator $D$
and a special conformal transformation operator $K^\kappa$. The action of the differential operators on the momenta is implicit on the 4th momentum, as discussed in the case of 3-point functions in previous works \cite{Coriano:2013jba, Bzowski:2013sza, Coriano:2018bbe}, with a modification of the Leibnitz rule.

\subsection{Trace and conformal anomalous Ward identities} 
The CWIs become anomalous as we move from $d$ spacetime dimensions to 4.
In $d$ dimensions the conformal symmetry of the correlator 4T is preserved and this property is reflected in the trace identity
\begin{equation}
\braket{T^{\mu}_{\ \ \mu}(x)}_g=g_{\mu\nu}\braket{T^{\mu\nu}(x)}_g=0,
\end{equation}
which generates, as we have already mentioned, a hierarchy of equations by functional differentiation of this result respect to the background metric $g$. 
Equivalently, the same equations can be derived from the condition of Weyl invariance of the effective action.\\
Following the same procedure as for the conservation WIs, we may derive the trace Ward identities for the four-point function 4T, in general $d$ dimensions, as
\begin{align}
\delta_{\mu_1\nu_1}\,\braket{T^{\mu_1\nu_1}(x_1)T^{\mu_2\nu_2}(x_2)T^{\mu_3\nu_3}(x_3)T^{\mu_4\nu_4}(x_4)}&=\notag\\
&\hspace{-3cm}=-2\Big[\delta_{x_1x_2}\braket{T^{\mu_2\nu_2}(x_1)T^{\mu_3\nu_3}(x_3)T^{\mu_4\nu_4}(x_4)}+(23)+(24)\Big],\label{traceD}
\end{align} 
that may be written in momentum space, after a Fourier transform, as
\begin{align}
&\delta_{\mu_1\nu_1}\braket{T^{\mu_1\nu_1}(p_1)T^{\mu_2\nu_2}(p_2)T^{\mu_3\nu_3}(p_3)T^{\mu_4\nu_4}(\bar p_4)}=-2\braket{T^{\mu_2\nu_2}(p_1+p_2)T^{\mu_3\nu_3}(p_3)T^{\mu_4\nu_4}(\bar p_4)}\notag\\[1.3ex]
&\hspace{2cm}-2\braket{T^{\mu_2\nu_2}(p_2)T^{\mu_3\nu_3}(p_1+p_3)T^{\mu_4\nu_4}(\bar p_4)}-2\braket{T^{\mu_2\nu_2}(p_2)T^{\mu_3\nu_3}(p_3)T^{\mu_4\nu_4}(p_1+\bar p_4)}.\label{traceP}
\end{align}
We have omitted an overall $\delta$ function, having replaced $p_4$ with $\bar{p}_4$.\\
In $d=4$  the equations need to be renormalized, by adding local covariant counterterms which will be generated from the action \eqref{counter}.\\
 If general covariance is respected by this procedure, the conservation WIs remain valid for the renormalized effective action and for its variations. This is reflected on the hierarchical structure of  the equations, which remain identical to the bare (naive) case. \\
Trace identities of the correlation functions involving at least three stress energy tensor operators are instead affected by the anomaly, due to the scaling violations induced by the regularization/renormalization procedure. 
In $d=4$ the corresponding anomalous Ward identities for the trace can be obtained by a functional variation of the equation \eqref{anomeq} with respect to the background metric. \\
In this case \eqref{traceD} is characterised by new contributions on its rhs, coming from the anomaly $\mathcal{A}(x)$
\begin{align}
&\delta_{\mu_1\nu_1}\braket{T^{\mu_1\nu_1}(p_1)T^{\mu_2\nu_2}(p_2)T^{\mu_3\nu_3}(p_3)T^{\mu_4\nu_4}(\bar p_4)}=-2\braket{T^{\mu_2\nu_2}(p_1+p_2)T^{\mu_3\nu_3}(p_3)T^{\mu_4\nu_4}(\bar p_4)}\notag\\[1.3ex]
&\hspace{2cm}-2\braket{T^{\mu_2\nu_2}(p_2)T^{\mu_3\nu_3}(p_1+p_3)T^{\mu_4\nu_4}(\bar p_4)}-2\braket{T^{\mu_2\nu_2}(p_2)T^{\mu_3\nu_3}(p_3)T^{\mu_4\nu_4}(p_1+\bar p_4)}\notag\\
&\hspace{2cm}+8\mathcal{A}^{\mu_2\nu_2\mu_3\nu_3\mu_4\nu_4}(p_2,p_3,\bar p_4),
\end{align}
where the trace anomaly functional $\mathcal{A}$ is given in Eq. \eqref{AF}, and the Fourier transform of its variation in the flat spacetime limit takes the form
\begin{align}
\mathcal{A}^{\mu_2\nu_2\mu_3\nu_3\mu_4\nu_4}(p_2,p_3,\bar p_4)\delta^4(P)=\int d^4 x_1d^4 x_2 d^4 x_3 d^4 x_4\,\frac{\delta^3\mathcal{A}(x_1)}{\delta g_{\mu_2\nu_2}(x_2)\delta g_{\mu_3\nu_3}(x_3)\delta g_{\mu_4\nu_4}(x_4)}\Bigg|_{flat, d=4}e^{i(p_1x_1+\dots+p_4x_4)}.
\end{align}
In the following we will also be using the simpler general definition 
\begin{equation}
\mathcal{A}^{\mu_1\nu_1\dots\mu_n\nu_n}(p_1,\dots,p_n)\equiv\bigg[\sqrt{-g}\bigg(b\,C^2+b'\,E\bigg)\bigg]^{\mu_1\nu_1\dots\mu_n\nu_n}(p_1,\dots,p_n)\label{defVarTrace}
\end{equation}
to denote the anomaly contributions to 3- and 4-point functions. Notice that these are genuine 4-dimensional terms, left over by the procedure of renormalization.

\subsection{The anomalous CWI's using conformal Killing vectors}
\label{heres}
The expressions of the anomalous conformal WIs can be derived in an alternative way following the formulation of  \cite{Coriano:2017mux}, that here we are going to extend to the 4-point function case.\\
The derivation of such identities relies uniquely on the effective action and can be obtained as follows. We illustrate it first in the TT case, and then move to the 4T.\\
We start from the conservation of the conformal current as derived in \eqref{iso}
\begin{equation}
\int d^d x \sqrt{g}\, \nabla^\alpha \left( \epsilon_\alpha \frac{2}{\sqrt{g}}\frac{\delta \sm}{\delta g_{\mu\alpha}}\right)=\int d^4 x \sqrt{g}\, \nabla_\mu\braket{\epsilon_\alpha T^{\alpha \nu}} =0.  
\end{equation}
In the TT case the derivation of the special CWIs is simplified, since there is no trace anomaly if the counterterm action is defined as in \eqref{counter}, a point that we will address in \secref{renren}. We rely on the fact that the conservation of the conformal current $J^\mu_{(K)}$ implies the conservation equation 
\begin{align}
0=\int\,d^dx\,\sqrt{-g}\, \,\nabla_\mu\,\braket{J^\mu_{(K)}(x)\,T^{\mu_1\nu_1}(x_1)}.
\end{align}
By making explicit the expression $J^\mu(x)=K_\nu(x)\,T^{\mu\nu}(x)$, with $\epsilon \to K$ in the flat limit,  the previous relation takes the form
 \begin{align}
0=\int\,d^dx\,\bigg(\partial_\mu K_\nu\,\braket{T^{\mu\nu}(x)\,T^{\mu_1\nu_1}(x_1)}+ K_\nu\,\partial_\mu\,\braket{T^{\mu\nu}(x)\,T^{\mu_1\nu_1}(x_1)}\bigg).\label{cons}
 \end{align}
We recall that $K_\nu$ satisfies the conformal Killing equation in flat space
\begin{align}
\label{flatc}
\partial_\mu K_\nu+\partial_\nu K_\mu=\frac{2}{d}\delta_{\mu\nu}\,\left(\partial\cdot K\right),
\end{align}
and by using this equation \eqref{cons} can be re-written in the form
\begin{align}
	0=\int\,d^dx\,\bigg(K_\nu\partial_\mu\,\braket{T^{\mu\nu}(x)\,T^{\mu_1\nu_1}(x_1)}+\frac{1}{d}\big(\partial\cdot K\big)\,\braket{T(x)\,T^{\mu_1\nu_1}(x_1)}\bigg).\label{newcons}
\end{align}
We can use in  this previous expression the conservation and trace Ward identities for the two-point function $\braket{TT}$, that in the flat spacetime limit are explicitly given by
\begin{align}
\partial_\mu\braket{T^{\mu\nu}(x)T^{\mu_1\nu_1}(x_1)}&=\bigg(\delta^{(\mu_1}_\mu\delta^{\nu_1)}_\lambda\partial^\nu\delta(x-x_1)-2\delta^{\nu(\mu_1}\delta^{\nu_1)}_\mu\partial_\lambda\delta(x-x_1)\bigg)\braket{T^{\lambda\mu}(x)},\label{consTT}\\
\delta_{\mu\nu}\braket{T^{\mu\nu}(x)T^{\mu_1\nu_1}(x_1)}&\equiv\braket{T(x)T^{\mu_1\nu_1}(x_1)}=-2\delta(x-x_1)\braket{T^{\mu_1\nu_1}(x)}\label{traceTT}
\end{align}
and the explicit expression of the Killing vector $K^{(C)}_\nu$ for the special conformal transformations 
\begin{equation}
\begin{split}
K^{(C)\,\kappa}_\mu&=2x^\kappa\,x_\mu-x^2\delta^\kappa_\mu\\
\partial\cdot  K^{(C)\,\kappa}&=2d\,x^\kappa
\end{split}\label{spc}
\end{equation}
where $\kappa=1,\dots,d$. By using \eqref{spc} in the integral \eqref{newcons}, we can rewrite that expression as
\begin{align}
	0=\int\,d^dx\,\bigg[\big(2x^\kappa\,x_\nu-x^2\delta^\kappa_\nu\big)\partial_\mu\,\braket{T^{\mu\nu}(x)\,T^{\mu_1\nu_1}(x_1)}+2\,x^\kappa\,\braket{T(x)\,T^{\mu_1\nu_1}(x_1)}\bigg],
\end{align}
A final integrating by parts finally gives the relations 
\begin{align}
&\left(2d\,x_1^\kappa+2x_1^\kappa\,x^{\mu}_1\frac{\partial}{\partial x_1^\mu}+x_1^2\frac{\partial}{\partial x_{1\kappa}}\right)\braket{T^{\mu_1\nu_1}(x_1)}\notag\\
&\quad+2\bigg(x_{1\lambda}\,\delta^{\mu_1\kappa}-x_1^{\mu_1}\delta^\kappa_\lambda\bigg)\braket{T^{\lambda\nu_1}(x_1)}+2\bigg(x_{1\lambda}\,\delta^{\nu_1\kappa}-x_1^{\nu_1}\delta^\kappa_\lambda\bigg)\braket{T^{\mu_1\lambda}(x_1)}=0
\end{align}
that are  the special CWIs for the 1-point function $\braket{T^{\mu_1\nu_1}(x_1)}$. 
\subsection{ 4-point functions}\label{heres2}
The derivation above can be extended to n-point functions, starting from the identity 
\begin{equation} 
\int d^d x \sqrt{g} \nabla_\alpha(x) \langle J^\alpha_c(x)T^{\mu_1\nu_1}(x_1)\ldots T^{\mu_n\nu_n}(x_n)
\label{div}
\rangle=0.
\end{equation}
We have used the conservation of the conformal current in d dimensions under variations of the metric, induced by the conformal Killing vectors.\\
In absence of an anomaly, the conservation of the current $J^\mu_c$ follows from the conservation of the stress energy tensor plus the zero trace condition, as discussed in \secref{two}. 
As in the example illustrated above, we consider \eqref{div} in the flat limit 
\begin{align}
\int dx^d \,\partial_\nu\bigg[K_\mu(x)\braket{T^{\mu\nu}(x)T^{\mu_1\nu_1}(x_1)\dots T^{\mu_4\nu_4}(x_4)}\bigg]=0,\label{Killing1}
\end{align}
where we are assuming that the surface terms vanish, due to the fast fall-off behaviour of the correlation function at infinity. Expanding \eqref{Killing1} we obtain an expression similar to \eqref{newcons}

\begin{align}
0=\int d^dx\left\{K_{\mu}(x)\partial_\nu\braket{T^{\mu\nu}(x)T^{\mu_1\nu_1}(x_1)\dots T^{\mu_4\nu_4}(x_4)}+\frac{1}{d}\big(\partial\cdot K\big)\delta_{\mu\nu}\braket{T^{\mu\nu}(x)T^{\mu_1\nu_1}(x_1)\dots T^{\mu_4\nu_4}(x_4)}
\right\}.\label{killing01}
\end{align}
Starting  from this expression, the dilatation CWI is obtained by the choice of the CKV characterising the dilatations  
\begin{equation}
K^{(D)}_\mu(x)=x_\mu,\qquad\partial\cdot K^{(D)}=d
\end{equation}
and \eqref{killing01} becomes
\begin{align}
0=\int d^d x\bigg\{x_\mu\,\partial_\nu\braket{T^{\mu\nu}(x)T^{\mu_1\nu_1}(x_1)\dots T^{\mu_4\nu_4}(x_4)}+\delta_{\mu\nu}\braket{T^{\mu\nu}(x)T^{\mu_1\nu_1}(x_1)\dots T^{\mu_4\nu_4}(x_4)}\label{anomDil}
\bigg\}.
\end{align}
At this stage, we use the conservation and trace Ward identities in $d=4$ for the $4$-point function written as
\begin{align}
&\partial_\nu\braket{T^{\mu\nu}(x)T^{\mu_1\nu_1}(x_1)\dots T^{\mu_4\nu_4}(x_4)}=\notag\\
=&-8\bigg\{\left[\Gamma^{\mu}_{\nu\lambda}(x)\right]^{\mu_1\nu_1\mu_2\nu_2\mu_3\nu_3}(x_1,x_2,x_3)\braket{T^{\lambda\nu}(x)T^{\mu_4\nu_4}(x_4)}+(14)+(24)+(34)\bigg\}\notag\\
&-4\bigg\{\left[\Gamma^{\mu}_{\nu\lambda}(x)\right]^{\mu_1\nu_1\mu_2\nu_2}(x_1,x_2)\braket{T^{\lambda\nu}(x)T^{\mu_3\nu_3}(x_3)T^{\mu_4\nu_4}(x_4)}+(13)+(23)+(14)+(24)+(34)\bigg\}\notag\\
&-2\bigg\{\left[\Gamma^{\mu}_{\nu\lambda}(x)\right]^{\mu_1\nu_1}(x_1)\braket{T^{\lambda\nu}(x)T^{\mu_2\nu_2}(x_2)T^{\mu_3\nu_3}(x_3)T^{\mu_4\nu_4}(x_4)}+(12)+(13)+(14)\bigg\}\label{5ptcons}
\end{align}
and
\begin{align}
\delta_{\mu\nu}\braket{T^{\mu\nu}(x)T^{\mu_1\nu_1}(x_1)\dots T^{\mu_4\nu_4}(x_4)}&=-2\bigg\{\delta_{xx_1}\braket{T^{\mu_1\nu_1}(x)T^{\mu_2\nu_2}(x_2)\dots T^{\mu_4\nu_4}(x_4)}+(12)+(13)+(14)\bigg\}\notag\\
&\hspace{2cm}+2^4\big[\mathcal{A}(x)\big]^{\mu_1\nu_1\dots\mu_4\nu_4}(x_1,\dots,x_4).\label{5pttrace}
\end{align}
to finally derive the dilatation WI from \eqref{anomDil} in the form
\begin{align}
\left(4d+\sum_{j=1}^4\,x_j^\alpha\frac{\partial}{\partial x_j^\alpha}\right)\braket{T^{\mu_1\nu_1}(x_1)\dots T^{\mu_4\nu_4}(x_4)}=2^4\int dx \big[\mathcal{A}(x)\big]^{\mu_1\nu_1\dots\mu_4\nu_4}(x_1,\dots,x_4)\label{DilAnom}
\end{align}
where $d=4$. It is worth mentioning that \eqref{DilAnom} is valid in any even spacetime dimension if we take into account the particular structure of the trace anomaly in that particular dimension.

The special CWIs correspond to the $d$ special conformal Killing vectors in flat space given in \eqref{spc}, as in the $TT$ case. Also in this case we derive the identity 
\begin{align}
0=&\int d^d x\bigg\{\big(2x^\kappa\,x_\mu-x^2\delta^\kappa_\mu\big)\,\partial_\nu\braket{T^{\mu\nu}(x)T^{\mu_1\nu_1}(x_1)\dots T^{\mu_4\nu_4}(x_4)}\notag\\
&\hspace{4cm}+2x^\kappa\delta_{\mu\nu}\braket{T^{\mu\nu}(x)T^{\mu_1\nu_1}(x_1)\dots T^{\mu_4\nu_4}(x_4)}\label{anomSpecWI}
\bigg\}.
\end{align}
By using the relations \eqref{5ptcons} and \eqref{5pttrace} and performing the integration over $x$ explicitly in the equation above,  the anomalous special CWIs for the $4$-point function take the form
\begin{align}
&\sum_{j=1}^4\left[2x_j^\kappa\left(d+x_j^\alpha\frac{\partial}{\partial x_j^\alpha}\right)-x_j^2\,\delta^{\kappa\alpha}\frac{\partial}{\partial x_j^\alpha}\right]\braket{T^{\mu_1\nu_1}(x_1)\dots T^{\mu_4\nu_4}(x_4)}\notag\\
&+2\sum_{j=1}^4\left(\delta^{\kappa\mu_j}x_{j\,\alpha}-\delta^\kappa_\alpha x_j^{\mu_j}\right)\braket{T^{\mu_1\nu_1}(x_1)\dots T^{\nu_j\alpha}(x_j)\dots T^{\mu_4\nu_4}(x_4)}\notag\\
&+2\sum_{j=1}^4\left(\delta^{\kappa\nu_j}x_{j\,\alpha}-\delta^\kappa_\alpha x_j^{\nu_j}\right)\braket{T^{\mu_1\nu_1}(x_1)\dots T^{\mu_j\alpha}(x_j)\dots T^{\mu_4\nu_4}(x_4)}=\notag\\
&=2^5\,\int dx\,x^\kappa\big[\mathcal{A}(x)\big]^{\mu_1\nu_1\dots\mu_4\nu_4}(x_1,\dots,x_4),
\end{align}
where the presence of the anomaly term, as discussed in \secref{two}, comes from the inclusion of the trace  WI, exactly as in the TT case. \\
At this stage, these equations can be transformed to momentum space, giving the final expressions of the CWIs in the form 
\begin{align}
\left(d-\sum_{j=1}^3\,p_j^\alpha\frac{\partial}{\partial p_j^\alpha}\right)\braket{T^{\mu_1\nu_1}(p_1)T^{\mu_2\nu_2}(p_2)T^{\mu_3\nu_3}(p_3)T^{\mu_4\nu_4}(\bar{p}_4)}=2^4\,\mathcal{A}^{\mu_1\nu_1\mu_2\nu_2\mu_3\nu_3\mu_4\nu_4}(p_1,p_2,p_3,\bar{p}_4)
\end{align} 
for the dilatation, and 
\begin{align}
&\sum_{j=1}^3\left(p_j^\kappa\frac{\partial^2}{\partial p_j^\alpha \partial p_{j\alpha}}-2p_j^\alpha\frac{\partial}{\partial p_j^\alpha\partial p_{j\kappa}}\right)\braket{T^{\mu_1\nu_1}(p_1)T^{\mu_2\nu_2}(p_2)T^{\mu_3\nu_3}(p_3) T^{\mu_4\nu_4}(\bar{p}_4)}\notag\\
&+2\sum_{j=1}^3\left(\delta^{\kappa\mu_j}\frac{\partial}{\partial p_{j\,\alpha}}-\delta^\kappa_\alpha \frac{\partial}{\partial p_j^{\mu_j}}\right)\braket{T^{\mu_1\nu_1}(p_1)\dots T^{\nu_j\alpha}(p_j)\dots T^{\mu_4\nu_4}(\bar{p}_4)}\notag\\
&+2\sum_{j=1}^3\left(\delta^{\kappa\nu_j}\frac{\partial}{\partial p_{j\,\alpha}}-\delta^\kappa_\alpha \frac{\partial}{\partial p_j^{\nu_j}}\right)\braket{T^{\mu_1\nu_1}(p_1)\dots T^{\mu_j\alpha}(p_j)\dots T^{\mu_4\nu_4}(\bar{p}_4)}\notag\\
&=-2^5\,\left[\frac{\partial}{\partial p_{4\kappa}}\,\mathcal{A}^{\mu_1\nu_1\mu_2\nu_2\mu_3\nu_3\mu_4\nu_4}(p_1,p_2,p_3,p_4)\right]_{\bar{p_4}=-p_1-p_2-p_3},
\end{align}
for the special CWI's, having used the definition \eqref{defVarTrace}. \\
When this procedure is applied to the $n$-point function, one finds that the anomalous CWIs are written as
\begin{align}
\left(d-\sum_{j=1}^{n-1}\,p_j^\alpha\frac{\partial}{\partial p_j^\alpha}\right)\braket{T^{\mu_1\nu_1}(p_1)\dots T^{\mu_n\nu_n}(\bar{p}_n)}=2^n\,\mathcal{A}^{\mu_1\nu_1\dots\mu_n\nu_n}(p_1,\dots,\bar{p}_n)
\end{align}
for the dilatation and 
	\begin{align}
		&\sum_{j=1}^{n-1}\left(p_j^\kappa\frac{\partial^2}{\partial p_j^\alpha \partial p_{j\alpha}}-2p_j^\alpha\frac{\partial}{\partial p_j^\alpha\partial p_{j\kappa}}\right)\braket{T^{\mu_1\nu_1}(p_1)\dots T^{\mu_n\nu_n}(\bar{p}_n)}\notag\\
		&+2\sum_{j=1}^{n-1}\left(\delta^{\kappa\mu_j}\frac{\partial}{\partial p_{j\,\alpha}}-\delta^\kappa_\alpha \frac{\partial}{\partial p_j^{\mu_j}}\right)\braket{T^{\mu_1\nu_1}(p_1)\dots T^{\nu_j\alpha}(p_j)\dots T^{\mu_n\nu_n}(\bar{p}_n)}\notag\\
		&+2\sum_{j=1}^{n-1}\left(\delta^{\kappa\nu_j}\frac{\partial}{\partial p_{j\,\alpha}}-\delta^\kappa_\alpha \frac{\partial}{\partial p_j^{\nu_j}}\right)\braket{T^{\mu_1\nu_1}(p_1)\dots T^{\mu_j\alpha}(p_j)\dots T^{\mu_n\nu_n}(\bar{p}_n)}\notag\\
		&=-2^{n+1}\left[\frac{\partial}{\partial p_{n\kappa}}\,\mathcal{A}^{\mu_1\nu_1\dots\mu_n\nu_n}(p_1,\dots,p_n)\right]_{p_n=\bar{p}_n},
	\end{align}
for the special conformal Ward identities, where $\bar{p}_n=-\sum_{i=1}^{n-1}p_i$ and we have used the definition \eqref{defVarTrace}. 

\section{Decomposition of the 4T} 
The analysis in momentum space allows to identify the contributions generated by the breaking of the conformal symmetry, after renormalization, in a direct manner. For this purpose we will be using the longitudinal transverse L/T decomposition of the correlator presented in \cite{Bzowski:2013sza} for 3-point functions, extending it to the 4T. This procedure has been investigated in detail for 3-point functions in \cite{Coriano:2018bsy} in the context of a perturbative approach \cite{Coriano:2018zdo}. The perturbative analysis in free field theory shows how renormalization acts on the two L/T subspaces, forcing the emergence of a trace in the longitudinal sector.   \\
Due to the constraint imposed by conformal symmetry (i.e. their CWI's), the correlation functions can be decomposed into a transverse-traceless and a semilocal part. The term semilocal refers to contributions which are obtained from the conservation and trace Ward identities. Of an external off- shell graviton only its spin-2 component will couple to transverse-traceless part. \\
In general, by decomposing the gravitational fluctuations into their transverse-traceless and spin-1 and spin-0 components one finds an interesting separation of the anomaly effective action which can be useful also in a phenomenological context. We will address this point in a forthcoming paper.  \\
The split of the energy momentum operator in terms of a transverse traceless ($tt$) part and of a longitudinal (local) part \cite{Bzowski:2013sza} is defined in the form
\begin{equation}
T^{\mu_i\nu_i}(p_i)\equiv t^{\mu_i\nu_i}(p_i)+t_{loc}^{\mu_i\nu_i}(p_i)\label{decT}
\end{equation}
with
\begin{align}
\label{loct}
t^{\mu_i\nu_i}(p_i)&=\Pi^{\mu_i\nu_i}_{\alpha_i\beta_i}(p)\,T^{\alpha_i \beta_i}(p_i)\\
t_{loc}^{\mu_i\nu_i}(p_i)&=\Sigma^{\mu_i\nu_i}_{\alpha_i\beta_i}(p)\,T^{\alpha_i \beta_i}(p_i).
\end{align}
We have introduced the transverse-traceless ($\Pi$), transverse-trace $(\tau)$ and 
longitudinal ($\mathcal{I}$) projectors, given respectively by 
\begin{align}
\label{prozero}
\pi^{\mu}_{\alpha} & = \delta^{\mu}_{\alpha} - \frac{p^{\mu} p_{\alpha}}{p^2},  \qquad \tilde{\pi}^{\mu}_{\alpha} =\frac{1}{d-1}\pi^{\mu}_{\alpha} \\\
\Pi^{\mu \nu}_{\alpha \beta} & = \frac{1}{2} \left( \pi^{\mu}_{\alpha} \pi^{\nu}_{\beta} + \pi^{\mu}_{\beta} \pi^{\nu}_{\alpha} \right) - \frac{1}{d - 1} \pi^{\mu \nu}\pi_{\alpha \beta}\label{TTproj}, 
\end{align}
\begin{align}
\mathcal{J}^{\mu\nu}_{\alpha\beta}&=\frac{1}{p^2}p_{\beta}\left( p^{\mu}\delta^{\nu}_{\alpha} +p^{\nu}\delta^{\mu}_{\alpha} -
\frac{p_{\alpha}}{d-1}( \delta^{\mu\nu} +(d-2)\frac{p^\mu p^\nu}{p^2})    \right)\\
\mathcal{I}^{\mu\nu}_{\alpha\beta}&=\frac{1}{2}\left(\mathcal{J}^{\mu\nu}_{\alpha\beta} +\mathcal{J}^{\mu\nu}_{\beta\alpha}\right) \qquad \tau^{\mu\nu}_{\alpha\beta} =\tilde{\pi}^{\mu \nu}\delta_{\alpha \beta}\\
\mathcal{I}^{\mu\nu}_{\alpha}&=\frac{1}{p^2}\left( p^{\mu}\delta^{\nu}_{\alpha} +
p^{\nu}\delta^{\mu}_{\alpha} -
\frac{p_{\alpha}}{d-1}( \delta^{\mu\nu} +(d-2)\frac{p^\mu p^\nu}{p^2}  \right)\label{proone}\\
\mathcal{I}^{\mu\nu}_{\alpha\beta}&=\frac{1}{2}\left(p_\beta \mathcal{I}^{\mu\nu}_{\alpha}
+ p_\alpha \mathcal{I}^{\mu\nu}_{\beta}\right)\label{protwo}
\end{align}
with 
\begin{align}
\delta^{(\mu}_\alpha\delta^{\nu)}_{\beta}&=\Pi^{\mu \nu}_{\alpha \beta} +\Sigma^{\mu\nu}_{\alpha\beta} \\
\Sigma^{\mu_i\nu_i}_{\alpha_i\beta_i}&\equiv\mathcal{I}^{\mu_i\nu_i}_{\alpha_i\beta_i} +\tau^{\mu_i\nu_i}_{\alpha_i\beta_i}\notag\\
&=\frac{1}{p_i^2}\Big[2\delta^{(\nu_i}_{(\alpha_i}p_i^{\mu_i)}p_{i\,\beta_i)}-\frac{p_{i\alpha_i}p_{i\beta_i}}{(d-1)}\left(\delta^{\mu_i\nu_i}+(d-2)\frac{p_i^{\mu_i}p_i^{\nu_i}}{p_i^2}\right)\Big]+\frac{1}{(d-1)}\pi^{\mu_i\nu_i}(p_i)\delta_{\alpha_i\beta_i}\label{Lproj}.
\end{align}
Notice that we have combined together the operators $\mathcal{I}$ and $\tau$ into a projector 
$\Sigma$ which defines the local components of a given tensor $T$, according to \eqref{loct}, which are proportional both to a given momentum $p$ (the longitudinal contribution) and to the trace parts. Both $\Pi$ and $\tau$ are transverse by construction, while $\mathcal{I}$ is longitudinal and of zero trace.

The projectors induce a decomposition respect to a specific momentum $p_i$.
By using \eqref{decT}, the entire correlator is written as
\begin{align}
&\braket{T^{\mu_1\nu_1}(p_1)T^{\mu_2\nu_2}(p_2)T^{\mu_3\nu_3}(p_3)T^{\mu_4\nu_4}(\bar{p}_4)}=\notag\\
&=\braket{t^{\mu_1\nu_1}(p_1)t^{\mu_2\nu_2}(p_2)t^{\mu_3\nu_3}(p_3)t^{\mu_4\nu_4}(\bar{p}_4)}+\braket{T^{\mu_1\nu_1}(p_1)T^{\mu_2\nu_2}(p_2)T^{\mu_3\nu_3}(p_3)T^{\mu_4\nu_4}(\bar{p}_4)}_{loc}\label{decTTTT}
\end{align} 
where the first contribution is the transverse-traceless part which satisfies  by construction the conditions
\begin{equation}
\begin{split}
p_{i\,\mu_i}\braket{t^{\mu_1\nu_1}(p_1)t^{\mu_2\nu_2}(p_2)t^{\mu_3\nu_3}(p_3)t^{\mu_4\nu_4}(\bar{p}_4)}&=0,\qquad i=1,2,3,4\,,\\
\delta_{\mu_i\nu_i}\braket{t^{\mu_1\nu_1}(p_1)t^{\mu_2\nu_2}(p_2)t^{\mu_3\nu_3}(p_3)t^{\mu_4\nu_4}(\bar{p}_4)}&=0,\qquad i=1,2,3,4.
\end{split}\label{ttproperties}
\end{equation}
It is clear now that only the second term in \eqref{decTTTT} contributes entirely to the conservation WIs. Thus, the proper new information on the form factors of the 4-point function is entirely encoded in its transverse-traceless ($tt$) part, since the remaining longitudinal + trace contributions,  corresponding to the local term, are related to lower point functions. 

\subsection{Projecting the Conformal Ward Identities}
The action of $D$ and $\mathcal{K}$ on the 4T in \eqref{Dequ} and \eqref{Keq}, after the projection on the transverse traceless component, simplifies.
We start by considering at the dilatation operator $D$ which has the property of leaving unchanged the two subspaces identified by the $\Pi$ and $\Sigma$ projectors. 
These properties can be summarized by the relations
\begin{equation}
\begin{split}
		&\Sigma^{\mu_i\nu_i}_{\rho_i\sigma_i}(p_i)\,D\,\braket{t^{\mu_1\nu_1}(p_1)t^{\mu_2\nu_2}(p_2)t^{\mu_3\nu_3}(p_3)t^{\mu_4\nu_4}(p_4)}=0,\\
		&\Pi^{\mu_i\nu_i}_{\rho_i\sigma_i}(p_i)\,D\,\braket{T^{\mu_1\nu_1}(p_1)T^{\mu_2\nu_2}(p_2)T^{\mu_3\nu_3}(p_3)T^{\mu_4\nu_4}(p_4)}_{loc}=0,
\end{split}	\qquad\qquad  i=1,\dots,4.\label{projD}
\end{equation}
having used the properties of orthogonality and idempotence of the projectors. \\
For this reason, if one wants to project the dilatation equation \eqref{Dequ}, by using the transverse-traceless and longitudinal projectors, there are only two ways of doing it. These are the cases where we have either four $\Pi$'s or four $\Sigma$'s, due to the relations in \eqref{projD}. Therefore, the only relevant projected dilatation WIs are
\begin{align}
	\Pi^{\mu_1\nu_1}_{\alpha_1\beta_1}(p_1)\Pi^{\mu_2\nu_2}_{\alpha_2\beta_2}(p_2)\Pi^{\mu_3\nu_3}_{\alpha_3\beta_3}(p_3)\Pi^{\mu_4\nu_4}_{\alpha_4\beta_4}(\bar p_4)\,D\,\braket{T^{\alpha_1\beta_1}(p_1)T^{\alpha_2\beta_2}(p_2)T^{\alpha_3\beta_3}(p_3)T^{\alpha_4\beta_4}(\bar{p}_4)}&=0\\
	\Sigma^{\mu_1\nu_1}_{\alpha_1\beta_1}(p_1)\Sigma^{\mu_2\nu_2}_{\alpha_2\beta_2}(p_2)\Sigma^{\mu_3\nu_3}_{\alpha_3\beta_3}(p_3)\Sigma^{\mu_4\nu_4}_{\alpha_4\beta_4}(\bar p_4)\,D\,\braket{T^{\alpha_1\beta_1}(p_1)T^{\alpha_2\beta_2}(p_2)T^{\alpha_3\beta_3}(p_3)T^{\alpha_4\beta_4}(\bar{p}_4)}&=0
\end{align}
that can be simplified as
\begin{align}
	\Pi^{\mu_1\nu_1}_{\alpha_1\beta_1}(p_1)\Pi^{\mu_2\nu_2}_{\alpha_2\beta_2}(p_2)\Pi^{\mu_3\nu_3}_{\alpha_3\beta_3}(p_3)\Pi^{\mu_4\nu_4}_{\alpha_4\beta_4}(\bar p_4)\,D\,\braket{t^{\alpha_1\beta_1}(p_1)t^{\alpha_2\beta_2}(p_2)t^{\alpha_3\beta_3}(p_3)t^{\alpha_4\beta_4}(\bar{p}_4)}&=0,\label{Dtttt}\\[2ex]
	\Sigma^{\mu_1\nu_1}_{\alpha_1\beta_1}(p_1)\Sigma^{\mu_2\nu_2}_{\alpha_2\beta_2}(p_2)\Sigma^{\mu_3\nu_3}_{\alpha_3\beta_3}(p_3)\Sigma^{\mu_4\nu_4}_{\alpha_4\beta_4}(\bar p_4)\,D\,\braket{T^{\alpha_1\beta_1}(p_1)T^{\alpha_2\beta_2}(p_2)T^{\alpha_3\beta_3}(p_3)T^{\alpha_4\beta_4}(\bar{p}_4)}_{loc}&=0\label{DTTTTloc},
\end{align}
once we insert the decomposition of the 4T and use Eqs. \eqref{projD}. It is worth mentioning that \eqref{DTTTTloc} does not impose any additional constraints on the $4$-point function. This because the longitudinal part of the correlator is explicitly given in terms of lower point functions, and \eqref{DTTTTloc} is related to the dilatation WIs of the $3$- and $2$-point functions. The constraints on the 4-point function will be derived from \eqref{Dtttt}, which is related to the transverse traceless part of the correlator. 

Turning our attention towards the special CWIs, we observe that the action of the $\mathcal{K}^\kappa$ operator on the transverse-traceless part gives a result that it is still transverse and traceless
\begin{equation}
	\begin{split}
		&p_{i\mu_i}\,\mathcal{K}^\kappa\braket{t^{\mu_1\nu_1}(p_1)t^{\mu_2\nu_2}(p_2)t^{\mu_3\nu_3}(p_3)t^{\mu_4\nu_4}(p_4)}=0,\\
		&\delta_{\mu_i\nu_i}\,\mathcal{K}^\kappa\braket{t^{\mu_1\nu_1}(p_1)t^{\mu_2\nu_2}(p_2)t^{\mu_3\nu_3}(p_3)t^{\mu_4\nu_4}(p_4)}=0
	\end{split}
	\qquad i=1,\dots,4 \label{propKTTT}
\end{equation}
This can be shown exactly as in  the case of the 3T discussed in previous works. 
This property of $\mathcal{K}^\kappa$ allows us to identify the  relevant subspace where the special CWIs act. \\
As already pointed out, the transverse-traceless part of the 4T is the only part of the  correlator which is really unknown. Indeed, the longitudinal components can be expressed in terms of 2- and 3-point functions, by using the conservation and trace WIs.  Therefore, the  special CWIs will be constraining that unknown part, which can be parametrized in terms of a certain number of independent for factors, as in the 3T case. \\
By using the properties of the projectors $\Sigma$ and $\Pi$, from \eqref{propKTTT} one derives the relation
\begin{equation}
	\Sigma^{\rho_i\sigma_i}_{\mu_i\nu_i}(p_i)K^\kappa\braket{t^{\mu_1\nu_1}(p_1)t^{\mu_2\nu_2}(p_2)t^{\mu_3\nu_3}(p_3)t^{\mu_4\nu_4}(p_4)}=0, \qquad i=1,\dots,4,
\end{equation}
since the action of $K^\kappa$, as just mentioned, is endomorphic on the transverse traceless sector. 
With this result in mind, using the projectors $\Pi$ and $\Sigma$ we project \eqref{Keq} into all the possible subspaces and observe that when at least one $\Sigma$ is present, the equations reduce to the form
\begin{align}
	0&=\Sigma^{\rho_i\sigma_i}_{\mu_i\nu_i}(p_i)\mathcal{K}^\kappa\braket{T^{\mu_1\nu_1}(p_1)T^{\mu_2\nu_2}(p_2)T^{\mu_3\nu_3}(p_3)T^{\mu_4\nu_4}(p_4)}\notag\\
	&=\Sigma^{\rho_i\sigma_i}_{\mu_i\nu_i}(p_i)\mathcal{K}^\kappa\bigg[\braket{t^{\mu_1\nu_1}(p_1)t^{\mu_2\nu_2}(p_2)t^{\mu_3\nu_3}(p_3)t^{\mu_4\nu_4}(p_4)}+\braket{T^{\mu_1\nu_1}(p_1)T^{\mu_2\nu_2}(p_2)T^{\mu_3\nu_3}(p_3)T^{\mu_4\nu_4}(p_4)}_{loc}\bigg]\notag\\
	&=\Sigma^{\rho_i\sigma_i}_{\mu_i\nu_i}(p_i)\mathcal{K}^\kappa\bigg(\braket{T^{\mu_1\nu_1}(p_1)T^{\mu_2\nu_2}(p_2)T^{\mu_3\nu_3}(p_3)T^{\mu_4\nu_4}(p_4)}_{loc}\bigg), \qquad i=1,\dots,4.
\end{align}
The equations that result
\begin{equation}
\label{ccc}
	\Sigma^{\rho_i\sigma_i}_{\mu_i\nu_i}(p_i)\mathcal{K}^\kappa\bigg(\braket{T^{\mu_1\nu_1}(p_1)T^{\mu_2\nu_2}(p_2)T^{\mu_3\nu_3}(p_3)T^{\mu_4\nu_4}(p_4)}_{loc}\bigg)=0, \qquad  i=1,\dots,4
\end{equation}
will involve only 3- and 2-point functions, via the canonical Ward identities. For this reason, being the structure of the equations hierarchical, if we have already solved for the correlators of lower orders, no new constraint will be induced by \eqref{ccc}.\\
 The only  significant constraint will be derived when acting with 4 $\Pi$ projectors. For this reason we are interested in studying the equation
\begin{equation}
\label{ddd}
\Pi^{\mu_1\nu_1}_{\alpha_1\beta_1}(p_1)\Pi^{\mu_2\nu_2}_{\alpha_2\beta_2}(p_2)\Pi^{\mu_3\nu_3}_{\alpha_3\beta_3}(p_3)\Pi^{\mu_4\nu_4}_{\alpha_4\beta_4}(\bar p_4)\,\mathcal{K}^\kappa\,\braket{T^{\alpha_1\beta_1}(p_1)T^{\alpha_2\beta_2}(p_2)T^{\alpha_3\beta_3}(p_3)T^{\alpha_4\beta_4}(\bar{p}_4)}=0.
\end{equation}
If we insert the decomposition \eqref{decTTTT} into the equation above, one can prove that terms containing two or more $t_{loc}$ operators will vanish when projected on the transverse traceless component, for instance
\begin{align}
	\Pi^{\mu_1\nu_1}_{\alpha_1\beta_1}(p_1)\Pi^{\mu_2\nu_2}_{\alpha_2\beta_2}(p_2)\Pi^{\mu_3\nu_3}_{\alpha_3\beta_3}(p_3)\Pi^{\mu_4\nu_4}_{\alpha_4\beta_4}(\bar p_4)\,\mathcal{K}^{\kappa}\,\braket{t^{\alpha_1\beta_1}_{loc}(p_1)t^{\alpha_2\beta_2}_{loc}(p_2)T^{\alpha_3\beta_3}(p_3)T^{\alpha_4\beta_4}(\bar{p}_4)}&=0,
\end{align}
where the definitions \eqref{TTproj} and \eqref{Lproj} have been used. 

Proceeding with the reconstruction program, we need the action of the special conformal transformations on the correlators with a single $t_{loc}$, after projecting on the transverse traceless sector. After a lengthy but straightforward calculations we obtain
\begin{align}
	&\Pi^{\mu_1\nu_1}_{\alpha_1\beta_1}(p_1)\Pi^{\mu_2\nu_2}_{\alpha_2\beta_2}(p_2)\Pi^{\mu_3\nu_3}_{\alpha_3\beta_3}(p_3)\Pi^{\mu_4\nu_4}_{\alpha_4\beta_4}(\bar p_4)\,\,\mathcal{K}^{\kappa}\braket{T^{\alpha_1\beta_1}(p_1)T^{\alpha_2\beta_2}(p_2)T^{\alpha_3\beta_3}(p_3)t_{loc}^{\alpha_4\beta_4}(\bar{p}_4)}\notag\\
	&=4\Pi^{\mu_1\nu_1}_{\alpha_1\beta_1}(p_1)\Pi^{\mu_2\nu_2}_{\alpha_2\beta_2}(p_2)\Pi^{\mu_3\nu_3}_{\alpha_3\beta_3}(p_3)\Pi^{\mu_4\nu_4}_{\alpha_4\beta_4}(\bar p_4)\notag\\
	&\bigg\{\bigg[(d-1)\delta^{\kappa\alpha_4}+L_1^{\kappa\alpha_4}+L_2^{\kappa\alpha_4}+L_3^{\kappa\alpha_4}
	\bigg]\left(\frac{p_{4\rho_4}}{p_4^2}\braket{T^{\alpha_1\beta_1}(p_1)T^{\alpha_2\beta_2}(p_2)T^{\alpha_3\beta_3}(p_3)T^{\rho_4\beta_4}(\bar{p}_4)}\right)\notag\\
	&\hspace{1cm}+\left(S^{\kappa\alpha_4}\right)^{\alpha_1}_{\hspace{2ex}\rho_1}\left(\frac{p_{4\rho_4}}{p_4^2}\braket{T^{\rho_1\beta_1}(p_1)T^{\alpha_2\beta_2}(p_2)T^{\alpha_3\beta_3}(p_3)T^{\rho_4\beta_4}(\bar{p}_4)}\right)\notag\\
	&\hspace{1cm}+\left(S^{\kappa\alpha_4}\right)^{\alpha_2}_{\hspace{2ex}\rho_2}\left(\frac{p_{4\rho_4}}{p_4^2}\braket{T^{\alpha_1\beta_1}(p_1)T^{\rho_2\beta_2}(p_2)T^{\alpha_3\beta_3}(p_3)T^{\rho_4\beta_4}(\bar{p}_4)}\right)\notag\\
	&\hspace{1cm}+\left(S^{\kappa\alpha_4}\right)^{\alpha_3}_{\hspace{2ex}\rho_3}\left(\frac{p_{4\rho_4}}{p_4^2}\braket{T^{\alpha_1\beta_1}(p_1)T^{\alpha_2\beta_2}(p_2)T^{\rho_3\beta_3}(p_3)T^{\rho_4\beta_4}(\bar{p}_4)}\right)\bigg\},
\end{align}
where 
\begin{align}
	L_i^{\kappa\alpha_4}\equiv p_i^{\alpha_4}\frac{\partial}{\partial p_{i\,\kappa}}-p_i^\kappa\frac{\partial}{\partial p_{i\,\alpha_4}},\qquad i=1,2,3
\end{align}
are the $SO(4)$ (Lorentz) generators 
and $S_{\mu\nu}$ is the spin part, for which
\begin{align}
	\big(S^{\mu\nu}\big)^{\rho\sigma}=\delta^{\mu\rho}\delta^{\nu\sigma}-\delta^{\mu\sigma}\delta^{\nu\rho}.
\end{align}
By using the Lorentz Ward identities we obtain the expression
\begin{align}
	&\Pi^{\mu_1\nu_1}_{\alpha_1\beta_1}(p_1)\Pi^{\mu_2\nu_2}_{\alpha_2\beta_2}(p_2)\Pi^{\mu_3\nu_3}_{\alpha_3\beta_3}(p_3)\Pi^{\mu_4\nu_4}_{\alpha_4\beta_4}(\bar p_4)\,\,\mathcal{K}^{\kappa}\braket{T^{\alpha_1\beta_1}(p_1)T^{\alpha_2\beta_2}(p_2)T^{\alpha_3\beta_3}(p_3)t_{loc}^{\alpha_4\beta_4}(\bar{p}_4)}\notag\\
	&=\Pi^{\mu_1\nu_1}_{\alpha_1\beta_1}(p_1)\Pi^{\mu_2\nu_2}_{\alpha_2\beta_2}(p_2)\Pi^{\mu_3\nu_3}_{\alpha_3\beta_3}(p_3)\Pi^{\mu_4\nu_4}_{\alpha_4\beta_4}(\bar p_4)\left(\frac{4\,d}{p_4^2}\,\delta^{\alpha_4\kappa}\,p_{4\rho_4}\braket{T^{\alpha_1\beta_1}(p_1)T^{\alpha_2\beta_2}(p_2)T^{\alpha_3\beta_3}(p_3)T^{\rho_4\beta_4}(\bar{p}_4)}\right).
\end{align}
Analogous results are obtained for the other terms involving one $t_{loc}$ operator. \\
In summary, when we project \eqref{Keq} on the transverse traceless components we find
\begin{align}
	0=&\Pi^{\mu_1\nu_1}_{\alpha_1\beta_1}(p_1)\Pi^{\mu_2\nu_2}_{\alpha_2\beta_2}(p_2)\Pi^{\mu_3\nu_3}_{\alpha_3\beta_3}(p_3)\Pi^{\mu_4\nu_4}_{\alpha_4\beta_4}(\bar p_4)\,\,\mathcal{K}^{\kappa}\braket{T^{\alpha_1\beta_1}(p_1)T^{\alpha_2\beta_2}(p_2)T^{\alpha_3\beta_3}(p_3)T^{\alpha_4\beta_4}(\bar{p}_4)}\notag\\
	&=\Pi^{\mu_1\nu_1}_{\alpha_1\beta_1}(p_1)\Pi^{\mu_2\nu_2}_{\alpha_2\beta_2}(p_2)\Pi^{\mu_3\nu_3}_{\alpha_3\beta_3}(p_3)\Pi^{\mu_4\nu_4}_{\alpha_4\beta_4}(\bar p_4)\Bigg[\mathcal{K}^{\kappa}\braket{t^{\alpha_1\beta_1}(p_1)t^{\alpha_2\beta_2}(p_2)t^{\alpha_3\beta_3}(p_3)t^{\alpha_4\beta_4}(\bar{p}_4)}\notag\\
	&\hspace{2cm}+\frac{4\,d}{p_1^2}\,\delta^{\alpha_1\kappa}\,p_{1\rho_1}\braket{T^{\rho_1\beta_1}(p_1)T^{\alpha_2\beta_2}(p_2)T^{\alpha_3\beta_3}(p_3)T^{\alpha_4\beta_4}(\bar{p}_4)}\notag\\
	&\hspace{2cm}+\frac{4\,d}{p_2^2}\,\delta^{\alpha_2\kappa}\,p_{2\rho_2}\braket{T^{\alpha_1\beta_1}(p_1)T^{\rho_2\beta_2}(p_2)T^{\alpha_3\beta_3}(p_3)T^{\alpha_4\beta_4}(\bar{p}_4)}\notag\\
	&\hspace{2cm}+\frac{4\,d}{p_3^2}\,\delta^{\alpha_3\kappa}\,p_{3\rho_3}\braket{T^{\alpha_1\beta_1}(p_1)T^{\alpha_2\beta_2}(p_2)T^{\rho_3\beta_3}(p_3)T^{\alpha_4\beta_4}(\bar{p}_4)}\notag\\
	&\hspace{2cm}+\frac{4\,d}{p_4^2}\,\delta^{\alpha_4\kappa}\,p_{4\rho_4}\braket{T^{\alpha_1\beta_1}(p_1)T^{\alpha_2\beta_2}(p_2)T^{\alpha_3\beta_3}(p_3)T^{\rho_4\beta_4}(\bar{p}_4)}\Bigg]\label{KTTTTdec}.
\end{align}
It is worth mentioning that the last four terms in the previous equations are completely expressible in  terms of lower point functions via the longitudinal WIs. Then, \eqref{KTTTTdec} imposes some constraints on the transverse traceless part of the 4T and connects this part of the correlator with the lower point functions 3T and $TT$. 

\section{Identifying the divergent form factors and their reduction in $d=4$}
The general expression of the $tt$-contributions can be identified by imposing the transversality and trace-free \eqref{ttproperties} conditions on all the possible tensor structures which are allowed by the symmetries of the correlation function. In this section we are going first to proceed with the classification of the divergent ones which are present in the decomposition and are affected by the renormalization.  \\
We introduce a general decomposition of the counterterm for the 4T in general $d$ dimensions, in order to identify such form factors. Their number gets reduced once we move to $d=4$, due to the possibility of expressing the Kronecker $\delta^{\mu\nu}$ in terms of 3 of the 4 momenta of the correlator, and of a linearly independent 4-vector $n^\mu$. The latter is defined via a generalization of the external product by the $\epsilon^{\mu\nu\rho\sigma}$, as we are going to illustrate below (see \cite{Bzowski:2013sza} for the case $d=3$)

The decomposition can be generically written in the form
\begin{align}
&\braket{t^{\mu_1\nu_1}(p_1)t^{\mu_2\nu_2}(p_2)t^{\mu_3\nu_3}(p_3)t^{\mu_4\nu_4}(\bar{p}_4)}\equiv \Pi^{\mu_1\nu_1}_{\alpha_1\beta_1}(p_1)\Pi^{\mu_2\nu_2}_{\alpha_2\beta_2}(p_2)\Pi^{\mu_3\nu_3}_{\alpha_3\beta_3}(p_3)\Pi^{\mu_4\nu_4}_{\alpha_4\beta_4}(\bar{p}_4)\notag\\
&\times\Big\{\sum_{a,b\in\{2,3\}}\, \sum_{c,d\in\{3,4\}}\,\sum_{e,f\in\{4,1\}}\,\sum_{g,h\in\{1,2\}}A^{(8p)}_{a\,b\,c\,d\,e\,f\,g\,h}\ \ p_{a}^{\alpha_1}\,p_{b}^{\beta_1}\,p_{c}^{\alpha_2}\,p_{d}^{\beta_2}\,p_{e}^{\alpha_3}\,p_{f}^{\beta_3}\,p_{g}^{\alpha_4}\,p_{h}^{\beta_4}\notag\\
&\qquad+\Bigg[\delta^{\alpha_1\alpha_2}\sum_{a\in\{2,3\}}\sum_{b\in\{3,4\}}\sum_{c,d\in\{4,1\}}\sum_{e,f\in\{1,2\}} A^{(6p)}_{a\,b\,c\,d\,e\,f}\,p_{a}^{\beta_1}\,p_{b}^{\beta_2}\,p_{c}^{\alpha_3}\,p_{d}^{\beta_3}\,p_{e}^{\alpha_4}\,p_{f}^{\beta_4}+\text{(permutations)}\Bigg]\notag\\
&\qquad+\Bigg[\delta^{\alpha_1\alpha_2}\delta^{\beta_1\beta_2}\sum_{a,b\in\{4,1\}}\sum_{c,d\in\{1,2\}} A^{(I,4p)}_{a\,b\,c\,d}\,p_{a}^{\alpha_3}\,p_{b}^{\beta_3}\,p_{c}^{\alpha_4}\,p_{d}^{\beta_4}+\text{(permutations)}
\Bigg]\notag\\
&\qquad+\Bigg[\delta^{\alpha_1\alpha_2}\delta^{\beta_2\beta_3}\sum_{a\in\{2,3\}}\sum_{b\in\{4,1\}}\sum_{c,d\in\{1,2\}} A^{(II,4p)}_{a\,b\,c\,d}\,p_{a}^{\beta_1}\,p_{b}^{\alpha_3}\,p_{c}^{\alpha_4}\,p_{d}^{\beta_4}+\text{(permutations)}
\Bigg]\notag\\
&\qquad+\Bigg[\delta^{\alpha_1\alpha_2}\delta^{\alpha_3\alpha_4}\sum_{a\in\{2,3\}}\sum_{b\in\{3,4\}}\sum_{c\in\{4,1\}}\sum_{d\in\{1,2\}} A^{(III,4p)}_{a\,b\,c\,d}\,p_{a}^{\beta_1}\,p_{b}^{\beta_2}\,p_{c}^{\beta_3}\,p_{d}^{\beta_4}+\text{(permutations)}
\Bigg]\notag\\
&\qquad+\Bigg[\delta^{\alpha_1\beta_2}\delta^{\alpha_2\beta_3}\delta^{\alpha_3\beta_1}\sum_{a,b\in\{1,2\}} A^{(I,2p)}_{a\,b}\,p_{a}^{\alpha_4}\,p_{b}^{\beta_4}+\text{(permutations)}
\Bigg]\notag\\
&\qquad+\Bigg[\delta^{\alpha_1\beta_2}\delta^{\alpha_2\beta_1}\delta^{\alpha_3\alpha_4}\sum_{a\in\{4,1\}}\sum_{b\in\{1,2\}} A^{(II,2p)}_{a\,b}\,p_{a}^{\beta_3}\,p_{b}^{\beta_4}+\text{(permutations)}
\Bigg]\notag\\
&\qquad+\Bigg[\delta^{\alpha_1\beta_2}\delta^{\alpha_2\beta_4}\delta^{\alpha_4\beta_3}\sum_{a\in\{2,3\}}\sum_{b\in\{4,1\}} A^{(III,2p)}_{a\,b}\,p_{a}^{\beta_1}\,p_{b}^{\alpha_3}+\text{(permutations)}
\Bigg]\notag\\
&\qquad+\Bigg[\delta^{\alpha_1\beta_2}\delta^{\alpha_2\beta_3}\delta^{\alpha_3\beta_4}\delta^{\alpha_4\beta_1} A^{(I,0p)}+\text{(permutations)}\Bigg]+\Bigg[\delta^{\alpha_1\beta_2}\delta^{\alpha_2\beta_1}\delta^{\alpha_3\beta_4}\delta^{\alpha_4\beta_3} A^{(II,0p)}+\text{(permutations)}\Bigg]
\Bigg\}\label{tttt}
\end{align}  
in terms of form factors $A$. We have explicitly labeled the form factors with an index counting the number of momenta that each of them multiplies in the decomposition. This notation will be useful in our discussion below. 
The choice of the independent momenta of the expansion, similarly to the case of 3-point functions, can be different for each set of uncontracted tensor indices. We will choose  
\begin{align}
&(\alpha_1,\beta_1)\leftrightarrow p_2,p_3,\quad(\alpha_2,\beta_2)\leftrightarrow p_3,p_4\notag\\
&(\alpha_3,\beta_3)\leftrightarrow p_4,p_1,\quad(\alpha_4,\beta_4)\leftrightarrow p_1,p_2.
\end{align}
as basis of the expansion for each pair of indices shown above.
The linear dependence of $p_4$, which we will impose at a later stage, is not in contradiction with this choice, which allows to reduce the number of form factors, due to the presence of a single $tt$ projector for each external momentum. This strategy has been introduced in \cite{Bzowski:2013sza} for 3-point functions and it allows to reduce the number of form factors. These, in eq. \eqref{tttt} are functions of the six kinematic invariants
\begin{align}
p_i=\sqrt{p_i^2},\quad\,i=1,2,3,4,\qquad
s=\sqrt{(p_1+p_2)^2},\ t=\sqrt{(p_2+p_3)^2} 
\end{align}
or equivalently, in a completely symmetric formulation, they are functions of the six invariants
\begin{align}
s_{ij}=\sqrt{p_i\cdot p_j},\quad \,i\ne j\in\{1,2,3,4\}. 
\end{align}
As already mentioned, the local part of the 4T can be expressed entirely in terms of three- and two-point 
functions due to the transverse and trace WIs. The explicit form of the local contribution is indeed given by the expression
\begin{align}
&\braket{T^{\mu_1\nu_1}(p_1)T^{\mu_2\nu_2}(p_2)T^{\mu_3\nu_3}(p_3)T^{\mu_4\nu_4}(\bar{p}_4)}_{loc}=\notag\\ &=\Bigg[\braket{t_{loc}^{\mu_1\nu_1}(p_1)T^{\mu_2\nu_2}(p_2)T^{\mu_3\nu_3}(p_3)T^{\mu_4\nu_4}(\bar{p}_4)}+(1\,2)+(1\, 3)+(1\, 4)\Bigg]\notag\\
&-\Bigg[\braket{t_{loc}^{\mu_1\nu_1}(p_1)t_{loc}^{\mu_2\nu_2}(p_2)T^{\mu_3\nu_3}(p_3)T^{\mu_4\nu_4}(\bar{p}_4)}+(1\,3)+(1\,4)+(2\,3)+(2\,4)+(1\,3)(2\,4)\Bigg]\notag\\
&+\Bigg[\braket{t_{loc}^{\mu_1\nu_1}(p_1)t_{loc}^{\mu_2\nu_2}(p_2)t_{loc}^{\mu_3\nu_3}(p_3)T^{\mu_4\nu_4}(\bar{p}_4)}+(1\,4)+(2\,4)+(3\,4)\Bigg]\notag\\
&-\braket{t_{loc}^{\mu_1\nu_1}(p_1)t_{loc}^{\mu_2\nu_2}(p_2)t_{loc}^{\mu_3\nu_3}(p_3)t_{loc}^{\mu_4\nu_4}(\bar{p}_4)}\label{longTer},
\end{align}
where the insertion of $t_{loc}$ gives 
\begin{align}
&\braket{t_{loc}^{\mu_1\nu_1}(p_1)T^{\mu_2\nu_2}(p_2)T^{\mu_3\nu_3}(p_3)T^{\mu_4\nu_4}(\bar{p}_4)}\notag\\
&=\frac{1}{p_1^2}\Big[2\delta^{(\nu_1}_{(\alpha_1}p_1^{\mu_1)}p_{1\,\beta_1)}-\frac{p_{1\alpha_1}p_{1\beta_1}}{(d-1)}\left(\delta^{\mu_1\nu_1}+(d-2)\frac{p_1^{\mu_1}p_1^{\nu_1}}{p_1^2}\right)\Big]\braket{T^{\alpha_1\beta_1}(p_1)T^{\mu_2\nu_2}(p_2)T^{\mu_3\nu_3}(p_3)T^{\mu_4\nu_4}(\bar{p}_4)}\notag\\
&\qquad+\frac{1}{(d-1)}\pi^{\mu_1\nu_1}(p_1)\delta_{\alpha_1\beta_1}\braket{T^{\alpha_1\beta_1}(p_1)T^{\mu_2\nu_2}(p_2)T^{\mu_3\nu_3}(p_3)T^{\mu_4\nu_4}(\bar{p}_4)}\label{tloc1}.
\end{align}
Notice that the right-hand-side of \eqref{tloc1} is entirely expressed in terms of lower-point correlation functions, due to the WI's \eqref{transverseP} and \eqref{traceP}. Similar relations hold for all the other contributions contained in \eqref{longTer}.

\subsection{Divergences and Renormalization}
In order to investigate the implications of the CWIs on the anomaly contributions of the 4T, 
we turn to 4 spacetime dimensions and discuss the anomaly form of such equations. We start from the dilatation WIs.\\
The scale invariance of the correlator is expressed through the Dilatation Ward Identity \eqref{Dequ},
which in terms of the corresponding form factors takes to scalar equations of the form
\begin{align}
\left(d-n_p-\sum_{i=1}^3\,p_i^\mu\frac{\partial}{\partial p_i^\mu}\right)A^{(np)}_{a\dots}=0,\label{DA}
\end{align}
where $n_p$ is the number of momenta multiplying the form factors in the decomposition \eqref{tttt}. Eqs.  \eqref{DA} characterize the scaling behaviour of the form factors, and allow to identify quite easily those among all which will be manifestly divergent in the UV. For instance, the form factor  corresponding to eight momenta in \eqref{tttt} has degree $d-8$ and is finite in $d=4$. This simple dimensional counting can be done for all the form factors allowed by the symmetry of the correlator.\\
We have summarised the UV behaviour in the table below
\begin{center}
	\begin{tabular}{  c | c | c | c | c | c }
		 Form Factor &$A^{(8p)}$&$A^{(6p)}$&$A^{(4p)}$&$A^{(2p)}$&$A^{(0p)}$\\ 
		\hline Degree &$d-8$ & $d-6$ & $d-4$ & $d-2$ &$d$		
		\\ 	\hline
		UV divergent in $d=4$ & \xmark & \xmark & \cmark & \cmark &\cmark \\
	\end{tabular}
\end{center}

The expected form factors that will manifest divergences in $d=4$ are those of the form $A^{(4p)}$, $A^{(2p)}$ and $A^{(0p)}$ in \eqref{tttt}, which will show up as single poles in the regulator $\varepsilon$. The procedure of renormalization, obtained by the inclusion of the counterterm \eqref{counter}, will remove these divergences and will generate an anomaly. An explicit check of this cancellation is contained in \cite{Serino:2020pyu}, in the case of a conformally coupled 
free scalar theory.
\subsection{Explicit form of the divergences}
Being the anomaly generated by the renormalization procedure, it is possible to derive the structure of the anomaly contributions and the form of the anomalous CWIs' by applying the reconstruction procedure to the counterterms. \\
One can also work out the explicit structure of the counterterms for each of the divergent form factors $A^{(np)}$ identified above. Their renormalization is obviously guaranteed by the general counterterm Lagrangian \eqref{counter}. 
For example, considering the decomposition identified in \eqref{tttt}, the corresponding counterterms to the $A^{(np)}$ form factors, expanded in series of power in $\varepsilon=d+4$, can be determined in the form 
\begin{align}
A_{4411,\,count}^{(I,4p)}=&\frac{2}{\varepsilon}\,b'
+O(\varepsilon^0)\\
A_{2411,\,count}^{(II,4p)}=&\frac{4}{\varepsilon}\,b'+O(\varepsilon^0)\\
A_{2341,\,count}^{(III,4p)}=&-\frac{5}{8\,\varepsilon}\,\left(b-6\,b'\right)+O(\varepsilon^0)\\
A_{11,\,count}^{(I,2p)}=&-\frac{7}{16\,\varepsilon}\bigg[b\left(p_1^2-4p_2^2+p_3^2+3s^2+3t^2\right) - b' \left(p_1^2+4
p_2^2+p_3^2+4 p_4^2-s^2-t^2\right)\bigg]+O(\varepsilon^0)\\
A_{41\,,count}^{(II,2p)}=&-\frac{7}{8\,\varepsilon}\bigg[\frac{b}{6}\left(7p_1^2+7p_2^2+6p_3^2+6p_4^2-9s^2\right)
+b' \left(t^2-s^2\right)\bigg]+O(\varepsilon^0)\\
A_{34\,,count}^{(III,2p)}=&\frac{1}{2\,\varepsilon}\bigg[b\left(p_1^2-p_2^2-2p_3^2-6p_4^2-s^2\right)+ b' \left(3 p_1^2+p_2^2+2p_3^2-2 p_4^2+s^2\right)\bigg]+O(\varepsilon^0)\\
A^{(I,0p)}_{counter}=&-\frac{1}{\varepsilon}\Bigg\{\frac{b}{8}\bigg[12 p_1^4+p_1^2 \Big(2
(p_2^2+p_4^2)+8 p_3^2-10 \left(s^2+t^2\right)\Big)+12p_2^4+p_2^2 \Big(2 p_3^2+8 p_4^2-10 \left(s^2+t^2\right)\Big)\notag\\
&\hspace{1cm}+p_4^2 \Big(2 p_3^2-10\left(s^2+t^2\right)\Big)
+12 p_3^4-10\,p_3^2 \left(s^2+t^2\right)
+12p_4^4+10 s^4+4s^2 t^2+10t^4\bigg]\notag\\
&\hspace{0.8cm}+\frac{b'}{4}\bigg[-p_1^2
\left(p_2^2+p_4^2+3 \left(s^2+t^2\right)\right)+2
\left(p_1^2\right)^2-p_2^2 \left(p_3^2+3
\left(s^2+t^2\right)\right)+2
\left(p_2^2\right)^2\notag\\
&\hspace{1.5cm}-p_4^2 \left(p_3^2+3
\left(s^2+t^2\right)\right)+p_3^2 \left(2 p_3^2-3
\left(s^2+t^2\right)\right)+2 \left(p_4^2\right)^2+3
s^4+2 s^2 t^2+3 t^4\bigg]\Bigg\}+O(\varepsilon^0)\\
A^{(II,0p)}_{count}=&\frac{1}{24\varepsilon}\Bigg\{b \bigg[-\left(p_1^2+p_2^2\right) \left(p_3^2+p_4^2+9 s^2\right)+6 p_1^4+6 p_2^4-9 p_3^2 s^2+6 p_3^4-9 p_4^2 s^2+6 p_4^4+15 s^4\bigg]\notag\\
&\hspace{-0.5cm}-6 b' \bigg[p_1^2\,p_2^2+p_4^2\,p_3^2 -s^2 t^2-2 s^4-t^4-p_1^4-p_2^4-p_3^4-p_4^4+\left(p_1^2+p_2^2+p_3^2+p_4^2\right) \left(2 s^2+t^2\right)\bigg]\Bigg\}+O(\varepsilon^0),
\end{align}
where the finite contributions contain the scale dependence $\mu$. 
The scalar case, discussed in \cite{Serino:2020pyu}, can be obtained by assigning specific values to the $b$ and $b'$ coefficients of the counterterms \eqref{counter}, expressed in terms of a scalar field content.
\subsection{A simplified decomposition}
Using \eqref{deltaepsilon} and the fact that the transverse traceless sector is identified by a decomposition which is contracted with four $\Pi^{\m \n}_{\a \b}$ projectors, and for $i\neq j \neq k \neq l=1,2,3,4$ generates terms with different numbers of external momenta. We illustrate the procedure for a couple of cases. For instance, the form factor which in \eqref{tttt} is characterised by a a single Kronecker $\delta$ and 6 tensorial momenta, will be re-expressed in terms of tensorial expressions either with 8 momenta or of $n n$ terms times 6 momenta. 
At the end, the decomposition will consist of terms proportional to tensor structures which 
are independent of $n^\a_i$; those which are proportional to $n^\a_i n^\b_j$ and will include  a term proportional to $n^{\a_i}n^{\b_j}n^{\b_k}n^{\b_l}$. The last type of  terms do not appear in the case of the 3-point functions of stress energy tensors in $d=3$ \cite{Bzowski:2013sza}.
The new decomposition can be written as
\begin{align}\label{newDecomp}
&\braket{t^{\mu_1\nu_1}(p_1)t^{\mu_2\nu_2}(p_2)t^{\mu_3\nu_3}(p_3)t^{\mu_4\nu_4}(\bar{p}_4)}\equiv \Pi^{\mu_1\nu_1}_{\alpha_1\beta_1}(p_1)\Pi^{\mu_2\nu_2}_{\alpha_2\beta_2}(p_2)\Pi^{\mu_3\nu_3}_{\alpha_3\beta_3}(p_3)\Pi^{\mu_4\nu_4}_{\alpha_4\beta_4}(\bar{p}_4)\notag\\
&\times\Big\{\sum_{a,b\in\{2,3\}}\, \sum_{c,d\in\{3,4\}}\,\sum_{e,f\in\{4,1\}}\,\sum_{g,h\in\{1,2\}}B^{(8p)}_{a\,b\,c\,d\,e\,f\,g\,h}\ \ p_{a}^{\alpha_1}\,p_{b}^{\beta_1}\,p_{c}^{\alpha_2}\,p_{d}^{\beta_2}\,p_{e}^{\alpha_3}\,p_{f}^{\beta_3}\,p_{g}^{\alpha_4}\,p_{h}^{\beta_4}\notag\\
&\qquad+\Bigg[\frac{1}{n^2}\,n^{\alpha_1}n^{\alpha_2}\sum_{a\in\{2,3\}}\sum_{b\in\{3,4\}}\sum_{c,d\in\{4,1\}}\sum_{e,f\in\{1,2\}} B^{(6p)}_{a\,b\,c\,d\,e\,f}\,p_{a}^{\beta_1}\,p_{b}^{\beta_2}\,p_{c}^{\alpha_3}\,p_{d}^{\beta_3}\,p_{e}^{\alpha_4}\,p_{f}^{\beta_4}+\text{(permutations)}\Bigg]\notag\\
&\qquad+\Bigg[\frac{1}{n^4}\,n^{\alpha_1}n^{\alpha_2}n^{\alpha_3}n^{\alpha_4}\sum_{a\in\{2,3\}}\sum_{b\in\{3,4\}}\sum_{c\in\{4,1\}}\sum_{d\in\{1,2\}} B^{(4p)}_{a\,b\,c\,d}\,p_{a}^{\beta_1}\,p_{b}^{\beta_2}\,p_{c}^{\beta_3}\,p_{d}^{\beta_4}+\text{(permutations)}
\Bigg]
\end{align}
in terms of form factors $B$.  By power counting, we see that in \eqref{newDecomp}, the only form factors that will manifest divergences in $d=4$ are those of form $B^{(4p)}$. The $B$'s will be written in terms of the form factors $A$'s of the general decomposition, simplifying the divergent terms and leading a structure with only one type of form factor divergent, i. e. $B^{(4p)}$.
\section{Renormalization and anomaly in $d=2$ and $d=4$: the $TT$ case } 
\label{renren}
 It is clear, from the analysis presented in the previous sections, that $V_E$ induces only 
a finite renormalization on the $n$-point function. However, as we will be showing here using the example of the correlator $TT$ in $d=2$, it is consistent to introduce a topological counterterm to renormalize a correlator, even if this does not need one. \\
In $d=2$ the TT does not exhibit any singularity, since a $1/\varepsilon$ behaviour in a kinematical prefactor is accompanied by a tensor structure which vanishes as $d\to 2$. We describe the steps in this case, and move to the case of $d=4$ after that.\\
In $d=2$ the $TT$ does not exhibit any divergence. 
A one-loop calculation yields the correlator of two conserved traceless stress tensors as
\begin{align}
	\braket{T^{\mu_1\nu_1}(p)T^{\mu_2\nu_2}(-p)}= \frac{ c(d)}{(d-2)}\left(p^2\right)^{d/2}\Pi^{\mu_1\nu_1\mu_2\nu_2}_{(d)}(p)\label{TT2dim}
\end{align}
where $\Pi_{(d)}$ is the transverse traceless projector defined in general $d$ dimensions and the constant $c(d)$ is defined as
\begin{align}
	c(d)=4\,c_T \left(\frac{\pi}{4}\right)^{d/2}\frac{(d-1)\Gamma\left(2-\frac{d}{2}\right)}{\Gamma\left(d+2\right)}
\end{align}
with $c_T$ depending on the matter field realization of the conformal invariant action. It is worth noticing  that the constant $c(d)$ is finite for any $d>1$. 

In dimensional regularization \eqref{TT2dim} the correlator in $d=2(1+\varepsilon)$ takes the following form
\begin{align}
	\braket{T^{\mu_1\nu_1}(p)T^{\mu_2\nu_2}(-p)}= \frac{ c(2+2\varepsilon)}{2\varepsilon}\left(p^2\right)^{1+\varepsilon}\Pi^{\mu_1\nu_1\mu_2\nu_2}_{(d=2+2\varepsilon)}(p)\label{TTreg}
\end{align}
with the appearance of the UV divergence as $1/\varepsilon$ pole after the expansion around $d=2$.
However the limit $\varepsilon\to 0$ is ambiguous due to the fact that the transverse traceless projector in two dimensions goes to zero. This properties is the consequence of the tensor degeneracy when $n=d$ for an $n$-point function. The way out of this indeterminacy is to 
introduce a $n$-$p$ decomposition of the Kronecker $\delta_{\mu\nu}$  and perform the limit 
 to $d=2$ of the entire tensor structure in order to prove the finiteness of the result, as discussed in \cite{Deser:1993yx, Bzowski:2017poo}.\\
Indeed, for the case of the two point function in $d=2$, one could define the independent momentum $n^\mu$ using the Levi-Civita tensor $\epsilon^{\mu\nu}$ as
\begin{align}
	n^{\mu}=\epsilon^{\mu\nu}p_\nu,\label{indMom}
\end{align}
 orthogonal to the other momentum $p_\mu$. Having such two independent momenta $p_\mu,\,n_\mu$, then the metric is not an independent tensor and we can rewrite it as
\begin{equation}
	\delta^{\mu\nu}=\sum_{j,k=1}^2q_j^\mu\,q_k^\nu\big(Z^{-1}\big)_{kj},
\end{equation}
with $q_1^\mu=p^\mu$ and $q_2^\mu=n^\mu$, and $Z$ is the gram matrix, i. e. $Z_{ij}=(q_i\cdot q_j)^2_{i,j=1}$. The Gram matrix is trivially $Z=p^2 \mathbb 1$, because $n^2=p^2$. Then, the metric tensor can be written as
\begin{align}
	\delta^{\mu\nu}=\frac{1}{p^2}\left(p^\mu p^\nu+n^\mu n^\nu\right),
\end{align}
for which the transverse projector takes the form
\begin{equation}
	\pi^{\mu\nu}_{(d=2)}(p)\equiv\delta^{\mu\nu}-\frac{p^{\mu}p^{\nu}}{p^2}=\frac{n^\mu n^\nu}{p^2}
\end{equation}
and the transverse traceless projector in $d=2$ vanishes as
\begin{equation}
	\Pi^{\mu_1\nu_1\mu_2\nu_2}_{(d=2)}(p)=\pi^{\mu_1(\mu_2}\pi^{\nu_2)\nu_1}-\pi^{\mu_1\nu_1}\pi^{\mu_2\nu_2}=0.
\end{equation}
Starting from the transverse traceless projector defined in $d$ dimensions and taking into account this degeneracy in $d=2$, we write the projector around $d=2$ as 
\begin{equation}
	\Pi^{\mu_1\nu_1\mu_2\nu_2}_{d=2+2\varepsilon}(p)=\pi^{\mu_1(\mu_2}\pi^{\nu_2)\nu_1}-\frac{1}{1+2\varepsilon}\pi^{\mu_1\nu_1}\pi^{\mu_2\nu_2}=\frac{2\vare}{1+2\varepsilon}\frac{n^{\mu_1}n^{\nu_1}n^{\mu_2}n^{\nu_2}}{p^4},
\label{ths}
\end{equation}
making the limit $\varepsilon\to0$ well defined. Notice that we are expanding the parametric 
dependence of the projector in $\varepsilon$ and then we are re-expressing the tensor structure in terms of the non-degenerate $n$-$p$ basis. \\
 Inserting \eqref{ths} in \eqref{TT2dim} one derives a finite result as $\varepsilon\to 0$ \begin{align}
	\braket{T^{\mu_1\nu_1}(p)T^{\mu_2\nu_2}(-p)}_{(d=2)}= \frac{ c(2+2\varepsilon)}{1+2\varepsilon}\left(p^2\right)^{\varepsilon-1}\,n^{\mu_1}n^{\nu_1}n^{\mu_2}n^{\nu_2}\,\,\overset{\varepsilon\to0}{=}\,\,c(2)\frac{n^{\mu_1}n^{\nu_1}n^{\mu_2}n^{\nu_2}}{p^2}
\end{align} 
which violates the tracelessness condition, making manifest the presence of the trace anomaly relation. The approach outlined above, is not relying on the introduction of any counterterm, but we are simply exploiting the degeneracy of the tensor structures in 
$d=2$ in order to to extract a renormalized expression of the correlation function. Notice, 
beside the generation of a trace anomaly in the renormalization process, also the absence of any scale dependence in the result. We are going to reproduce the same features of this result using a different approach, which will require the inclusion of a topological counterterm. 

\subsection{Dimensional Regularization}
This result can be obtained differently following the standard procedure of dimensional regularization scheme and renormalization. The same approach will be used in $d=4$.\\
 We start from the regulated expression of the $TT$ obtained by an expansion of \eqref{TTreg} in power of $\varepsilon$, that gives
\begin{align}
	\braket{T^{\mu_1\nu_1}(p)T^{\mu_2\nu_2}(-p)}_{Reg}=& \frac{ c(2)}{2\varepsilon}\left(p^2\right)\,\Pi^{\mu_1\nu_1\mu_2\nu_2}_{(2+2\varepsilon)}(p)+\frac{c(2)}{2}\Pi^{\mu_1\nu_1\mu_2\nu_2}_{(2)}\,p^2\log p^2+p^2c^\prime(2)\,\Pi^{\mu_1\nu_1\mu_2\nu_2}_{(2)}+O(\varepsilon),\label{RegTT}
\end{align}
In this case we are not making use of the tensor degeneracies in $d=2$ yet. We introduce the only possible counterterm action definable in $d=2$ as
\begin{align}
	S_{ct}=-\frac{1}{\varepsilon}\,\beta_c\,\int d^dx\, \sqrt{-g}\,\mu^{d-2}\,R
\end{align}
which will be included in the process of renormalization. By considering two functional variations of this action with respect to the metric, taking the flat space limit and after a Fourier transform we obtain the contribution
\begin{align}
	\braket{T^{\mu_1\nu_1}(p)T^{\mu_2\nu_2}(-p)}_{Count}=-\frac{\beta_c\,p^2\,\mu^{d-2}}{2\varepsilon}\,\left(\Pi^{\mu_1\nu_1\mu_2\nu_2}_{(d)}(p)-\frac{(d-2)}{(d-1)}\pi^{\mu_1\nu_1}(p)\pi^{\mu_2\nu_2}(p)\right), 
\end{align}
whose expansion in power of $\varepsilon$ is 
\begin{align}
	\braket{T^{\mu_1\nu_1}(p)T^{\mu_2\nu_2}(-p)}_{Count}=-\frac{\beta_c\,p^2}{2\varepsilon}\,\left(\Pi^{\mu_1\nu_1\mu_2\nu_2}_{(2+2\varepsilon)}(p)-\frac{2\varepsilon}{(1+2\varepsilon)}\pi^{\mu_1\nu_1}(p)\pi^{\mu_2\nu_2}(p)\right)\left(1+\varepsilon \log\mu^2\right).\label{CountTT}
\end{align}

Adding the counterterm contribution \eqref{CountTT} to the regularized correlator \eqref{RegTT}, and with the choice of $\beta_c=c(2)$ we remove the divergence. We obtain a finite renormalized result in the limit $\varepsilon\to0$ given by
\begin{align}
	\braket{T^{\mu_1\nu_1}(p)T^{\mu_2\nu_2}(-p)}_{(d=2)}^{Ren}=&\frac{c(2)}{2}\Pi^{\mu_1\nu_1\mu_2\nu_2}_{(2)}\,p^2\log\left( \frac{p^2}{\mu^2}\right)+p^2c^\prime(2)\,\Pi^{\mu_1\nu_1\mu_2\nu_2}_{(2)}+c(2)\,p^2\pi^{\mu_1\nu_1}(p)\pi^{\mu_2\nu_2}(p).\label{TTRen}
\end{align}
This result is not traceless, but is characterized by non zero anomalous trace 
\begin{equation}
	\braket{T^{\mu_1}_{\ \ \ \mu_1}(p)T^{\mu_2\nu_2}(-p)}_{(d=2)}^{Ren}=c(2)\,p^2\,\pi^{\mu_2\nu_2}(p)
\end{equation}
which coincides with the result of taking functional derivative of trace anomaly of the two point function in two dimensions
\begin{align}
	\mathcal{A}=c(2)\,\sqrt{-g}\,R.
\end{align}
While we have reproduced the correct structure of the anomaly, as in the previous section, 
we need to recover by this scheme the scale independence of the result in 
\eqref{TTRen}, which is not apparent from that equation. Notice that $c(2)$ identifies a topological contribution and therefore should be independent of any scale. \\
The apparent anomalous scale-dependence of the two point function is
\begin{align}
\label{rr}
	\mu\frac{\partial}{\partial\mu}	\braket{T^{\mu_1\nu_1}(p)T^{\mu_2\nu_2}(-p)}=-c(2)\,p^2\,\Pi^{\mu_1\nu_1\mu_2\nu_2}_{(2)}.
\end{align}
that shows an apparent contradiction. It is sufficient, at this stage, to use the degeneracy of the tensor structure $\Pi^{\mu_1\nu_1\mu_2\nu_2}_{(2)}$ to immediately realize that the 
the rhs of \eqref{rr} is zero. 
The use of the counterterm, even if it is zero in the dimensions on which we are going to project, of course, allows to obtain the correct expression of the renormalized correlator with iooen indices, while at the same time induces the correct expression of its anomaly.\\
 The advantage in following this procedure is evident especially for multi-point functions. 
\subsection{Longitudinal projectors in $d=4$}
To illustrate the emergence of longitudinal projectors in multi-point correlation functions in $d=4$, we start from the case of the $TT$, then move to the 3T and conclude our discussion with the 4T. 
The simplest context in which to discuss the renormalization of the $TT$ is in free field theory, and include three independent sectors with $n_S$ scalars, $n_F$ fermions and $n_G$ gauge fields. A direct perturbative computation gives 
\begin{align}
\braket{T^{\mu_1\nu_1}(p)T^{\mu_2\nu_2}(-p)}=&-\frac{\p^2\,p^4}{4(d-1)(d+1)}\,B_0(p^2)\,\Pi^{\mu_1\nu_1\mu_2\nu_2}(p)\Big[2(d-1)n_F+(2d^2-3d-8)n_G+n_S\Big]\notag\\
&+\frac{\p^2\,p^4\,n_G}{8(d-1)^2}(d-4)^2(d-2)\p^{\mu_1\nu_1}(p)\p^{\mu_2\nu_2}(p)\,B_0(p^2)\label{TTddim}
\end{align}
where $B_0(p^2)$ is the scalar 2-point function defined as
\begin{equation}
B_0(p^2)=\frac{1}{\pi^\frac{d}{2}}\,\int\,d^d\ell\,\frac{1}{\ell^2\,(\ell-p)^2}
\end{equation}
and \eqref{TTddim} shows the separation of the result into a transverse-traceless $(\Pi)$ and longitudinal part $(\pi^{\mu_1\nu_1})$. Around $d=4$, the projectors are expanded according to the relation 
\begin{equation}
\label{pexp}
\Pi^{\,\mu_1\nu_1\mu_2\nu_2}(p)=\Pi^{(4)\,\mu_1\nu_1\mu_2\nu_2}(p)-\frac{2}{9}\varepsilon\,\pi^{\mu_1\nu_1}(p)\,\pi^{\mu_2\nu_2}(p)+O(\varepsilon^2), 
\end{equation}
where the expansion is performed on the parametric dependence of the projector $\Pi(d)$. 
As usual in DR, the tensor indices are continued to $d$ dimensions and contracted with a d-dimensional Euclidean metric $(\delta^\mu_\mu=d)$.

Using \eqref{pexp} in \eqref{TTddim}, the latter takes the form
\begin{align}
\label{result}
\braket{T^{\mu_1\nu_1}(p)T^{\mu_2\nu_2}(-p)}&=-\frac{\p^2\,p^4}{4}\,\bigg(\frac{1}{\varepsilon}+\bar{B}_0(p^2)\bigg)\,\bigg(\Pi^{(4)\,\mu_1\nu_1\mu_2\nu_2}(p)-\frac{2}{9}\varepsilon\,\pi^{\mu_1\nu_1}(p)\,\pi^{\mu_2\nu_2}(p)+O(\varepsilon^2)\bigg)\notag\\
&\hspace{-2cm}\times\Bigg[\bigg(\frac{2}{5}+\frac{4}{25}\varepsilon+O(\varepsilon^2)\bigg)n_F+\bigg(\frac{4}{5}-\frac{22}{25}\varepsilon+O(\varepsilon^2)\bigg)n_G+\bigg(\frac{1}{15}+\frac{16}{225}\varepsilon+O(\varepsilon^2)\bigg)n_S\Bigg]\notag\\
&\hspace{-1cm}+\frac{\p^2\,p^4\,n_G}{8}\p^{\mu_1\nu_1}(p)\p^{\mu_2\nu_2}(p)\,\bigg(\frac{1}{\varepsilon}+\bar{B}_0(p^2)\bigg)\bigg[\frac{8}{9}\varepsilon^2+\frac{8}{27}\varepsilon^3+O(\varepsilon^4)\bigg]
\end{align}
where $\Pi^{ (4)\,\,\mu_1\nu_1\mu_2\nu_2}(p)$ is the transverse and traceless projector in $d=4$ and $\bar{B}_0(p^2)= 2 - \log(p^2)$ is the finite part in $d=4$ of the scalar integral in the $\overline{MS}$ scheme. The last term of \eqref{result}, generated by the addition of a non-conformal sector ($\sim n_G$), vanishes separately as $\varepsilon\to 0$. Finally, combining all the terms we obtain the regulated ($reg$) expression of the $TT$ around $d=4$ in the form

\begin{align}
\label{regular}
\braket{T^{\mu_1\nu_1}(p)T^{\mu_2\nu_2}(-p)}_{reg}&=-\frac{\p^2\,p^4}{60\,\varepsilon}\Pi^{(4)\,\mu_1\nu_1\mu_2\nu_2}(p)\left(6 n_F + 12 n_G + n_S\right)\notag\\
&\hspace{-3cm}+\frac{\p^2\,p^4}{270}\p^{\mu_1\nu_1}(p)\p^{\mu_2\nu_2}(p)\left(6 n_F + 12 n_G + n_S\right)-\frac{\p^2\,p^4}{300}\bar{B}_0(p^2)\Pi^{\mu_1\nu_1\mu_2\nu_2}(p)\left(30n_F+60n_G+5n_S\right)\notag\\
&\hspace{-3cm}-\frac{\p^2\,p^4}{900}\Pi^{\mu_1\nu_1\mu_2\nu_2}(p)\left(36n_F-198 n_G+16n_S\right)+O(\varepsilon)
\end{align}
The divergence in the previous expression can be removed through the one loop counterterm Lagrangian \eqref{nTcount}. In fact, the second functional derivative of $S_{count}$ with respect to the background metric gives 
\begin{align}
\braket{T^{\mu_1\nu_1}(p)T^{\mu_2\nu_2}(-p)}_{count}&\equiv -\sdfrac{\mu^{-\varepsilon}}{\varepsilon}\bigg(4b\,\big[\sqrt{-g}\,C^2\big]^{\m_1\n_1\m_2\n_2}(p,-p)\bigg)=-\frac{8(d-3)\,\mu^{-\varepsilon}\,b}{(d-2)\,\varepsilon}p^4\Pi^{(d)\,\mu_1\nu_1\mu_2\nu_2}(p)
\end{align}
having used the relation $V_{E}^{\m_1\n_1\m_2\n_2}(p,-p)=0$. In particular, expanding around $d=4$ and using again \eqref{pexp} we obtain
\begin{align}
&\braket{T^{\mu_1\nu_1}(p)T^{\mu_2\nu_2}(-p)}_{count}=\notag\\
&\hspace{1cm}-\frac{8\,b\,p^4}{\varepsilon}\bigg(\Pi^{(4)\,\mu_1\nu_1\mu_2\nu_2}(p)-\frac{2}{9}\varepsilon\,\pi^{\mu_1\nu_1}(p)\,\pi^{\mu_2\nu_2}(p)+O(\varepsilon^2)\bigg)\,\bigg(\frac{1}{2}-\frac{\varepsilon}{2}\left(\frac{1}{2}+\log\mu\right)+O(\varepsilon^2)\bigg)\notag\\
&\hspace{1cm}=-\sdfrac{4\,b}{\varepsilon}p^4\,\Pi^{(4)\,\mu_1\nu_1\mu_2\nu_2}(p)+4\,b\, p^4\bigg[\Pi^{(4)\,\mu_1\nu_1\mu_2\nu_2}(p)+\frac{2}{9}\p^{\mu_1\nu_1}(p)\p^{\mu_2\nu_2}(p)\bigg]+O(\varepsilon)
\end{align}
which cancels the divergence arising in the two point function, if one chooses the parameter $b$ as 
\begin{equation}
\label{choiceparm1}
b=-\frac{3\pi^2}{720}n_S-\frac{9\pi^2}{360}n_F-\frac{18\pi^2}{360}n_G,
\end{equation}
The renormalized 2-point function using  \eqref{choiceparm1} then takes the form  \begin{align}
\braket{T^{\mu_1\nu_1}(p)T^{\mu_2\nu_2}(-p)}_{Ren}&=\braket{T^{\mu_1\nu_1}(p)T^{\mu_2\nu_2}(-p)}+\braket{T^{\mu_1\nu_1}(p)T^{\mu_2\nu_2}(-p)}_{count}\notag\\
&=-\frac{\p^2\,p^4}{60}\bar{B}_0\left(\frac{p^2}{\mu^2}\right)\Pi^{\mu_1\nu_1\mu_2\nu_2}(p)\left(6n_F+12n_G+n_S\right)\notag\\
&\quad-\frac{\p^2\,p^4}{900}\Pi^{\mu_1\nu_1\mu_2\nu_2}(p)\big(126n_F-18n_G+31n_S\big).
\label{tren}
\end{align}
Notice that the final renormalized expression is transverse and traceless. Obviously, this result holds in the case in which we choose a counterterm in such a way that \eqref{oneq} is satisfied.  If we had chosen the $C^2$ counterterm action to satisfy \eqref{twoq}, we would have found the relation 
\begin{equation}
\delta_{\mu_1\nu_1}\langle T^{\mu_1\nu_1}(p_1)T^{\mu_1\nu_2}(-p_1)\rangle=
\mathcal{A}^{\mu_2\nu_2}(p_1),
\end{equation}
where on the right hand side we have the contribution of the $\square R$ term that can be removed by adding a local term $R^2$ in the effective action, obtaining a finite renormalization procedure. 
We will see that the same choice of parameters $b$ given in \eqref{choiceparm1} and for $b'$ as
\begin{equation}\label{choiceparm2}
	b'=\frac{\pi^2}{720}n_S+\frac{11\pi^2}{720}n_F+\frac{31\pi^2}{360}n_G.
\end{equation}
removes the divergences in the three point function, as we are going to discuss below. In fact, as we have seen, the expression of $b$ is related to the renormalization of the two point function, instead $b'$ is intrinsically related to the renormalization of the 3-point function.

It is worth mentioning that the renormalized result of the two point function \eqref{tren} does not contain any trace anomaly contributions but, due to the explicit presence of a $\mu$-dependence, it acquires anomalous dilatation Ward Identities written as
\begin{align}
\mu\frac{\partial}{\partial\mu}\braket{T^{\mu_1\nu_1}(p)T^{\mu_2\nu_2}(-p)}_{Ren}=-2\left[\frac{\pi^2\,p^4}{60}\,\Pi^{\mu_1\nu_1\mu_2\nu_2}(p)\left(6n_F+12n_G+n_S\right)\right].
\end{align}
\subsection{Three-point function}
We can now move to the 3-point function. 
In this case we start showing how the derivation of the anomaly part of the correlator can 
be worked out directly from the CWI's, in a simplified way.\\
As pointed out in \cite{Coriano:2018bsy,Bzowski:2013sza, Bzowski:2017poo }, the 3T correlator can be separated into its transverse-traceless part and in its longitudinal one by using the projectors \eqref{prozero}\eqref{proone}\eqref{protwo} as
\begin{align}
&\braket{T^{\mu_1\nu_1}(p_1)T^{\mu_2\nu_2}(p_2)T^{\mu_3\nu_3}(\bar{p}_3)}=\braket{t^{\mu_1\nu_1}(p_1)t^{\mu_2\nu_2}(p_2)t^{\mu_3\nu_3}(\bar{p}_3)}\notag\\
&\hspace{1cm}+\bigg(\braket{t_{loc}^{\mu_1\nu_1}(p_1)T^{\mu_2\nu_2}(p_2)T^{\mu_3\nu_3}(\bar{p}_3)}+(\text{cyclic perm.})\bigg)\notag\\
&-\bigg(\braket{t_{loc}^{\mu_1\nu_1}(p_1)t_{loc}^{\mu_2\nu_2}(p_2)T^{\mu_3\nu_3}(\bar{p}_3)}+(\text{cyclic perm.})\bigg)+\braket{t_{loc}^{\mu_1\nu_1}(p_1)t_{loc}^{\mu_2\nu_2}(p_2)t_{loc}^{\mu_3\nu_3}(\bar{p}_3)},\label{TTT}
\end{align}
where the transverse-traceless part is decomposed in terms of form factors and independent tensor structures as
\begin{align}
&\braket{t^{\mu_1\nu_1}(p_1)t^{\mu_2\nu_2}(p_2)t^{\mu_3\nu_3}(\bar{p}_3)}=\Pi^{\mu_1\nu_1}_{\a_1\b_1}(p_1)\Pi^{\mu_2\nu_2}_{\a_2\b_2}(p_2)\Pi^{\mu_3\nu_3}_{\a_3\b_3}(\bar{p}_3)\notag\\
&\times\Big\{A_1\,p_2^{\a_1} p_2^{\b_1} \bar{p}_3^{\a_2} p_3^{\b_2} p_1^{\a_3} p_1^{\b_3}+ A_2\,\d^{\b_1\b_2} p_2^{\a_1} p_3^{\a_2} p_1^{\a_3} p_1^{\b_3} 
+ A_2\,(p_1 \leftrightarrow p_3)\, \d^{\b_2\b_3}  p_3^{\a_2} p_1^{\a_3} p_2^{\a_1} p_2^{\b_1} \notag\\
&\hspace{0.8cm}+ A_2\,(p_2\leftrightarrow p_3)\, \d^{\b_3\b_1} p_1^{\a_3} p_2^{\a_1}  p_3^{\a_2} p_3^{\b_2}+ A_3\,\d^{\a_1\a_2} \d^{\b_1\b_2}  p_1^{\a_3} p_1^{\b_3} + A_3(p_1\leftrightarrow p_3)\,\d^{\a_2\a_3} \d^{\b_2\b_3}  p_2^{\a_1} p_2^{\b_1} \notag\\
&\hspace{1.2cm}
+ A_3(p_2\leftrightarrow p_3)\,\d^{\a_3\a_1} \d^{\b_3\b_1}  p_3^{\a_2} p_3^{\b_2} + A_4\,\d^{\a_1\a_3} \d^{\a_2\b_3}  p_2^{\b_1} p_3^{\b_2} + A_4(p_1\leftrightarrow p_3)\, \d^{\a_2\a_1} \d^{\a_3\b_1}  p_3^{\b_2} p_1^{\b_3} \notag\\
&\hspace{3.5cm}+ A_4(p_2\leftrightarrow p_3)\, \d^{\a_3\a_2} \d^{\a_1\b_2}  p_1^{\b_3} p_2^{\b_1} + A_5  \d^{\a_1\b_2}  \d^{\a_2\b_3}  \d^{\a_3\b_1}\Big\},\label{ttt}
\end{align}
while the longitudinal part is expressed in terms of lower point functions by using the CWIs, and takes the form
\begin{align}
p_{1\n_1}\braket{T^{\mu_1\nu_1}(p_1)\,T^{\m_2\n_2}(p_2)\,T^{\m_3\n_3}({p_3})}&=-p_2^{\m_1}\braket{T^{\m_2\n_2}(p_1+p_2)T^{\m_3\n_3}({p_3})}-{p_3^{\m_1}}\braket{T^{\m_2\n_2}(p_2)T^{\m_3\n_3}(p_1+{p_3})}\notag\\
&\hspace{-1.5cm}+p_{2\a}\left[\d^{\m_1\n_2}\braket{T^{\mu_2\a}(p_1+p_2)T^{\m_3\n_3}({p_3})}+\d^{\m_1\m_2}\braket{T^{\nu_2\a}(p_1+p_2)T^{\m_3\n_3}({p_3})}\right]\notag\\
&\hspace{-1.5cm}+{p_{3\a}}\left[\d^{\m_1\n_3}\braket{T^{\mu_3\a}(p_1+{p_3})T^{\m_2\n_2}(p_2)}+\d^{\m_1\m_3}\braket{T^{\nu_3\a}(p_1+{p_3})T^{\m_2\n_2}(p_2)}\right].
\label{long}
\end{align}

In $d=4$ this correlator manifests divergences in the forms of single poles in $1/\varepsilon$ ($\varepsilon=(4-d)$), as for any CFT affected by the trace anomaly. \\
These divergences are present in both the transverse-traceless and longitudinal parts. As discussed in detail in \cite{Coriano:2018bsy}, the counterterm \eqref{nTcount}, for the $3$-point case, renormalizes the correlator \eqref{TTT} by canceling all the divergences with the same choice of the coefficients \eqref{choiceparm1} and \eqref{choiceparm2} but, at the same time, gives extra contributions in the final renormalized 3T from the local/longitudinal part. These extra contributions defines the anomalous part of the correlator. In the case of $n=4$ \eqref{cct} specializes in the obvious form
\begin{align}
&\bigg[\braket{T^{\mu_1\nu_1}T^{\mu_2\nu_2}T^{\mu_3\nu_3}}_{bare}+\braket{T^{\mu_1\nu_1}T^{\mu_2\nu_2}T^{\mu_3\nu_3}}_{count}\bigg]_{d\to4}=\notag\\
&=\braket{T^{\mu_1\nu_1}T^{\mu_2\nu_2}T^{\mu_3\nu_3}}^{(d=4)}_{f}+\braket{T^{\mu_1\nu_1}T^{\mu_2\nu_2}T^{\mu_3\nu_3}}^{(d=4)}_{anomaly}.
\end{align} 

 The anomaly part is given as
\begin{align}
&\braket{T^{\mu_1\nu_1}(p_1)T^{\mu_2\nu_2}(p_2)T^{\mu_3\nu_3}(p_3)}^{(d=4)}_{anomaly}=\notag\\
&=\left(\frac{4}{3}\,\p^{\mu_1\nu_1}(p_1)\,\mathcal{A}^{\mu_2\nu_2\mu_3\nu_3}(p_2,\bar{p}_3)+(\text{perm.})\right)\notag\\
&\hspace{0.5cm}-\left(\frac{4}{9}\,\p^{\mu_1\nu_1}(p_1)\,\p^{\mu_2\nu_2}(p_2)\,\d_{\a_2\b_2}\,\mathcal{A}^{\a_2\b_2\mu_3\nu_3}(p_2,\bar{p}_3)+(\text{perm.})\right)\notag\\
&\hspace{1cm}+\frac{4}{27}\,\p^{\mu_1\nu_1}(p_1)\,\p^{\mu_2\nu_2}(p_2)\,\p^{\mu_3\nu_3}(\bar{p}_3)\,\d_{\a_2\b_2}\d_{\a_3\b_3}\,\mathcal{A}^{\a_2\b_2\a_3\b_3}(p_2,\bar{p}_3),\label{3TAnomaly}
\end{align}
which is the expression depicted in \figref{FeynTTTX}. The equation above has indeed a clear and simple interpretation in terms of anomaly poles extracted from the $\pi^{\mu\nu}$ projectors attached to each of the external graviton legs. We can re-express it in the form
\begin{align}
\braket{T^{\mu_1\nu_1}(p_1)T^{\mu_2\nu_2}(p_2)T^{\mu_3\nu_3}(\bar{p}_3)}^{(4)}_{anomaly}&=\left(\frac{\p^{\mu_1\nu_1}(p_1)}{3}\braket{T(p_1)T^{\mu_2\nu_2}(p_2)T^{\mu_3\nu_3}(\bar{p}_3)}^{(4)}_{anomaly}+(\text{perm.})\right)\notag\\
&\hspace{-1cm}-\left(\frac{\p^{\mu_1\nu_1}(p_1)}{3}\frac{\p^{\mu_2\nu_2}(p_2)}{3}\braket{T(p_1)T(p_2)T^{\mu_3\nu_3}(\bar{p}_3)}^{(4)}_{anomaly}+(\text{perm.})\right)\notag\\
&\hspace{-1cm}+\frac{\p^{\mu_1\nu_1}(p_1)}{3}\frac{\p^{\mu_2\nu_2}(p_2)}{3}\frac{\p^{\mu_3\nu_3}(\bar{p}_3)}{3}\braket{T(p_1)T(p_2)T(\bar{p}_3)}^{(4)}_{anomaly},\label{3TAnomaly2}
\end{align}
from which it is clear that a contribution such as 
\begin{align}
\braket{T(p_1)T^{\mu_2\nu_2}(p_2)T^{\mu_3\nu_3}(\bar{p}_3)}^{(4)}_{anomaly}
&=\mathcal{A}^{\mu_2\nu_2\mu_3\nu_3}(p_2,\bar{p}_3)\notag\\
&=4b'[E]^{\mu_2\nu_2\mu_3\nu_3}(p_2,\bar{p}_3)+4b[C^2]^{\mu_2\nu_2\mu_3\nu_3}(p_2,\bar{p}_3)
\end{align} 
where we trace one of the three stress energy tensors, is obtained by differentiating the anomaly functional twice. Similar results hold for the other contributions with double
\begin{align}
\braket{T(p_1)T(p_2)T^{\mu_3\nu_3}(\bar{p}_3)}^{(4)}_{anomaly}&=\d_{\a_2\b_2}\,\mathcal{A}^{\a_2\b_2\mu_3\nu_3}(p_2,\bar{p}_3)\notag\\
&=\delta_{\alpha_2\beta_2}\left(4b'[E]^{\a_2\b_2\mu_3\nu_3}(p_2,\bar{p}_3)+4b[C^2]^{\a_2\b_2\mu_3\nu_3}(p_2,\bar{p}_3)\right)
\end{align}
and triple traces
\begin{align}
\braket{T(p_1)T(p_2)T(\bar{p}_3)}^{(4)}_{anomaly}&=\d_{\a_2\b_2}\d_{\a_3\b_3}\,\mathcal{A}^{\a_2\b_2\a_3\b_3}(p_2,\bar{p}_3)\notag\\
&=\d_{\a_2\b_2}\d_{\a_3\b_3}\left(4b'[E]^{\a_2\b_2\a_3\b_3}(p_2,\bar{p}_3)+4b[C^2]^{\a_2\b_2\a_3\b_3}(p_2,\bar{p}_3)\right).
\end{align}
It is clear from these expressions that all the possible anomaly contributions generated in the flat limit are associated with single, double and triple traces of the 3T, where on each external graviton leg we are allowing for a scalar exchange due to the presence of the transverse $\pi$  projector, as clear from \figref{FeynTTTX}. This result is strongly reminiscent of the emergence of an anomaly pole in the AVV chiral anomaly diagram for a $J_5 JJ $ correlator, with one axial-vector $(J_5)$ and two vector $(J)$ currents, that manifests a similar pattern.\\
 Indeed, an analogy with the behaviour of the 3T is present if we decompose the $AAA$ anomaly diagram as 
 \begin{equation}
AAA\to  1/3(AVV + VAV + VVA)
 \end{equation}
using its permutational symmetry on the axial-vector lines.\\
In both cases one encounters a scalar or a pseudoscalar mode, respectively, via a bilinear mixing term attached to the external graviton or gauge lines. This mode is directly coupled to the (chiral/conformal) anomaly. Therefore, we encounter a feature that unifies both the conformal and the chiral cases.\\
Notice that the spin-2 part of the gravitational fluctuations do not couple to such bilinear mixing term in the 3T case, which therefore mediates only spin-1 and spin-0 interactions. \\
 It is also clear that this massless scalar interaction, in the 3T, is not removed by the inclusion of other Weyl-invariant terms which are obviously present in the complete expression of the correlator, which are not identified by our method. \\
For this correlator the anomalous trace WI takes the form 
\begin{align}
&\delta_{\mu_1\nu_1}\braket{ T^{\mu_1\nu_1}(p_1)T^{\mu_2\nu_2}(p_2)T^{\mu_3\nu_3}(p_3)}\notag\\
&\hspace{2cm}=
4 \, \mathcal A^{\mu_2\nu_2\mu_3\nu_3}(p_2,p_3)
- 2 \, \braket{ T^{\mu_2\nu_2}(p_1+p_2)T^{\mu_3\nu_3}(p_3)} - 2 \, \braket{ T^{\mu_2\nu_2}(p_2)T^{\mu_3\nu_3}(p_1+p_3)}\nn
&\hspace{2cm}=
4 \, \bigg[ b\,\big[C^2\big]^{\mu_2\nu_2\mu_3\nu_3}(p_2,p_3)+ b'\, \big[E\big]^{\mu_2\nu_2\mu_3\nu_3}(p_2,p_3) \bigg]\nn
&\hspace{3cm}- 2 \, \braket{ T^{\mu_2\nu_2}(p_1+p_2)T^{\mu_3\nu_3}(p_3)} - 2 \, \braket{ T^{\mu_2\nu_2}(p_2)T^{\mu_3\nu_3}(p_1+p_3)}. \label{munu3PFanomaly}
\end{align}

\subsection{The four-point function}
\label{fourp}
We will now come to illustrate the reconstruction procedure for the renormalized 4T, showing how the separation of the vertex into a transverse-traceless part, a longitudinal one and an anomaly contribution takes place after renormalization. Clearly, by construction, the transverse traceless sector of the 4T is renormalized by adding the contribution coming from the counterterm 

\begin{align}
&\bigg[\braket{t^{\mu_1\nu_1}t^{\mu_2\nu_2}t^{\mu_3\nu_3}t^{\mu_4\nu_4}}_{bare}+\braket{t^{\mu_1\nu_1}t^{\mu_2\nu_2}t^{\mu_3\nu_3}t^{\mu_4\nu_4}}_{count}\bigg]_{d\to4}=\braket{t^{\mu_1\nu_1}t^{\mu_2\nu_2}t^{\mu_3\nu_3}t^{\mu_4\nu_4}}^{(d=4)}_{Ren}.
\end{align} 
If we consider a Lagrangian realization, the renormalization of this part and the corresponding form factors are ensured by the choice of the coefficients $b$ and $b'$ as in \eqref{choiceparm1} and \eqref{choiceparm2}, where $n_I$, $I=S,F,G$, are the number of scalar, fermion and gauge fields running into the virtual corrections of this correlator. For a general CFT, not directly related to a specific free field theory realization, the $b$ and $b'$ should be interpreted as fundamental constants of that theory, and are arbitrary.\\
  Now, we turn to the longitudinal part of the correlator, which is the most interesting component when the case $d=4$ is considered, due to the appearance of the anomaly. For instance, we study the bare part $\braket{t_{loc}TTT}$ in \eqref{longTer} that is explicitly written as
\begin{align}
&\braket{t_{loc}^{\mu_1\nu_1}(p_1)T^{\mu_2\nu_2}(p_2)T^{\mu_3\nu_3}(p_3)T^{\mu_4\nu_4}(p_4)}_{bare}=\notag\\
&=\mathcal{I}^{\mu_1\nu_1}_{\alpha_1}(p_1)\Bigg\{\left[4\, \mathcal{B}^{\alpha_1\hspace{0.4cm}\mu_2\nu_2\mu_3\nu_3}_{\hspace{0.3cm}\lambda\beta_1}(p_2,p_3)\braket{T^{\lambda\beta_1}(p_1+p_2+p_3)T^{\mu_4\nu_4}(\bar{p}_4)}+(34)+(24)\right]\notag\\
&\hspace{2cm}+\left[2 \, \mathcal{C}^{\alpha_1\hspace{0.4cm}\mu_2\nu_2}_{\hspace{0.3cm}\lambda\beta_1}(p_2)\braket{T^{\lambda\beta_1}(p_1+p_2)T^{\mu_3\nu_3}(p_3)T^{\mu_4\nu_4}(\bar p_4)}+(23)+ (24)\right]\Bigg\}\notag\\
&\hspace{2cm}-\frac{2}{(d-1)}\pi^{\mu_1\nu_1}(p_1)\bigg[\braket{T^{\mu_2\nu_2}(p_1+p_2)T^{\mu_3\nu_3}(p_3)T^{\mu_4\nu_4}(\bar p_4)}+(2 3)+(24)\bigg].\label{tTTT}
\end{align}
This contribution in $d=4$ manifests some divergences due to the presence of the $3$- and $2$- point functions on its rhs. \\
A similar equation holds for the counterterm in \eqref{TTTTcount}, which can be decomposed as well into the transverse-traceless part and the longitudinal one. The contribution that renormalises \eqref{tTTT} is
\begin{align}
&\braket{t_{loc}^{\mu_1\nu_1}(p_1)T^{\mu_2\nu_2}(p_2)T^{\mu_3\nu_3}(p_3)T^{\mu_4\nu_4}(\bar p_4)}_{count}=\notag\\
&=\mathcal{I}^{\mu_1\nu_1}_{\alpha_1}(p_1)\Bigg\{\left[4\, \mathcal{B}^{\alpha_1\hspace{0.4cm}\mu_2\nu_2\mu_3\nu_3}_{\hspace{0.3cm}\lambda\beta_1}(p_2,p_3)\braket{T^{\lambda\beta_1}(p_1+p_2+p_3)T^{\mu_4\nu_4}(\bar{p}_4)}_{count}+(34)+(24)\right]\notag\\
&\hspace{2cm}+\left[2 \, \mathcal{C}^{\alpha_1\hspace{0.4cm}\mu_2\nu_2}_{\hspace{0.3cm}\lambda\beta_1}(p_2)\braket{T^{\lambda\beta_1}(p_1+p_2)T^{\mu_3\nu_3}(p_3)T^{\mu_4\nu_4}(\bar p_4)}_{count}+(23)+ (24)\right]\Bigg\}\notag\\
&\hspace{2cm}-\frac{2}{(d-1)}\pi^{\mu_1\nu_1}(p_1)\bigg[\braket{T^{\mu_2\nu_2}(p_1+p_2)T^{\mu_3\nu_3}(p_3)T^{\mu_4\nu_4}(\bar p_4)}_{count}+(2 3) +(2 4)\bigg]\notag\\
&\quad+\frac{8\,(d-4)\,\mu^{-\varepsilon}}{(d-1)\varepsilon}\pi^{\mu_1\nu_1}(p_1)\Bigg\{b\left[\sqrt{-g}C^2\right]^{\mu_2\nu_2\mu_3\nu_3\mu_4\nu_4}(p_2,p_3,p_4)+b'\left[\sqrt{-g}E\right]^{\mu_2\nu_2\mu_3\nu_3\mu_4\nu_4}(p_2,p_3,p_4)
\Bigg\}\label{tloctttcount}
\end{align}
where we have taken into account the definition \eqref{nTcount}. It is worth mentioning that, as for the $1/\varepsilon\, V_E$ counterterm,  near $d=4$ one has to use the analogous expansion of the $d$-dimensional counterterms for $V_{C^2}$
\begin{align}
V_{C^2}^{\m_1\n_1\m_2\n_2\m_3\n_3\mu_4\nu_4}(p_1,p_2,p_3,\bar{p}_4)\simeq\bigg[ V_{C^2}^{\m_1\n_1\m_2\n_2\m_3\n_3\mu_4\nu_4}(p_1,p_2,p_3,\bar{p}_4)\bigg]_{d=4}+ \varepsilon {V'}_{C^2}^{\m_1\n_1\m_2\n_2\m_3\n_3\mu_4\nu_4}(p_1,p_2,p_3,\bar{p}_4)
\end{align}
and the expansion of the extra term present in \eqref{tloctttcount}, for instance, gives
\begin{align}
&\frac{8\,(d-4)\,\mu^{-\varepsilon}}{(d-1)\varepsilon}\pi^{\mu_1\nu_1}(p_1)\Bigg\{b\left[\sqrt{-g}C^2\right]^{\mu_2\nu_2\mu_3\nu_3\mu_4\nu_4}(p_2,p_3,p_4)+b'\left[\sqrt{-g}E\right]^{\mu_2\nu_2\mu_3\nu_3\mu_4\nu_4}(p_2,p_3,p_4)\Bigg\}\notag\\
&\simeq -\frac{8}{3}\pi^{\mu_1\nu_1}(p_1)\Bigg\{b\left[\sqrt{-g}C^2\right]^{\mu_2\nu_2\mu_3\nu_3\mu_4\nu_4}(p_2,p_3,p_4)+b'\left[\sqrt{-g}E\right]^{\mu_2\nu_2\mu_3\nu_3\mu_4\nu_4}(p_2,p_3,p_4)\Bigg\}+O(\varepsilon).
\end{align}
After adding the counterterms to the bare correlator and expanding around $d=4$ we obtain 
\begin{align}
&\bigg[\braket{t_{loc}^{\mu_1\nu_1}T^{\mu_2\nu_2}T^{\mu_3\nu_3}T^{\mu_4\nu_4}}_{bare}+\braket{t_{loc}^{\mu_1\nu_1}T^{\mu_2\nu_2}T^{\mu_3\nu_3}T^{\mu_4\nu_4}}_{count}\bigg]_{d\to4}=\notag\\
&=\braket{t_{loc}^{\mu_1\nu_1}T^{\mu_2\nu_2}T^{\mu_3\nu_3}T^{\mu_4\nu_4}}^{(d=4)}_{Ren}+\braket{t_{loc}^{\mu_1\nu_1}T^{\mu_2\nu_2}T^{\mu_3\nu_3}T^{\mu_4\nu_4}}^{(d=4)}_{anomaly}
\end{align}
with the inclusion of an extra anomalous contribution in the final expression. In particular, this contribution takes the explicit form
\begin{align}
&\braket{t_{loc}^{\mu_1\nu_1}T^{\mu_2\nu_2}T^{\mu_3\nu_3}T^{\mu_4\nu_4}}^{(d=4)}_{anomaly}=\frac{8\,\pi^{\mu_1\nu_1}(p_1)}{3}\,\mathcal{A}^{\mu_2\nu_2\mu_3\nu_3\mu_4\nu_4}(p_2,p_3,\bar p_4)\notag\\
&\hspace{0.5cm}+\mathcal{I}^{\mu_1\nu_1}_{\alpha_1}(p_1)\left[2 \, \mathcal{C}^{\alpha_1\hspace{0.4cm}\mu_2\nu_2}_{\hspace{0.3cm}\lambda\beta_1}(p_2)\braket{T^{\lambda\beta_1}(p_1+p_2)T^{\mu_3\nu_3}(p_3)T^{\mu_4\nu_4}(\bar p_4)}_{anomaly}+(23)+(24)\right]\notag\\
&\hspace{0.5cm}-\frac{2}{3}\pi^{\mu_1\nu_1}(p_1)\bigg[\braket{T^{\mu_2\nu_2}(p_1+p_2)T^{\mu_3\nu_3}(p_3)T^{\mu_4\nu_4}(\bar p_4)}_{anomaly}+(2 3)+(2 4)\bigg].\label{tlocTTTanom}
\end{align}
It is worth noticing that the anomaly of the $3$-point function \eqref{3TAnomaly} contributes to the anomaly part of the $4$-point function in \eqref{tlocTTTanom}, as expected. The same equation can be written in the simpler form 
\begin{align}
&\braket{t_{loc}^{\mu_1\nu_1}T^{\mu_2\nu_2}T^{\mu_3\nu_3}T^{\mu_4\nu_4}}^{(d=4)}_{anomaly}=\frac{\pi^{\mu_1\nu_1}(p_1)}{3}\,\braket{T(p_1)T^{\mu_2\nu_2}(p_2)T^{\mu_3\nu_3}(p_3)T^{\mu_4\nu_4}(\bar{p}_4)}_{anomaly}\notag\\
&\hspace{4cm}+\mathcal{I}^{\mu_1\nu_1}_{\alpha_1}(p_1)\,p_{1\beta_1}\,\braket{T^{\alpha_1\beta_1}(p_1)T^{\mu_2\nu_2}(p_2)T^{\mu_3\nu_3}(p_3)T^{\mu_4\nu_4}(\bar{p}_4)}_{anomaly}\label{tlocTTTanom2}
\end{align}
with 
\begin{align}
&\braket{T(p_1)T^{\mu_2\nu_2}(p_2)T^{\mu_3\nu_3}(p_3)T^{\mu_4\nu_4}(\bar{p}_4)}_{anomaly}\equiv\delta_{\mu_1\nu_1}\braket{T^{\mu_1\nu_1}(p_1)T^{\mu_2\nu_2}(p_2)T^{\mu_3\nu_3}(p_3)T^{\mu_4\nu_4}(\bar{p}_4)}_{anomaly}\notag\\
&\qquad=8\,\mathcal{A}^{\mu_2\nu_2\mu_3\nu_3\mu_4\nu_4}(p_2,p_3,\bar p_4)-2\braket{T^{\mu_2\nu_2}(p_1+p_2)T^{\mu_3\nu_3}(p_3)T^{\mu_4\nu_4}(\bar p_4)}_{anomaly}\notag\\
&\qquad\quad-2\braket{T^{\mu_2\nu_2}(p_2)T^{\mu_3\nu_3}(p_1+p_3)T^{\mu_4\nu_4}(\bar p_4)}_{anomaly}-2\braket{T^{\mu_2\nu_2}(p_2)T^{\mu_3\nu_3}(p_3)T^{\mu_4\nu_4}(p_1+\bar p_4)}_{anomaly},
\end{align}
and 
\begin{align}
&\,p_{1\beta_1}\,\braket{T^{\alpha_1\beta_1}(p_1)T^{\mu_2\nu_2}(p_2)T^{\mu_3\nu_3}(p_3)T^{\mu_4\nu_4}(\bar{p}_4)}_{anomaly}=\notag\\
&\qquad=2 \, \mathcal{C}^{\alpha_1\hspace{0.4cm}\mu_2\nu_2}_{\hspace{0.3cm}\lambda\beta_1}(p_2)\braket{T^{\lambda\beta_1}(p_1+p_2)T^{\mu_3\nu_3}(p_3)T^{\mu_4\nu_4}(\bar p_4)}_{anomaly}+(23)+(24),
\end{align} 
having taken into account the anomaly part of the $3$-point function defined in \eqref{3TAnomaly2}.\\
We proceed by studying the other local terms of the correlator that are responsible for the generation of extra contributions to its renormalized expression, thereby contributing to the anomaly in $d=4$. These are a term with two and three $t_{loc}$ contributions 
\begin{align}
&\bigg[\braket{t_{loc}^{\mu_1\nu_1}t_{loc}^{\mu_2\nu_2}T^{\mu_3\nu_3}T^{\mu_4\nu_4}}_{bare}+\braket{t_{loc}^{\mu_1\nu_1}t_{loc}^{\mu_2\nu_2}T^{\mu_3\nu_3}T^{\mu_4\nu_4}}_{count}\bigg]_{d\to4}=\notag\\
&=\braket{t_{loc}^{\mu_1\nu_1}t_{loc}^{\mu_2\nu_2}T^{\mu_3\nu_3}T^{\mu_4\nu_4}}^{(d=4)}_{Ren}+\braket{t_{loc}^{\mu_1\nu_1}t_{loc}^{\mu_2\nu_2}T^{\mu_3\nu_3}T^{\mu_4\nu_4}}^{(d=4)}_{anomaly}
\end{align}
where
\begin{align}
&\braket{t_{loc}^{\mu_1\nu_1}(p_1)t_{loc}^{\mu_2\nu_2}(p_2)T^{\mu_3\nu_3}(p_3)T^{\mu_4\nu_4}(p_4)}_{anomaly}^{(d=4)}\notag\\
&\quad=2\,\mathcal{I}^{\mu_2\nu_2}_{\alpha_2}(p_2)\,\mathcal{I}^{\mu_1\nu_1}_{\alpha_1}(p_1)\,p_{2\beta_2}\mathcal{C}^{\alpha_1\hspace{0.4cm}\alpha_2\beta_2}_{\hspace{0.3cm}\lambda\beta_1}(p_2)\,\braket{T^{\lambda\beta_1}(p_1+p_2)T^{\mu_3\nu_3}(p_3)T^{\mu_4\nu_4}(p_4)}_{anom}\notag\\
&+\bigg\{\frac{1}{3}\,\mathcal{I}^{\mu_1\nu_1}_{\alpha_1}(p_1)\,\pi^{\mu_2\nu_2}(p_2)\,\,p_{1\beta_1}\bigg[-2\,\braket{T^{\alpha_1\beta_1}(p_1+p_2)T^{\mu_3\nu_3}(p_3)T^{\mu_4\nu_4}(p_4)}_{anom}\notag\\
&\hspace{1cm}+8\,\mathcal{A}^{\alpha_1\beta_1\mu_3\nu_3\mu_4\nu_4}(p_1,p_3,p_4)\bigg]+(12)\bigg\}+\frac{8}{9}\pi^{\mu_1\nu_1}(p_1)\pi^{\mu_2\nu_2}(p_2)\bigg\{\delta_{\alpha_2\beta_2}\,\mathcal{A}^{\alpha_2\beta_2\mu_3\nu_3\mu_4\nu_4}(p_2,p_3,p_4)\notag\\
&\hspace{2cm}-\mathcal{A}^{\mu_3\nu_3\mu_4\nu_4}(p_3,p_4)-\mathcal{A}^{\mu_3\nu_3\mu_4\nu_4}(p_1+p_3,p_4)-\mathcal{A}^{\mu_3\nu_3\mu_4\nu_4}(p_3,p_1+p_4)\bigg\},
\end{align}
that can be re-expressed as
\begin{align}
&\braket{t_{loc}^{\mu_1\nu_1}(p_1)t_{loc}^{\mu_2\nu_2}(p_2)T^{\mu_3\nu_3}(p_3)T^{\mu_4\nu_4}(p_4)}_{anomaly}^{(d=4)}\notag\\
&\quad=\frac{\pi^{\mu_1\nu_1}(p_1)}{3}\frac{\pi^{\mu_2\nu_2}(p_2)}{3}\braket{T(p_1)T(p_2)T^{\mu_3\nu_3}(p_3)T^{\mu_4\nu_4}(\bar{p}_4)}_{anomaly}\notag\\
&\qquad+\,\mathcal{I}^{\mu_2\nu_2}_{\alpha_2}(p_2)\,\mathcal{I}^{\mu_1\nu_1}_{\alpha_1}(p_1)\,p_{2\beta_2}\,p_{1\beta_1}\braket{T^{\alpha_1\beta_1}(p_1)T^{\alpha_2\beta_2}(p_2)T^{\mu_3\nu_3}(p_3)T^{\mu_4\nu_4}(p_4)}_{anom}\notag\\
&\qquad+\mathcal{I}^{\mu_1\nu_1}_{\alpha_1}(p_1)\,\frac{\pi^{\mu_2\nu_2}(p_2)}{3}\,\,p_{1\beta_1}\braket{T^{\alpha_1\beta_1}(p_1)T(p_2)T^{\mu_3\nu_3}(p_3)T^{\mu_4\nu_4}(\bar{p}_4)}_{anomaly}\notag\\
&\qquad+\frac{\pi^{\mu_1\nu_1}(p_1)}{3}\,\mathcal{I}^{\mu_2\nu_2}_{\alpha_2}(p_2)\,\,p_{2\beta_2}\braket{T(p_1)T^{\alpha_2\beta_2}(p_2)T^{\mu_3\nu_3}(p_3)T^{\mu_4\nu_4}(\bar{p}_4)}_{anomaly}.
\end{align}
In a similar fashion, the terms involving three $t_{loc}$ will contribute as
\begin{align}
&\bigg[\braket{t_{loc}^{\mu_1\nu_1}t_{loc}^{\mu_2\nu_2}t_{loc}^{\mu_3\nu_3}T^{\mu_4\nu_4}}_{bare}+\braket{t_{loc}^{\mu_1\nu_1}t_{loc}^{\mu_2\nu_2}t_{loc}^{\mu_3\nu_3}T^{\mu_4\nu_4}}_{count}\bigg]_{d\to4}=\notag\\
&=\braket{t_{loc}^{\mu_1\nu_1}t_{loc}^{\mu_2\nu_2}t_{loc}^{\mu_3\nu_3}T^{\mu_4\nu_4}}^{(d=4)}_{Ren}+\braket{t_{loc}^{\mu_1\nu_1}t_{loc}^{\mu_2\nu_2}t_{loc}^{\mu_3\nu_3}T^{\mu_4\nu_4}}^{(d=4)}_{anomaly}
\end{align}
with 
\begin{align}
&\braket{t_{loc}^{\mu_1\nu_1}(p_1)t_{loc}^{\mu_2\nu_2}(p_2)t_{loc}^{\mu_3\nu_3}(p_3)T^{\mu_4\nu_4}(p_4)}_{anomaly}^{(d=4)}=\notag\\
&=\bigg\{\frac{8}{3}\,\mathcal{I}^{\mu_1\nu_1}_{\alpha_1}(p_1)\,\mathcal{I}^{\mu_2\nu_2}_{\alpha_2}(p_2)\pi^{\mu_3\nu_3}(p_3)\bigg[\,p_{2\beta_2}\,\mathcal{C}^{\alpha_1\hspace{0.4cm}\alpha_2\beta_2}_{\hspace{0.3cm}\lambda\beta_1}(p_2)\,\mathcal{A}^{\lambda\beta_1\mu_4\nu_4}(p_1+p_2,p_4)\bigg]+(13)+(23)\bigg\}\notag\\
&+\bigg\{\frac{8}{9}\,\mathcal{I}^{\mu_1\nu_1}_{\alpha_1}(p_1)\pi^{\mu_2\nu_2}(p_2)\,\pi^{\mu_3\nu_3}(p_3)\bigg[p_{1\beta_1}\,\delta_{\alpha_2\beta_2}\,\mathcal{A}^{\alpha_1\beta_1\alpha_2\beta_2\mu_4\nu_4}(p_1,p_2,p_4)-p_{1\beta_1}\,\mathcal{A}^{\alpha_1\beta_1\mu_4\nu_4}(p_1,p_4)\notag\\
&-p_{1\beta_1}\,\mathcal{A}^{\alpha_1\beta_1\mu_4\nu_4}(p_1+p_3,p_4)-p_{1\beta_1}\,\mathcal{A}^{\alpha_1\beta_1\mu_4\nu_4}(p_1,p_3+p_4)\bigg]+(12)+(13)\bigg\}\notag\\
&+\frac{8}{27}\,\pi^{\mu_1\nu_1}(p_1)\pi^{\mu_2\nu_2}(p_2)\,\pi^{\mu_3\nu_3}(p_3)\bigg\{\,\delta_{\alpha_2\beta_2}\,\delta_{\alpha_3\beta_3}\,\mathcal{A}^{\alpha_2\beta_2\alpha_3\beta_3\mu_4\nu_4}(p_2,p_3,p_4)\notag\\
&\hspace{1cm}-\delta_{\alpha_3\beta_3}\,\mathcal{A}^{\alpha_3\beta_3\mu_4\nu_4}(p_3,p_4)-\delta_{\alpha_3\beta_3}\,\mathcal{A}^{\alpha_3\beta_3\mu_4\nu_4}(p_1+p_3,p_4)-\delta_{\alpha_3\beta_3}\,\mathcal{A}^{\alpha_3\beta_3\mu_4\nu_4}(p_3,p_1+p_4)\bigg\}
\end{align}
that we rewrite in the form
\begin{align}
&\braket{t_{loc}^{\mu_1\nu_1}(p_1)t_{loc}^{\mu_2\nu_2}(p_2)t_{loc}^{\mu_3\nu_3}(p_3)T^{\mu_4\nu_4}(p_4)}_{anomaly}^{(d=4)}=\notag\\[1.5ex]
&=\frac{\pi^{\mu_1\nu_1}(p_1)}{3}\frac{\pi^{\mu_2\nu_2}(p_2)}{3}\,\frac{\pi^{\mu_3\nu_3}(p_3)}{3}\braket{T(p_1)T(p_2)T(p_3)T^{\mu_4\nu_4}(\bar{p}_4)}_{anomaly}\notag\\
&+\bigg\{\,\mathcal{I}^{\mu_1\nu_1}_{\alpha_1}(p_1)\,\mathcal{I}^{\mu_2\nu_2}_{\alpha_2}(p_2)\frac{\pi^{\mu_3\nu_3}(p_3)}{3}\braket{T^{\alpha_1\beta_1}(p_1)T^{\alpha_2\beta_2}(p_2)T(p_3)T^{\mu_4\nu_4}(\bar{p}_4)}_{anomaly}+(13)+(23)\bigg\}\notag\\
&+\bigg\{\,\mathcal{I}^{\mu_1\nu_1}_{\alpha_1}(p_1)\frac{\pi^{\mu_2\nu_2}(p_2)}{3}\,\frac{\pi^{\mu_3\nu_3}(p_3)}{3}\,p_{1\beta_1}\braket{T^{\alpha_1\beta_1}(p_1)T(p_2)T(p_3)T^{\mu_4\nu_4}(\bar{p}_4)}_{anomaly}+(12)+(13)\bigg\}.
\end{align}
Finally, the last term that involves four insertions of the operator $t_{loc}$, after renormalization takes the form
\begin{align}
&\bigg[\braket{t_{loc}^{\mu_1\nu_1}t_{loc}^{\mu_2\nu_2}t_{loc}^{\mu_3\nu_3}t_{loc}^{\mu_4\nu_4}}_{bare}+\braket{t_{loc}^{\mu_1\nu_1}t_{loc}^{\mu_2\nu_2}t_{loc}^{\mu_3\nu_3}t_{loc}^{\mu_4\nu_4}}_{count}\bigg]_{d\to4}=\notag\\
&=\braket{t_{loc}^{\mu_1\nu_1}t_{loc}^{\mu_2\nu_2}t_{loc}^{\mu_3\nu_3}t_{loc}^{\mu_4\nu_4}}^{(d=4)}_{Ren}+\braket{t_{loc}^{\mu_1\nu_1}t_{loc}^{\mu_2\nu_2}t_{loc}^{\mu_3\nu_3}t_{loc}^{\mu_4\nu_4}}^{(d=4)}_{anomaly},
\end{align}
generating an extra term that contributes to the anomaly part of the full correlator as
\begin{align}
&\braket{t_{loc}^{\mu_1\nu_1}t_{loc}^{\mu_2\nu_2}t_{loc}^{\mu_3\nu_3}t_{loc}^{\mu_4\nu_4}}^{(d=4)}_{anomaly}\notag\\
&\quad=\bigg\{\frac{8}{9}\,\mathcal{I}^{\mu_1\nu_1}_{\alpha_1}(p_1)\,\mathcal{I}^{\mu_2\nu_2}_{\alpha_2}(p_2)\,\pi^{\mu_3\nu_3}(p_3)\,\pi^{\mu_4\nu_4}(p_4)\,\bigg[\,p_{2\beta_2}\,\mathcal{C}^{\alpha_1\hspace{0.4cm}\alpha_2\beta_2}_{\hspace{0.3cm}\lambda\beta_1}(p_2)\,\delta_{\alpha_4\beta_4}\,\mathcal{A}^{\lambda\beta_1\alpha_4\beta_4}(p_1+p_2,p_4)\bigg]\notag\\
&\hspace{2cm}+(13)+(23)+(14)+(24)+(13)(24)\bigg\}\notag\\
&\quad+\bigg\{\frac{8}{27}\,\mathcal{I}^{\mu_1\nu_1}_{\alpha_1}(p_1)\,\pi^{\mu_2\nu_2}(p_2)\,\pi^{\mu_3\nu_3}(p_3)\,\pi^{\mu_4\nu_4}(p_4)\,\bigg[\,p_{1\beta_1}\,\delta_{\alpha_2\beta_2}\delta_{\alpha_3\beta_3}\,\mathcal{A}^{\alpha_1\beta_1\alpha_2\beta_2\alpha_3\beta_3}(p_1,p_2,p_3)\notag\\
&\hspace{1cm}-\,p_{1\beta_1}\,\delta_{\alpha_3\beta_3}\,\mathcal{A}^{\alpha_1\beta_1\alpha_3\beta_3}(p_1,p_3)-\,p_{1\beta_1}\,\delta_{\alpha_3\beta_3}\,\mathcal{A}^{\alpha_1\beta_1\alpha_3\beta_3}(p_1,p_3+p_4)\notag\\
&\hspace{3cm}-\,p_{1\beta_1}\,\delta_{\alpha_3\beta_3}\,\mathcal{A}^{\alpha_1\beta_1\alpha_3\beta_3}(p_1+p_4,p_3)\bigg]+(12)+(13)+(14)\bigg\}\notag\\
&\quad+\frac{8}{81}\,\pi^{\mu_1\nu_1}(p_1)\,\pi^{\mu_2\nu_2}(p_2)\,\pi^{\mu_3\nu_3}(p_3)\,\pi^{\mu_4\nu_4}(p_4)\notag\\
&\hspace{1cm}\times\bigg[\delta_{\alpha_2\beta_2}\delta_{\alpha_3\beta_3}\delta_{\alpha_4\beta_4}\mathcal{A}^{\alpha_2\beta_2\alpha_3\beta_3\alpha_4\beta_4}(p_2,p_3,p_4)-\delta_{\alpha_2\beta_2}\delta_{\alpha_4\beta_4}\mathcal{A}^{\alpha_2\beta_2\alpha_4\beta_4}(p_1+p_2,p_4)\notag\\
&\hspace{3cm}-\delta_{\alpha_2\beta_2}\delta_{\alpha_4\beta_4}\mathcal{A}^{\alpha_2\beta_2\alpha_4\beta_4}(p_2,p_1+p_4)-\delta_{\alpha_3\beta_3}\delta_{\alpha_4\beta_4}\mathcal{A}^{\alpha_3\beta_3\alpha_4\beta_4}(p_1+p_3,p_4)
\bigg].
\end{align}
It can be re-expressed as
\begin{align}
	&\braket{t_{loc}^{\mu_1\nu_1}t_{loc}^{\mu_2\nu_2}t_{loc}^{\mu_3\nu_3}t_{loc}^{\mu_4\nu_4}}^{(d=4)}_{anomaly}\notag\\
	&\quad=\frac{\pi^{\mu_1\nu_1}(p_1)}{3}\,\frac{\pi^{\mu_2\nu_2}(p_2)}{3}\,\frac{\pi^{\mu_3\nu_3}(p_3)}{3}\,\frac{\pi^{\mu_4\nu_4}(p_4)}{3}\braket{T(p_1)T(p_2)T(p_3)T(\bar{p}_4)}_{anomaly}\notag\\
	&\qquad+\bigg\{\,\mathcal{I}^{\mu_1\nu_1}_{\alpha_1}(p_1)\,\mathcal{I}^{\mu_2\nu_2}_{\alpha_2}(p_2)\,\frac{\pi^{\mu_3\nu_3}(p_3)}{3}\,\frac{\pi^{\mu_4\nu_4}(p_4)}{3}\,p_{1\beta_1}\,p_{2\beta_2}\braket{T^{\alpha_1\beta_1}(p_1)T^{\alpha_2\beta_2}(p_2)T(p_3)T(p_4)}_{anomaly}\notag\\
	&\hspace{2cm}+(13)+(23)+(14)+(24)+(13)(24)\bigg\}\notag\\
	&\quad+\bigg\{\,\mathcal{I}^{\mu_1\nu_1}_{\alpha_1}(p_1)\,\frac{\pi^{\mu_2\nu_2}(p_2)}{3}\,\frac{\pi^{\mu_3\nu_3}(p_3)}{3}\,\frac{\pi^{\mu_4\nu_4}(p_4)}{3}p_{1\beta_1}\braket{T^{\alpha_1\beta_1}(p_1)T(p_2)T(p_3)T(\bar{p}_4)}_{anomaly}\notag\\
	&\hspace{2cm}+(12)+(13)+(14)\bigg\}.
\end{align}
In summary we have obtained for the renormalized 4T the general expression
\begin{align}
&\bigg[\braket{T^{\mu_1\nu_1}T^{\mu_2\nu_2}T^{\mu_3\nu_3}T^{\mu_4\nu_4}}_{bare}+\braket{T^{\mu_1\nu_1}T^{\mu_2\nu_2}T^{\mu_3\nu_3}T^{\mu_4\nu_4}}_{count}\bigg]_{d\to4}=\notag\\
&=\braket{T^{\mu_1\nu_1}T^{\mu_2\nu_2}T^{\mu_3\nu_3}T^{\mu_4\nu_4}}^{(d=4)}_{Ren}+\braket{T^{\mu_1\nu_1}T^{\mu_2\nu_2}T^{\mu_3\nu_3}T^{\mu_4\nu_4}}^{(d=4)}_{anomaly}
\end{align} 
where the anomaly part can be identified using all the results presented above as
\begin{align}
&\braket{T^{\mu_1\nu_1}(p_1)T^{\mu_2\nu_2}(p_2)T^{\mu_3\nu_3}(p_3)T^{\mu_4\nu_4}(\bar{p}_4)}^{(d=4)}_{anomaly}=\notag\\ &\hspace{1cm}=\Bigg[\braket{t_{loc}^{\mu_1\nu_1}(p_1)T^{\mu_2\nu_2}(p_2)T^{\mu_3\nu_3}(p_3)T^{\mu_4\nu_4}(\bar{p}_4)}^{(d=4)}_{anomaly}+(1\,2)+(1\, 3)+(1\, 4)\Bigg]\notag\\
&-\Bigg[\braket{t_{loc}^{\mu_1\nu_1}(p_1)t_{loc}^{\mu_2\nu_2}(p_2)T^{\mu_3\nu_3}(p_3)T^{\mu_4\nu_4}(\bar{p}_4)}^{(d=4)}_{anomaly}+(1\,3)+(1\,4)+(2\,3)+(2\,4)+(1\,3)(2\,4)\Bigg]\notag\\
&\hspace{1cm}+\Bigg[\braket{t_{loc}^{\mu_1\nu_1}(p_1)t_{loc}^{\mu_2\nu_2}(p_2)t_{loc}^{\mu_3\nu_3}(p_3)T^{\mu_4\nu_4}(\bar{p}_4)}^{(d=4)}_{anomaly}+(1\,4)+(2\,4)+(3\,4)\Bigg]\notag\\
&\hspace{2cm}-\braket{t_{loc}^{\mu_1\nu_1}(p_1)t_{loc}^{\mu_2\nu_2}(p_2)t_{loc}^{\mu_3\nu_3}(p_3)t_{loc}^{\mu_4\nu_4}(\bar{p}_4)}^{(d=4)}_{anomaly}\label{TTTTanomaly}.
\end{align}
We have shown how the anomaly part of the 4T is extracted through the procedure of renormalization. It is clear from this procedure that such component is exactly the one predicted by the 4-dimensional reconstruction method, using the anomalous Ward identities. 

  \subsection{Summary}
In summary, we write the anomaly part of the correlator in the form
\begin{align}
&\braket{T^{\mu_1\nu_1}(p_1)T^{\mu_2\nu_2}(p_2)T^{\mu_3\nu_3}(p_3)T^{\mu_4\nu_4}(\bar{p}_4)}^{(d=4)}_{anomaly}=\notag\\
&\qquad=\braket{T^{\mu_1\nu_1}(p_1)T^{\mu_2\nu_2}(p_2)T^{\mu_3\nu_3}(p_3)T^{\mu_4\nu_4}(\bar{p}_4)}_{poles}+\braket{T^{\mu_1\nu_1}(p_1)T^{\mu_2\nu_2}(p_2)T^{\mu_3\nu_3}(p_3)T^{\mu_4\nu_4}(\bar{p}_4)}_{0-residue}
\end{align}
where the first contribution is anomalous (Weyl-variant) and the second one
is traceless
\begin{equation}
	\delta_{\mu_i\nu_i}\braket{T^{\mu_1\nu_1}(p_1)T^{\mu_2\nu_2}(p_2)T^{\mu_3\nu_3}(p_3)T^{\mu_4\nu_4}(\bar{p}_4)}_{0-residue}=0,\qquad i=1,2,3,4. 
\end{equation}
We call it the "zero residue" or the "zero trace" (0T) part, since the operation of tracing the anomalous part removes the anomaly pole in the bilinear mixing terms, leaving a residue which is proportional to the anomaly. This part carries no pole.\\
This contribution is explicitly given by the expression
\begin{align}
\label{Weyl}
&\braket{T^{\mu_1\nu_1}(p_1)T^{\mu_2\nu_2}(p_2)T^{\mu_3\nu_3}(p_3)T^{\mu_4\nu_4}(\bar{p}_4)}_{0-residue}=\notag\\
&=\mathcal{I}^{\mu_1\nu_1}_{\alpha_1}(p_1)\,p_{1\beta_1}\,\braket{T^{\alpha_1\beta_1}(p_1)T^{\mu_2\nu_2}(p_2)T^{\mu_3\nu_3}(p_3)T^{\mu_4\nu_4}(\bar{p}_4)}_{anomaly}+(perm.)\notag\\
&-\bigg\{\bigg[\,\mathcal{I}^{\mu_2\nu_2}_{\alpha_2}(p_2)\,\mathcal{I}^{\mu_1\nu_1}_{\alpha_1}(p_1)\,p_{2\beta_2}\,p_{1\beta_1}\braket{T^{\alpha_1\beta_1}(p_1)T^{\alpha_2\beta_2}(p_2)T^{\mu_3\nu_3}(p_3)T^{\mu_4\nu_4}(p_4)}_{anom}\notag\\
&\qquad+\mathcal{I}^{\mu_1\nu_1}_{\alpha_1}(p_1)\,\frac{\pi^{\mu_2\nu_2}(p_2)}{3}\,\,p_{1\beta_1}\braket{T^{\alpha_1\beta_1}(p_1)T(p_2)T^{\mu_3\nu_3}(p_3)T^{\mu_4\nu_4}(\bar{p}_4)}_{anomaly}\notag\\
&\qquad+\frac{\pi^{\mu_1\nu_1}(p_1)}{3}\,\mathcal{I}^{\mu_2\nu_2}_{\alpha_2}(p_2)\,\,p_{2\beta_2}\braket{T(p_1)T^{\alpha_2\beta_2}(p_2)T^{\mu_3\nu_3}(p_3)T^{\mu_4\nu_4}(\bar{p}_4)}_{anomaly}\bigg]+(perm.)\bigg\}\notag\\
&+\bigg\{\bigg[\mathcal{I}^{\mu_1\nu_1}_{\alpha_1}(p_1)\,\mathcal{I}^{\mu_2\nu_2}_{\alpha_2}(p_2)\frac{\pi^{\mu_3\nu_3}(p_3)}{3}\braket{T^{\alpha_1\beta_1}(p_1)T^{\alpha_2\beta_2}(p_2)T(p_3)T^{\mu_4\nu_4}(\bar{p}_4)}_{anomaly}+(13)+(23)\bigg]+(perm.)\notag
\end{align}
\begin{align}
&+\bigg[\,\mathcal{I}^{\mu_1\nu_1}_{\alpha_1}(p_1)\frac{\pi^{\mu_2\nu_2}(p_2)}{3}\,\frac{\pi^{\mu_3\nu_3}(p_3)}{3}\,p_{1\beta_1}\braket{T^{\alpha_1\beta_1}(p_1)T(p_2)T(p_3)T^{\mu_4\nu_4}(\bar{p}_4)}_{anomaly}+(12)+(13)\bigg]+(perm.)\bigg\}\notag\\
&\qquad-\bigg\{\,\mathcal{I}^{\mu_1\nu_1}_{\alpha_1}(p_1)\,\mathcal{I}^{\mu_2\nu_2}_{\alpha_2}(p_2)\,\frac{\pi^{\mu_3\nu_3}(p_3)}{3}\,\frac{\pi^{\mu_4\nu_4}(p_4)}{3}\,p_{1\beta_1}\,p_{2\beta_2}\braket{T^{\alpha_1\beta_1}(p_1)T^{\alpha_2\beta_2}(p_2)T(p_3)T(p_4)}_{anomaly}\notag\\
&\hspace{1.5cm}+(13)+(23)+(14)+(24)+(13)(24)\bigg\}\notag\\
&\quad-\bigg\{\,\mathcal{I}^{\mu_1\nu_1}_{\alpha_1}(p_1)\,\frac{\pi^{\mu_2\nu_2}(p_2)}{3}\,\frac{\pi^{\mu_3\nu_3}(p_3)}{3}\,\frac{\pi^{\mu_4\nu_4}(p_4)}{3}p_{1\beta_1}\braket{T^{\alpha_1\beta_1}(p_1)T(p_2)T(p_3)T(\bar{p}_4)}_{anomaly}\notag\\
&\hspace{2cm}+(12)+(13)+(14)\bigg\}.
\end{align}
On the other hand, the anomaly part is then explicitly given as
\begin{align}
&\braket{T^{\mu_1\nu_1}(p_1)T^{\mu_2\nu_2}(p_2)T^{\mu_3\nu_3}(p_3)T^{\mu_4\nu_4}(\bar{p}_4)}_{poles}=\notag\\
&=\frac{\pi^{\mu_1\nu_1}(p_1)}{3}\,\braket{T(p_1)T^{\mu_2\nu_2}(p_2)T^{\mu_3\nu_3}(p_3)T^{\mu_4\nu_4}(\bar{p}_4)}_{anomaly}+(perm.)\notag\\
&-\frac{\pi^{\mu_1\nu_1}(p_1)}{3}\frac{\pi^{\mu_2\nu_2}(p_2)}{3}\,\braket{T(p_1)T(p_2)T^{\mu_3\nu_3}(p_3)T^{\mu_4\nu_4}(\bar{p}_4)}_{anomaly}+(perm.)\notag\\
&+\frac{\pi^{\mu_1\nu_1}(p_1)}{3}\frac{\pi^{\mu_2\nu_2}(p_2)}{3}\frac{\pi^{\mu_3\nu_3}(p_2)}{3}\,\braket{T(p_1)T(p_2)T(p_3)T^{\mu_4\nu_4}(\bar{p}_4)}_{anomaly}+(perm.)\notag\\
&-\frac{\pi^{\mu_1\nu_1}(p_1)}{3}\frac{\pi^{\mu_2\nu_2}(p_2)}{3}\frac{\pi^{\mu_3\nu_3}(p_3)}{3}\frac{\pi^{\mu_4\nu_4}(p_4)}{3}\,\braket{T(p_1)T(p_2)T(p_3)T(\bar{p}_4)}_{anomaly}.
\end{align}
\begin{figure}[t]
 \centering
	\raisebox{-1.5ex}{\includegraphics[scale=0.5]{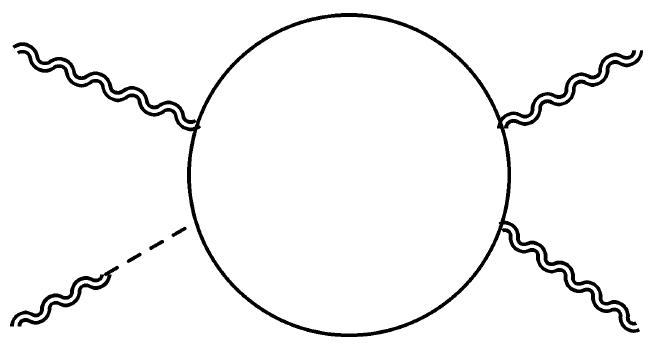}} 
	\raisebox{-1.5ex}{\includegraphics[scale=0.5]{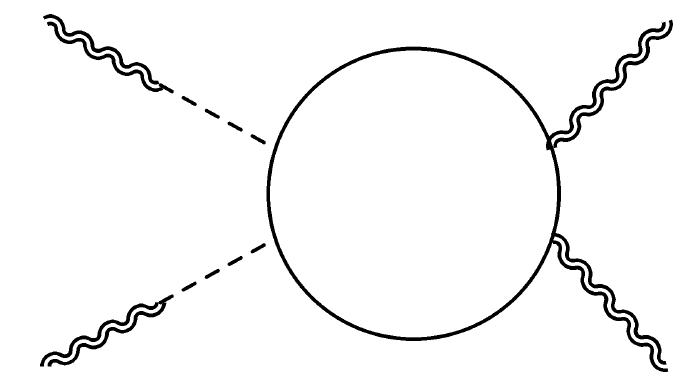}} 
	\raisebox{-1.5ex}{\includegraphics[scale=0.5]{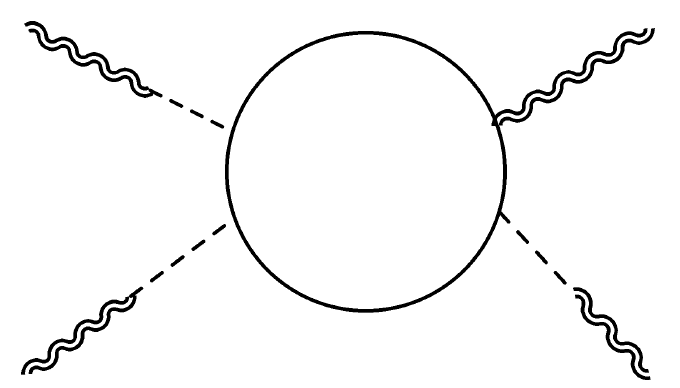}} 
	\raisebox{-1.5ex}{\includegraphics[scale=0.5]{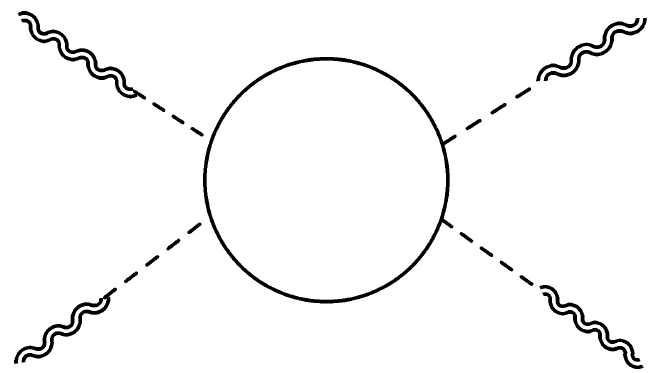}} 
	\caption{The Weyl-variant contributions from $\mathcal{S}_A$  to the renormalized vertex for the 4T with the corresponding bilinear mixings in $d=4$ \label{4T}}
\end{figure}
The picture that emerges from our analysis is shown in \figref{4T}, which generalizes the case of the 3T, with the inclusion of one extra bilinear graviton- spin-0 mixing term on the external legs. \\
However, we have shown though, that differently from the 3T, in the 4T there is an extra, Weyl-invariant  component in $\mathcal{S}_A$ that is also predicted by the reconstruction. It is also clear why this component is not present in the 3T. \\
Our result for the anomaly action, as predicted by the CWIs, can then be collected into the form 
\begin{align}
\label{res}
\mathcal{S}_A=&\int d^4 x_1 d^4 x_2 \langle T\cdot h(x_1)T\cdot h(x_2)\rangle +
\int d^4 x_1 d^4 x_2 d^4 x_3 \langle T\cdot h(x_1)T\cdot h(x_2)T\cdot h(x_3)\rangle_{pole}
\notag \\
&+ \int d^4 x_1 d^4 x_2 d^4 x_3 d^4 x_4 
\left( \langle T\cdot h(x_1)T\cdot h(x_2)T\cdot h(x_3)T\cdot h(x_4)\rangle_{pole} +\right.
\notag \\
&\qquad\qquad\qquad\hspace{2cm} \left. + \langle T\cdot h(x_1)T\cdot h(x_2)T\cdot h(x_3)T\cdot h(x_4)\rangle_{0T}\right),
\end{align}
where we have also included the (complete) $TT$, plus the extra traceless (0T) term appearing in the 4T, as identified in \eqref{Weyl}.\\
 It is quite clear from \eqref{res} that $\mathcal{S}_A$ can be organized at each order in the expansion in the gravitational fluctuations $h$, in terms of pole parts and of traceless contributions. In principle, all the traceless contributions can be omitted from the definition of $\mathcal{S}_A$ and the entire result can be expressed uniquely in terms of contributions affected by bilinear mixings (pole) terms, but this is matter of convention. The "true" and unique anomaly action can be defined only by the exact evalutation of the functional expansion which defines the complete effective action $\mathcal{S}(g)$ of which 
 $\mathcal{S}_A$ is part. \\
 Obviously, the key feature of $\mathcal{S}_A$ lays in its specific pole structure, which is characterised by the insertion of multiple $R\square^{-1}$ operators.
As we have  emphasized at various stages in this work, we don't allow massless tadpoles in our regularization scheme, and the linear terms in $h$, which otherwise would be present and dominant at the Planck scale, in a flat background are absent. These terms would be phenomenologically important if one were interested in extending our analysis to the cosmological case.\\
 For instance, one possible application would be to determine the contribution of the conformal anomaly both in the early and late stages of the cosmological evolution, addressing the issue of the dark energy dominance of its more recent epochs. This 
 can be performed using an extension of this procedure to more general backgrounds, starting from the Weyl-flat case.\\
The structure of the traceless contributions beyond the fourth order, induced by the CWIs at the level of the 5T and higher, can be worked out order by order, following the approach that we have illustrated, as we are going to describe in a related work.\\
Being the analysis formulated in the Euclidean case, it is clear that the extension of our result to Minkowski space requires an analytic continuation. It is quite obvious, though, that such continuation is trivial, since the residui at each pole are just polynomial in momentum space and do not involve any branch cut. The residui computed at each pole, for any nT, are just given by anomaly polynomials, obtained by functional differentiations of the anomaly functional to arbitrary high orders.\\
The logarithmic contributions which are part of the structure of the anomaly action, as shown in the case of the 3T, using the free field theory analysis, contribute only to the Weyl-invariant part of the effective action \cite{Coriano:2018bsy}. \\
It is then clear that $\mathcal{S}_A$ predicts the emergence of intermediate massless states coupled to such anomaly polynomials. The entire anomaly action then appears to be quite simple, and can be identified in a very direct way. 

 \section{Conclusions and Perspectives}
In this work we have shown that a previous analysis of the the 3T vertex, in which some special features of the anomaly action, emerging at $O(h^3)$ in the gravitational fluctuations, had been identified, can be extended to the 4-graviton vertex 4T. This result has been obtained exclusively by relying on a careful analysis of the CWIs that such correlators have to satisfy. \\
For this purpose, we have extended to the 4T the longitudinal/transverse decomposition, introduced in the case of the 3T, which is ideal for a detailed analysis of the conformal constraints. \\ 
The method that we have presented is general, and can be extended, as we are going to show in a separate work, to any order and, in principle, to any dimension, with the inclusion of the corresponding counterterm Lagrangian. We demonstrated that these results can be obtained without the need to proceed with a complete analysis of all the form factors which appear in the transverse traceless sector of such correlator.\\
We have shown that the anomaly action carries a very specific signature of the breaking of the conformal symmetry by the conformal anomaly, in agreement with a former analysis of the same action at cubic level. 
Respect to the case of the 3T, the solution of the conformal constraints, in the 4T, induces a new traceless  component in its expression, summarised in Fig \ref{4T}.  \\
There are several possible implications of these results, especially in the search 
for some key phenomenological signatures of the breaking of the conformal symmetry, which could be tested at experimental level. For example, several recent studies, involving the nonlocal Riegert action, have investigated this phenomenon in quite diverse areas such as in the production of gravitational waves \cite{Mottola:2016mpl} as well as in Dirac and Weyl semimetals \cite{Chernodub:2017jcp,Chernodub:2020dqo,Chernodub:2019tsx}.\\
Here we have presented a general approach in order to identify the structure of $\mathcal{S}_A$ which avoids the variational solution of the anomaly equation. It can be used, for instance, to investigate the analogies and the differences 
among various types of anomaly actions, derived by independents methods.\\ 
Our method can be generalized in several directions, for instance by considering metric fluctuations around a conformally flat space, where the dynamics of the dilaton on the renormalization process should emerge more clearly compared to the case discussed here. It would also be interesting to compare our results with those predicted by the nonlocal anomaly action \cite{Riegert:1987kt}. We hope to  address some of 
these important issues in future work.  
    
\vspace{1.0cm}

\centerline{\bf Acknowledgements}
We thank Fiorenzo Bastianelli, Maxim Chernodub, Paul Frampton and Mirko Serino for discussions. This work is partially supported by INFN of Italy within Iniziativa Specifica QFT-HEP. The work of M.M.M. is partially supported by an INFN fellowship at GGI. We thank Mario Cret\`i and Riccardo Tommasi for their checks of Section 2. D.T. would like to thank Fotis Koutroulis and Konstantinos Rigatos for  discussions. M. M. M. would like to thank Paolo Benincasa, Carlo Alberto Cremonini and Alessio Caddeo for discussions. 

\appendix
\newpage
\section{Contact terms}
\label{contact}
Contact terms generated by 4th variation of a fundamental action are generated in the form 
\begin{align}
&\left\langle\frac{\delta^4 S }{\delta g(d)\delta g(c)\delta g(b)\delta g(a)}\right\rangle =\frac{1}{2} \sqrt{\text{dg}(a)}
\left\langle \frac{\delta ^3 T(a)}{\delta g(b) \delta g(c) \text{$\delta
		$g}(d)}\right\rangle+\frac{1}{8} g(b) \sqrt{\text{dg}(c)} g(c) \delta (b,a) \delta (c,b) \left\langle
\frac{\text{$\delta $T}(a)}{\delta g(d)}\right\rangle \nonumber \\&
+\frac{1}{8} g(b)
\sqrt{\text{dg}(d)} g(d) \delta (b,a) \delta (d,b) \left\langle \frac{\text{$\delta
		$T}(a)}{\delta g(c)}\right\rangle\nonumber
+\frac{1}{8} g(c) \sqrt{\text{dg}(d)} g(d)
\delta (c,a) \delta (d,c) \left\langle \frac{\text{$\delta $T}(a)}{\text{$\delta
		$g}(b)}\right\rangle \nonumber \\&
+\frac{1}{8} \sqrt{\text{dg}(d)} g(d) \delta (b,a) \delta (c,b)
s(c,b) \delta (d,b) \langle T(a)\rangle
+\frac{1}{8} \sqrt{\text{dg}(c)} g(c) \delta (b,a)
\delta (c,b) \delta (d,b) s(d,b) \langle T(a)\rangle \nonumber \\&
+\frac{1}{8} g(b) \sqrt{\text{dg}(c)}
\delta (b,a) \delta (c,b) \delta (d,c) s(d,c) \langle T(a)\rangle
+\frac{1}{4}
\sqrt{\text{dg}(b)} g(b) \delta (b,a) \left\langle \frac{\delta ^2 T(a)}{\text{$\delta
		$g}(c) \delta g(d)}\right\rangle \nonumber \\&
	+\frac{1}{4} \sqrt{\text{dg}(c)} g(c) \delta
(c,a) \left\langle \frac{\delta ^2 T(a)}{\delta g(b) \text{$\delta
		$g}(d)}\right\rangle 
+\frac{1}{4} \sqrt{\text{dg}(d)} g(d) \delta (d,a) \left\langle
\frac{\delta ^2 T(a)}{\delta g(b) \delta g(c)}\right\rangle\nonumber \\&
+\frac{1}{16} g(b) g(c) \sqrt{\text{dg}(d)} g(d) \delta (b,a) \delta (c,b) \delta (d,c)
\langle T(a)\rangle  +\frac{1}{4} \sqrt{\text{dg}(b)} \delta (b,a) \delta (c,b) s(c,b)
\left\langle \frac{\text{$\delta $T}(a)}{\delta g(d)}\right\rangle \nonumber \\&
+\frac{1}{4}
\sqrt{\text{dg}(b)} \delta (b,a) \delta (d,b) s(d,b) \left\langle \frac{\text{$\delta
		$T}(a)}{\delta g(c)}\right\rangle
+\frac{1}{4} \sqrt{\text{dg}(c)} \delta (c,a)
\delta (d,c) s(d,c) \left\langle \frac{\text{$\delta $T}(a)}{\text{$\delta
		$g}(b)}\right\rangle\nonumber \\&
+\frac{1}{4} \sqrt{\text{dg}(b)} \delta (b,a) \delta (c,b) \delta
(d,b) \delta (d,c) \langle T(a)\rangle  s(d,c,b).
\end{align}
In the flat limit we obtain 
\begin{align}
\langle \frac{\delta^4 S}{\delta g(a)\delta g(b)\delta g(c)\delta g(d)} \rangle=&+\frac{1}{8} \delta ^{b}\delta ^{c} \delta (b,a) \delta (c,b) \left\langle \frac{\text{$\delta
		$T}(a)}{\delta g(d)}\right\rangle +\frac{1}{8} \delta ^{b}\delta ^{d} \delta (b,a) \delta
(d,b) \left\langle \frac{\text{$\delta $T}(a)}{\delta g(c)}\right\rangle
\nonumber \\&
+\frac{1}{8} \delta ^{d}\delta ^{c} \delta (c,a) \delta (d,c) \left\langle \frac{\text{$\delta
		$T}(a)}{\delta g(b)}\right\rangle
+\frac{1}{4} \delta (b,a) \delta (c,b) s(c,b)
\left\langle \frac{\text{$\delta $T}(a)}{\delta g(d)}\right\rangle\nonumber \\&
+\frac{1}{4}
\delta (b,a) \delta (d,b) s(d,b) \left\langle \frac{\text{$\delta $T}(a)}{\text{$\delta
		$g}(c)}\right\rangle +\frac{1}{4} \delta (c,a) \delta (d,c) s(d,c) \left\langle
\frac{\text{$\delta $T}(a)}{\delta g(b)}\right\rangle\nonumber \\& +\frac{1}{4} \delta ^d
\delta (d,a) \left\langle \frac{\delta ^2 T(a)}{\delta g(b) \text{$\delta
		$g}(c)}\right\rangle +\frac{1}{4} \delta ^c \delta (c,a) \left\langle \frac{\delta ^2
	T(a)}{\delta g(b) \delta g(d)}\right\rangle\nonumber \\&
+\frac{1}{4} \delta ^b
\delta (b,a) \left\langle \frac{\delta ^2 T(a)}{\delta g(c) \text{$\delta
		$g}(d)}\right\rangle +\frac{1}{2} \left\langle \frac{\delta ^3 T(a)}{\delta g(b)
	\delta g(c) \delta g(d)}\right\rangle.
\end{align}

\subsection{Construction of a new ($n$-$p$)  basis}
\label{np}
In $d=4$ we may construct a new orthogonal four-vector $n^\mu$ using the completely antisymmetric $\epsilon$ tensor and three of the four external momenta in the form 
\begin{equation}
n^{\mu}=\epsilon^{\m \a \b\g}p_{1,\a}p_{2,\b}p_{3,\g}.
\end{equation}
Discussions of this  basis can be found in \cite{Bzowski:2013sza,Serino:2020pyu,Bzowski:2017poo}.
We will refer to this basis as to the $n$-$p$ basis. Notice that this basis is the direct generalization of the orthogonalization procedure in $d=3$ of the usual external product of two 3-vectors $\mathbf{a},\mathbf{b}$ in Euclidean space $\mathbf{n}=\mathbf{a}\times \mathbf{b}$. In the process of renormalization, the $\delta^{\mu\nu}$ is taken as an independent symmetric tensor, along with the external momenta, which appears in the covariant expansion of the form factors for a generic $d$. After removing the singularities by the relevant counterterms, having performed all the contractions in $d$ dimensions and having obtained a finite expression, we can 
dimensionally reduced the indices of all the tensor components to $d=4$. This reduction allows us to use the $n$-$p$ basis as a basis of expansion. Notice that such simplifications are possible if we have in $d$ dimensions at least $d-1$ independent external momenta in a correlation function. For instance, in $d=3$, this simplification starts in the expansion of 3-point functions, and in $d=4$  from 4-point functions.

We are allowed to use this basis in order to expand $\delta^{\mu\nu}$ in the form 

\begin{equation}
\d^{\m \n}=\sum_{i,j}^4 p_i^{\m} p_j^{\n} (Z^{-1})_{j i},
\end{equation}
where $(Z^{-1})_{j i}$ is the inverse of the Gramm matrix, defined as $Z=[p_i\cdot p_j]_{i,j=1}^d$. In our case we have:
\begin{equation}
\begin{pmatrix}
p_1^2 & p_1\cdot p_2 &  p_1\cdot p_3 & 0\\
 p_2\cdot p_1 & p_2^2 & p_2\cdot p_3 & 0\\
 p_3 \cdot p_1 & p_3\cdot p_2 & p_3^2 &0\\
 0 & 0 &0 & n^2
\end{pmatrix}
\end{equation}
The zero entries in the matrix above come from the orthonormality relation between the vector $n^{\m}$ and the momenta.  We also have
\begin{equation}
n^2=-p_1^2 p_2^2 p_3^2+p_1^2 (p_2\cdot p_3)^2+p_2^2
   (p_1\cdot p_3)^2+p_3^2 (p_1\cdot p_2)^2-2 p_1\cdot p_2  p_1\cdot p_3
   p_2\cdot p_3
\end{equation}

This implies that when the transverse-traceless projector gets contracted with two vectors ${n^\mu}$, will generate a term whose tensorial structure involves two of the external momenta, such as ${p^\a_i}_j {p^\b_i}_k$ in the expression above, with the related 
scalar factors. 

\section{ Trace relations}\label{TraceRelations}
 An analysis similar to that discussed for \eqref{double} can be performed for the traces of the third functional derivatives of $V_E$ in $d$ dimensions. 
One finds, expanding the lhs of \eqref{third} 
\begin{align}
&8 g_{\mu_1\nu_1}(x_1)V_E^{\mu_1\nu_1}(x_1)\delta^d (x_3-x_1)\delta^d(x_2-x_1) +
8 g_{\mu_3\nu_3}(x_3)g_{\mu_1\nu_1}(x_1)\delta^d (x_1-x_2)V_E^{\mu_1\nu_1\mu_3\nu_3}(x_1,x_3) \notag \\
& +8 g_{\mu_1\nu_1}(x_1)g_{\mu_2\nu_2}(x_2)\delta^d (x_1-x_3)V_E^{\mu_1\nu_1\mu_2\nu_2}(x_1,x_2) + 
8 g_{\mu_1\nu_1}(x_1)g_{\mu_2\nu_2}(x_2) \delta^d(x_2-x_3)V_E^{\mu_1\nu_1\mu_2\nu_2}(x_1,x_2)\notag \\
& + 8g_{\mu_3\nu_3}(x_3)g_{\mu_2\nu_2}(x_2) g_{\mu_1\nu_1}(x_1)
V_E^{\mu_1\nu_1\mu_2\nu_2\mu_3\nu_3}(x_1,x_2,x_3) = \varepsilon^3{\sqrt{g(x_1)}}E(x_1)\delta^d(x_1-x_2)\delta^d(x_2-x_3).
\end{align}
By going to the flat limit and using \eqref{double} one obtains 
\begin{align}
\label{triple}
\delta^{d}_{\mu_1\nu_2}\delta^{d}_{\mu_2\nu_2}\delta^{d}_{\mu_3\nu_3}V_E^{\mu_1\nu_1\mu_2\nu_2\mu_3\nu_3}=0. 
\end{align}
Moving to the fourth derivative we have:
 \begin{equation}
 \label{fourth}
 \frac{\delta}{\delta \sigma(x_4)} \frac{\delta}{\delta \sigma(x_3)} \frac{\delta}{\delta \sigma(x_2)} \frac{\delta}{\delta \sigma(x_1)}V_E=\varepsilon^4 \sqrt{g}E \delta^d(x_1-x_2)\delta^d(x_3-x_4)\delta^d(x_2-x_3)
\end{equation}
Expanding the l.h.s of \eqref{fourth} we get
\begin{align}
&16 g_{\mu_1\nu_1}(x_1)V_E^{\mu_1\nu_1}(x_1)\delta^d (x_4-x_1)\delta^d (x_3-x_1)\delta^d(x_2-x_1) \notag\\
&+16 g_{\mu_4\nu_4}(x_4)g_{\mu_1\nu_1}(x_1)V_E^{\mu_1\nu_1\mu_4\nu_4}(x_1,x_4)\delta^d (x_1-x_2) \delta^d (x_1-x_3)\notag \\
& +16 g_{\mu_1\nu_1}(x_1)g_{\mu_3\nu_3}(x_3)V_E^{\mu_1\nu_1\mu_3\nu_3}(x_1,x_3)\big(\delta^d (x_1-x_4)\delta^d (x_1-x_2)+\delta^d (x_1-x_2)\delta^d (x_3-x_4)\big) + \notag \\&+
16 g_{\mu_1\nu_1}(x_1)g_{\mu_2\nu_2}(x_2) V_E^{\mu_1\nu_1\mu_2\nu_2}(x_1,x_2)\big(\delta^d(x_1-x_3)\delta^d(x_2-x_4)+\delta^d(x_1-x_3)\delta^d(x_1-x_4)\notag \\&+\delta^d(x_2-x_3)\delta^d(x_2-x_4)+\delta^d(x_2-x_3)\delta^d(x_1-x_4)\big)\notag\\
& + 16g_{\mu_4\nu_4}(x_4)g_{\mu_3\nu_3}(x_3) g_{\mu_1\nu_1}(x_1)
V_E^{\mu_1\nu_1\mu_2\nu_2\mu_3\nu_3}(x_1,x_3,x_4)\delta^d(x_1-x_2)\notag \\
& + 16g_{\mu_4\nu_4}(x_4)g_{\mu_2\nu_2}(x_2) g_{\mu_1\nu_1}(x_1)
V_E^{\mu_1\nu_1\mu_2\nu_2\mu_3\nu_3}(x_1,x_2,x_4)\big(\delta^d(x_1-x_3)+\delta^d(x_2-x_3)\big)\notag \\
& + 16g_{\mu_3\nu_3}(x_3)g_{\mu_2\nu_2}(x_2) g_{\mu_1\nu_1}(x_1)
V_E^{\mu_1\nu_1\mu_2\nu_2\mu_3\nu_3}(x_1,x_2,x_4)\big(\delta^d(x_1-x_4)+\delta^d(x_2-x_3)+\delta^d(x_3-x_4)\big)\notag \\& + 16g_{\mu_4\nu_4}(x_4)g_{\mu_3\nu_3}(x_3)g_{\mu_2\nu_2}(x_2) g_{\mu_1\nu_1}(x_1)
V_E^{\mu_1\nu_1\mu_2\nu_2\mu_3\nu_3 \mu_4\nu_4}(x_1,x_2,x_3,x_4) \notag\\
&= \varepsilon^4 \sqrt{g}E \delta^d(x_1-x_2)\delta^d(x_2-x_3)\delta^d(x_3-x_4).
\end{align}
By going to the flat limit and using \eqref{double} and \eqref{triple} one obtains 
\begin{align}
\label{quadro}
\delta^{d}_{\mu_1\nu_2}\delta^{d}_{\mu_2\nu_2}\delta^{d}_{\mu_3\nu_3}\delta^{d}_{\mu_4\nu_4}V_E^{\mu_1\nu_1\mu_2\nu_2\mu_3\nu_3\mu_4\nu_4}=0.
\end{align}
This analysis can be extended to any $n$, yielding zero for all the fully traced functional derivatives of $V_E$. Notice that in \eqref{general} we have used only Weyl variations, which combine changes in the metric accompanied by contractions, as clear from \eqref{var}.\\

\section{Definitions}\label{Definitions}
The Christoffel symbols are
\begin{equation}
\Gamma^\lambda_{\mu\nu}(x)=\frac{1}{2}g^{\lambda \kappa}(x)\big(\partial_\mu g_{\nu \kappa}+\partial_\nu g_{\mu \kappa}-\partial_\kappa g_{\mu \nu}\big).
\end{equation}
Our definition of the Riemann tensor is
\bea \label{Tensors}
{R^\lambda}_{\mu\kappa\nu}
&=&
\pd_\nu \Gamma^\lambda_{\mu\kappa} - \pd_\kappa \Gamma^\lambda_{\mu\nu}
+ \Gamma^\lambda_{\nu\eta}\Gamma^\eta_{\mu\kappa} - \Gamma^\lambda_{\kappa\eta}\Gamma^\eta_{\mu\nu}.
\eea
The Ricci tensor is defined by the contraction $R_{\mu\nu} = {R^{\lambda}}_{\mu\lambda\nu}$ 
and the scalar curvature by $R = g^{\mu\nu}R_{\mu\nu}$.\\
The traceless part of the Riemann tensor in $d$ dimension is the Weyl tensor, 

The functional variations with respect to the metric tensor are computed using the relations
\beqa\label{Tricks}
\delta \sqrt{-g} = -\frac{1}{2} \sqrt{-g}\, g_{\a\b}\,\delta g^{\a \b}\quad &&
\delta \sqrt{-g} = \frac{1}{2} \sqrt{-g}\, g^{\a\b}\,\delta g_{\a \b}  \nonumber \\
\delta g_{\mu\nu} = - g_{\mu\a} g_{\nu\b}\, \delta g^{\a\b} \quad&&
\delta g^{\mu\nu} = - g^{\mu\a} g^{\nu\b}\, \delta g_{\a\b}.\,
\eeqa
The following structure has been repeatedly used throughout the calculations
\beqa\label{Tricks2}
s^{\a\b\g\delta} \, \delta(z,x) &\equiv& - \frac{\d g^{\a\b}(z)}{\d g_{\g\d}(x)} =
\frac{1}{2}\left[\delta^{\a\g}\delta^{\b\delta} + \delta^{\a\delta}\delta^{\b\g}\right]\delta(z,x)\, .
\eeqa
Another useful quantity is:
\begin{equation}
[\sqrt{-g}]^{\m_1\n_1 \m_2 \n_2 }=-\frac{1}{4} g^{\mu_1\nu_2} g^{\mu_2\nu_1}+\frac{1}{4} g^{\mu_1\nu_1} g^{\mu_2\nu_2}-\frac{1}{4} g^{\mu_1\mu_2} g^{\nu_1\nu_2}
\end{equation}
Regarding the derivatives of the Christoffel symbols we get, in the flat space limit:
\begin{align}
&[\Gamma^{\a}_{\b \chi}]^{\m_2 \n_2}(p_2)=-\frac{i}{2}\delta^{\a \ka}\bigg(\delta^{(\m_2}_{\b}\delta^{\n_2)}_{\kappa}p_{2,\chi}+\delta^{(\m_2}_{\chi}\delta^{\n_2)}_{\kappa}p_{2,\beta}-\delta^{(\m_2}_{\b}\delta^{\n_2)}_{\chi}p_{2,\kappa}\bigg)=\delta^{\a \ka}[\tilde{\Gamma}_{\kappa\b \chi}]^{\m_2 \n_2}(p_2),\\
&[\Gamma^{\a}_{\b \chi}]^{\m_2 \n_2 \m_3 \n_3}(p_2,p_3)=[g^{\a\kappa}]^{\m_2\n_2}[\tilde{\Gamma}_{\kappa \b \chi}]^{\m_3 \n_3}(p_3)+[g^{\a\kappa}]^{\m_3\n_3}[\tilde{\Gamma}_{\kappa\b \chi}]^{\m_2 \n_2}(p_2),\\
&[\Gamma^{\a}_{\b \chi}]^{\m_2 \n_2 \m_3 \n_3 \m_4 \n_4}(p_2,p_3,p_4)=[g^{\a\kappa}]^{\m_2\n_2 \m_4 \n_4}[\tilde{\Gamma}_{\kappa \b \chi}]^{\m_3 \n_3}(p_3)+[g^{\a\kappa}]^{\m_2\n_2 \m_3 \n_3}[\tilde{\Gamma}_{\kappa \b \chi}]^{\m_4 \n_4}(p_4)\notag\\&\hspace{5cm}+[g^{\a\kappa}]^{\m_3\n_3 \m_4 \n_4}[\tilde{\Gamma}_{\kappa \b \chi}]^{\m_2 \n_2}(p_2).
\end{align}
The above notation can be understood as follows:
\begin{equation}
[\Gamma^{\a}_{\b \chi}]^{\m_2 \n_2...\m_n \n_n}(p_2,\dots,p_n)=\int d^d x_1...d^d x_n e^{-i (x_1 p_1+\dots+x_n p_n)}\left(\sdfrac{\d^n \Gamma^{\a}_{\b \chi}(x_1)}{\d g_{\m_n \n_n}(x_n)\dots\d g_{\m_2 \n_2}(x_2)}\right)_{g=\delta}
\end{equation}
Also  we have defined (in the flat space-time limit)
\begin{align}
[g^{\a\kappa}]^{\m_2\n_2 \m_3 \n_3}=&\frac{1}{4} \d^{\alpha \n_3} \d^{\kappa \n_2} \d^{\m_2\m_3}+\frac{1}{4} \d^{\alpha \n_2} \d^{\kappa \n_3} \d^{\m_2\m_3}+\frac{1}{4} \d^{\alpha \n_2} \d^{\kappa \m_3} \d^{\m_2 \nu_3}+\frac{1}{4} \d^{\alpha \m_3} \d^{\kappa \n_2} \d^{\m_2\n_3}+\frac{1}{4} \d^{\alpha \n_3}
\d^{\kappa \m_2} \d^{\m_3\n_2}\notag\\&+\frac{1}{4} \d^{\alpha \m_2} \d^{\kappa \n_3} \d^{\mu_3\n_2}+\frac{1}{4} \d^{\alpha \m_3} \d^{\kappa \m_2} \d^{\n_2\n_3}+\frac{1}{4} \d^{\alpha
	\m_2} \d^{\kappa \m_3} \d^{\n_2\n_3}.
\end{align}
Using the above we can compute the derivatives of the Riemann, Ricci and the scalar curvature. Up to second derivative the expressions can be found at \cite{Coriano:2018bsy}. For our convenience, we reproduce the formulae here.
In order to simplify the notation, we introduce the tensor components
\begin{align}
A^{\m_1\n_1\m\n}&\equiv\d^{\mu_1\nu_1}\d^{\m\n}-2\d^{\m(\m_1}\d^{\n_1)\n}\nn
B^{\m_1\n_1\m\n}&\equiv\d^{\mu_1\nu_1}\d^{\m\n}-\d^{\m(\m_1}\d^{\n_1)\n}\nn
C^{\m_1\n_1\m_2\n_2\m\n}&\equiv\d^{\m(\m_1}\d^{\n_1)(\m_2}\d^{\n_2)\n}+\d^{\m(\m_2}\d^{\n_2)(\m_1}\d^{\n_1)\n}\nn
\tilde{C}^{\m_1\n_1\m_2\n_2\m\n}&\equiv\d^{\m(\m_1}\d^{\n_1)(\m_2}\d^{\n_2)\n}\nn
D^{\m_1\n_1\m_2\n_2\m\n}&\equiv\d^{\m_1\n_1}\d^{\m(\m_2}\d^{\n_2)\n}+\d^{\m_2\n_2}\d^{\m(\m_1}\d^{\n_1)\n}\nn
E^{\m_1\n_1\m_2\n_2\m\n}&\equiv\d^{\m_1\n_1}B^{\m_2\n_2\m\n}+C^{\m_1\n_1\m_2\n_2\m\n},\nn
F^{\a_1\a_2\m\n}&\equiv\d^{\a_1[\m}\d^{\n]\a_2}\nn
\tilde{F}^{\a_1\a_2\m\n}&\equiv\d^{\a_1(\n}\d^{\m)\a_2}\nn
\tilde{F}^{\a_1\a_2}_{\m\n}&\equiv\d^{(\a_1}_{\n}\d_{\m}^{\a_2)}\nn
G^{\m_1\n_1\a_1\a_2\m\n}&\equiv\d^{\m[\n}\d^{\a_2](\m_1}\d^{\n_1)\a_1}+\d^{\a_1[\a_2}\d^{\n](\m_1}\d^{\n_1)\m}\nn
H^{\m_1\n_1\m_2\n_1\a_1\a_2\m\n}&\equiv A^{\m_1\n_1\m\a_1}\tilde{F}^{\m_2\n_2\n\a_2}-A^{\m_2\n_2\m\a_1}\tilde{F}^{\m_1\n_1\n\a_2}\nn
I^{\m_1\n_1\m_2\n_2\a_1\a_1\m\n}&\equiv\d^{\m_1\n_1}D^{\m\a_1\n\a_2\m_2\n_2}-\sdfrac{1}{2}\d^{\a_1\m}\d^{\a_2\n}A^{\m_1\n_1\m_2\n_2},
\end{align}
where we indicate with the circle brackets the symmetrization of the indices and with the square brackets the anti-symmetrization of the indices, as follows
\begin{align}
\d^{\m(\m_1}\d^{\n_1)\n}&\equiv\sdfrac{1}{2}\bigg(\d^{\m\m_1}\d^{\n_1\n}+\d^{\m\n_1}\d^{\m_1\n}\bigg)\nn
\d^{\m[\m_1}\d^{\n_1]\n}&\equiv\sdfrac{1}{2}\bigg(\d^{\m\m_1}\d^{\n_1\n}-\d^{\m\n_1}\d^{\m_1\n}\bigg).
\end{align}
The metric variation are consider in the flat space-time limit and the first variation of the square of the metric, Riemann, Ricci and the scalar curvature are given as
\begin{align}
\big[\sqrt{-g}\big]^{\m_i\n_i}&=\sdfrac{1}{2}\d^{\m_i\n_i}\nn
\big[R_{\m\a\n\b}\big]^{\m_i\n_i}(p_i)&=\frac{1}{2}\,\bigg(\d_\a^{(\m_i}\d^{\n_i)}_\b\,p_{i\m}\,p_{i\n}
+\d_\m^{(\m_i}\d^{\n_i)}_\n\,p_{i\a}\,p_{i\b}-\d_\m^{(\m_i}\d^{\n_i)}_\b\,p_{i\a}\,p_{i\n}-\d_\a^{(\m_i}\d^{\n_i)}_\n\,p_{i\m}\,p_{i\b}\bigg)\nn
\big[R^{\m\a\n\b}\big]^{\m_i\n_i}(p_i)&=\frac{1}{2}\,\bigg(\d^{\a(\m_i}\d^{\n_i)\b}\,p_i^\m\,p_i^\n
+\d^{\m(\m_i}\d^{\n_i)\n}\,p_i^\a\,p_i^\b-\d^{\m(\m_i}\d^{\n_i)\b}\,p_i^\a\,p_i^\n-\d^{\a(\m_i}\d^{\n_i)\n}\,p_i^\m\,p_i^\b\bigg)\nn
\big[R_{\m\n}\big]^{\m_i\n_i}(p_i)&=\frac{1}{2}\,\bigg(\d_\m^{(\m_i}\d^{\n_i)}_\n\,p_{i}^2
+\d^{\m_i\n_i}\,p_{i\m}\,p_{i\n}-p_i^{(\m_i}\d^{\n_i)}_\m\,p_{i\n}-p_i^{(\m_i}\d^{\n_i)}_\n\,p_{i\m}\bigg)\nn
\big[R^{\m\n}\big]^{\m_i\n_i}(p_i)&=\frac{1}{2}\,\bigg(\d^{\m(\m_i}\d^{\n_i)\n}\,p_{i}^2
+\d^{\m_i\n_i}\,p_i^\m\,p_i^\n-p_i^{(\m_i}\d^{\n_i)\m}\,p_i^\n-p_i^{(\m_i}\d^{\n_i)\n}\,p_i^\m\bigg)\nn
\big[R\big]^{\m_i\n_i}(p_i)&=\bigg(\d^{\m_i\n_i}\,p_{i}^2-p_i^{(\m_i}p_i^{\n_i)}\bigg)\nn
\big[\square R\big]^{\m_i\n_i}(p_i)&=p_i^2\,\bigg(p_i^{(\m_i}p_i^{\n_i)}-\d^{\m_i\n_i}\,p_{i}^2\bigg).
\end{align}
Their second variations can be calculated in the form
\begin{align}
\big[R^{\,\b}_{\ \ \n\r\s}\big]^{\m_1\n_1\m_2\n_2}(p_1,p_2)&=\Big[-\frac{1}{2}\,\tilde F^{\m_1\n_1\b\epsilon}p_{1\s}\big(\tilde F^{\m_2\n_2}_{\e\n}p_{2\r}+\tilde F^{\m_2\n_2}_{\e\r}p_{2\n}-\tilde F^{\m_2\n_2}_{\n\r}p_{2\e}\big)\notag\nn
&\hspace{-2.5cm}-\frac{1}{2}\big(\tilde C^{\m_1\n_1\m_2\n_2\b}_{\hspace{1.4cm}\r}\ p_{2\n}-\tilde F^{\m_1\n_1\b\e}\,\tilde F^{\m_2\n_2}_{\n\r}\,p_{2\e}\big)p_{2\s}\notag\nn
&\hspace{-2.5cm}-\frac{1}{4}\big(\tilde F^{\m_1\n_1}_{\a\n}p_{1\s}+\tilde F^{\m_1\n_1}_{\a\s}p_{1\n}-\tilde F^{\m_1\n_1}_{\s\n}p_{1\a}\big)\big(\tilde F^{\m_2\n_2\b\a}\,p_{2\r}+\tilde F^{\m_2\n_2\b}_{\qquad\r}\,p_{2}^\a-\tilde F^{\m_2\n_2\a}_{\qquad\r}\,p_{2}^{\b}\big)\Big]-(\s\leftrightarrow \r)\nn
\big[R_{\m\n\r\s}\big]^{\m_1\n_1\m_2\n_2}(p_1,p_2)&=\d^{(\m_1}_\m\d^{\n_1)}_\b\big[R^{\,\b}_{\ \ \n\r\s}\big]^{\m_2\n_2}(p_2)+\d_{\m\b}\big[R^{\,\b}_{\ \ \n\r\s}\big]^{\m_1\n_1\m_2\n_2}(p_1,p_2)\nn
\end{align}
\begin{align}
\big[R^{\m\n\r\s}\big]^{\m_1\n_1\m_2\n_2}(p_1,p_2)&=\d^{\a\n}\d^{\b\r}\d^{\s\g}\big[R^{\,\m}_{\ \ \a\b\g}\big]^{\m_1\n_1\m_2\n_2}(p_1,p_2)\notag\nn
&\hspace{-1.5cm}-\big(\d^{\a(\m_1}\d^{\n_1)\n}\d^{\b\r}\d^{\s\g}+\d^{\a\n}\d^{\b(\m_1}\d^{\n_1)\r}\d^{\s\g}+\d^{\a\n}\d^{\b\r}\d^{\s(\m_1}\d^{\n_1)\g}\big)\big[R^{\,\b}_{\ \ \n\r\s}\big]^{\m_2\n_2}(p_2)
\end{align}
\begin{align}
\big[R_{\n\s}\big]^{\m_1\n_1\m_2\n_2}(p_1,p_2)&=-\frac{1}{2}\tilde F^{\m_1\n_1\m_2\n_2}\left(p_{1\s}p_{2\n}-\sdfrac{1}{2}p_{1\n}p_{2\s}+p_{2\n}p_{2\s}\right)-\frac{1}{4}\d^{\m_2\n_2}\left(\tilde F^{\m_1\n_1}_{\a\n}\,p_{1\s}+\tilde F^{\m_1\n_1}_{\a\s}\,p_{1\n}\right)\,p_2^\a\notag\\
&\hspace{-1.5cm}+\frac{1}{2}\big(\tilde C^{\m_1\n_1\m_2\n_2\e}_{\hspace{1.4cm}\n}\,p_{2\s}+\tilde C^{\m_1\n_1\m_2\n_2\e}_{\hspace{1.4cm}\s}\,p_{2\n}\big)\,(p_1+p_2)_{\e}+\frac{1}{2}F^{\m_2\n_2}_{\a\s}\tilde F^{\m_1\n_1}_{\b\n}\,p_1^\a\,p_2^\b\notag\\
&\hspace{-1.5cm}-\frac{1}{2}\tilde F^{\m_2\n_2}_{\n\s}\,\tilde F^{\m_1\n_1\a\b}(p_1+p_2)_\a\,p_{2\b}-\frac{1}{2}\left(\tilde C^{\m_1\n_1\m_2\n_2}_{\hspace{1.25cm}\n\s}-\frac{1}{2}\d^{\m_2\n_2}\,\tilde F^{\m_1\n_1}_{\n\s}\right)\,p_1\cdot p_2\nn
\end{align}
\begin{align}
\big[R^{\n\s}\big]^{\m_1\n_1\m_2\n_2}(p_1,p_2)&=\d^{\n\a}\d^{\s\b}\big[R_{\a\b}\big]^{\m_1\n_1\m_2\n_2}(p_1,p_2)-\big(\d^{\n(\m_1}\d^{\n_1)\a}\d^{\s\b}+\d^{\n\a}\d^{\s(\m_1}\d^{\n_1)\b}\big)\big[R_{\a\b}\big]^{\m_2\n_2}(p_2)\\
\big[R\big]^{\m_1\n_1\m_2\n_2}(p_1,p_2)&=-\left(p_2^2+\sdfrac{1}{4}p_1\cdot p_2\right)\,\tilde F^{\m_1\n_1\m_2\n_2}+\frac{1}{4}\,A^{\m_1\n_1\m_2\n_2}\,p_1\cdot p_2\notag\\
&\hspace{-1.5cm}+\tilde C^{\m_1\n_1\m_2\n_2\a\b}\,(p_{1\a}+2p_{2\a})p_{2\b}-\d^{\m_2\n_2}\tilde F^{\m_1\n_1\a\b}\,(p_{1\a}+p_{2\a})p_{2\b}+\sdfrac{1}{2}\tilde C^{\m_2\n_2\m_1\n_1\a\b}\,p_{1\a}p_{2\b}
\end{align}
\begin{align}
\big[\square R\big]^{\m_1\n_1\m_2\n_2}(p_1,p_2)&=\tilde F^{\m_1\n_1\m_2\n_2}\,\bigg[p_2^2(p_1+p_2)^2+\sdfrac{3}{2}(p_2^2+p_1\cdot p_2)\bigg]+\sdfrac{1}{2}\d^{\m_1\n_1}\tilde F^{\m_2\n_2\a\b}(p_1\cdot p_2)\,p_{2\a}p_{2\b}\notag\\
&\hspace{-1.5cm}-\sdfrac{1}{2}\d^{\m_1\n_1}\d^{\m_2\n_2}(p_1\cdot p_2)\bigg[(p_1+p_2)^2-p_1\cdot p_2\bigg]+\d^{\m_2\n_2}F^{\m_1\n_1\a\b}p_{2\a}(p_1+p_2)_\b\bigg[(p_1+p_2)^2+p_2^2\bigg]\notag\\
&\hspace{-1.5cm}-\tilde F^{\m_2\n_2\a\b}p_{2\a}p_{2\b}\,\tilde F^{\m_1\n_1\gamma\d}p_{2\g}(p_1+p_2)_\d-(p_1+p_2)^2\,\tilde C^{\m_1\n_1\m_2\n_2\a\b}\bigg[2p_{2\a}p_{2\b}+p_{1\a}p_{2\b}+\sdfrac{1}{2}p_{2\a}p_{1\b}\bigg].
\end{align}

For the third derivatives the expressions can be written as
\begin{align}
&[R^{\a}_{\beta \chi \de}]^{\m_2 \n_2 \m_3 \n_3 \m_4 \n_4}(p_2,p_3,p_4)=i(p_2+p_3+p_4)_{\delta}[\Gamma^{\a}_{\b \chi}]^{\m_2 \n_2 \m_3 \n_3 \m_4 \n_4}(p_2,p_3,p_4)\notag\\&-i(p_2+p_3+p_4)_{\chi}[\Gamma^{\a}_{\b \delta}]^{\m_2 \n_2 \m_3 \n_3 \m_4 \n_4}(p_2,p_3,p_4)+[\Gamma^{\a}_{\delta \eta}]^{\m_2 \n_2  \m_4 \n_4}(p_2,p_4)[\Gamma^{\eta}_{\beta \chi}]^{ \m_3 \n_3 }(p_3)\notag\\&
+[\Gamma^{\a}_{\delta \eta}]^{\m_3 \n_3  }(p_3)[\Gamma^{\eta}_{\beta \chi}]^{ \m_2 \n_2 \m_4 \n_4 }(p_2,p_4)-[\Gamma^{\a}_{\chi \eta}]^{\m_2 \n_2  \m_3 \n_3}(p_2,p_3)[\Gamma^{\eta}_{\beta \delta}]^{ \m_4 \n_4 }(p_4)\notag\\&
-[\Gamma^{\a}_{\chi \eta}]^{\m_4 \n_4  }(p_4)[\Gamma^{\eta}_{\beta \delta}]^{ \m_2 \n_2 \m_3 \n_3 }(p_2,p_3)+\text{permutations}
\end{align}
 It follows that
\begin{align}
&[R_{\beta \delta}]^{\m_2 \n_2 \m_3 \n_3 \m_4 \n_4}(p_2,p_3,p_4)=[R^{\a}_{\beta \a\de}]^{\m_2 \n_2 \m_3 \n_3 \m_4 \n_4}(p_2,p_3,p_4)\\
&[R]^{\m_2 \n_2 \m_3 \n_3 \m_4 \n_4}(p_2,p_3,p_4)=\delta^{\b \d}[R_{\beta \delta}]^{\m_2 \n_2 \m_3 \n_3 \m_4 \n_4}(p_2,p_3,p_4).
\end{align}
\section{Metric variations of the anomaly}\label{Mvc}
From the anomaly functional
\begin{equation}
\mathcal{A}(x)=\sqrt{-g(x)}(b \,C^2(x)+b' E(x))=\sqrt{-g(x)} \chi(x)
\end{equation}
expanding around a flat space-time and transforming to momentum space we obtain \begin{equation}
[\mathcal{A}(p_1,p_2)]^{\m_1 \n_1 \m_2 \n_2}=\chi(p_1,p_2)^{\m_1 \n_1 \m_2 \n_2},
\end{equation}
where 
\begin{equation}
\chi(p_1,p_2)^{\m_1 \n_1 \m_2 \n_2}=b [C^2]^{\m_1\n_1\m_2\n_2}(p_1,p_2)+b' [E]^{\m_1\n_1\m_2\n_2}(p_1,p_2).
\end{equation}
The second-order derivatives of the Weyl tensor and the Euler density in flat space-time are
\begin{align}
\big[C^2\big]^{\m_1\n_1\m_2\n_2}(p_1,p_2)&=2\bigg([R_{abcd}]^{\m_1\n_1}(p_1)[R^{abcd}]^{\m_2\n_2}(p_2)-\sdfrac{4}{d-2}[R_{ab}]^{\m_1\n_1}(p_1)[R^{ab}]^{\m_2 \n_2}(p_2)\notag \\&+\sdfrac{2}{(d-2)(d-1)}[R]^{\m_1\n_1}(p_1)[R]^{\m_2\n_2 }(p_2)\bigg)
\end{align}
\begin{align}
\big[E\big]^{\m_1\n_1\m_2\n_2}(p_1,p_2)&=2\bigg([R_{abcd}]^{\m_1\n_1}(p_1)[R^{abcd}]^{\m_2\n_2}(p_2)-4[R_{ab}]^{\m_1\n_1}(p_1)[R^{ab}]^{\m_2 \n_2}(p_2)\notag \\&+[R]^{\m_1\n_1}(p_1)[R]^{\m_2\n_2 }(p_2)\bigg).
\end{align}
The third derivative of the anomaly is given by 
\begin{equation}
[\mathcal{A}(p_1,p_2,p_3)]^{\m_1 \n_1 \m_2 \n_2 \m_3 \n_3}=\big[\sqrt{-g(x)}\big]^{\m_3 \n_3}\chi(p_1,p_2)^{\m_1 \n_1 \m_2 \n_2}+\chi(p_1,p_2,p_3)^{\m_1 \n_1 \m_2 \n_2 \m_3 \n_3}+\text{permutations},
\end{equation}
where 
\begin{equation}
\chi(p_1,p_2,p_3)^{\m_1 \n_1 \m_2 \n_2 \m_3\n_3}=b [C^2]^{\m_1\n_1\m_2\n_2 \m_3\n_3}(p_1,p_2)+b' [E]^{\m_1\n_1\m_2\n_2 \m_3\n_3}(p_1,p_2).
\end{equation}
The derivatives of the relevant tensors are
\begin{align}
\big[C^2\big]^{\m_1\n_1\m_2\n_2\m_3\n_3}(p_1,p_2,p_3)&=\Bigg\{[R_{abcd}]^{\m_1\n_1\m_2\n_2}(p_1,p_2)[R^{abcd}]^{\m_3\n_3}(p_3)\notag\\
&\hspace{-3cm}+[R_{abcd}]^{\m_2\n_2}(p_2)[R^{abcd}]^{\m_1\n_1\m_3\n_3}(p_1,p_3)-\sdfrac{4}{d-2}[R_{ab}]^{\m_1\n_1\m_2\n_2}(p_1,p_2)[R^{ab}]^{\m_3\n_3}(p_3)\notag\\
&\hspace{-3cm}-\sdfrac{4}{d-2}[R_{ab}]^{\m_2\n_2}(p_2)[R^{ab}]^{\m_1\n_1\m_3\n_3}(p_1,p_3)+\sdfrac{2}{(d-2)(d-1)}[R]^{\m_1\n_1\m_2\n_2}(p_1,p_2)[R]^{\m_3\n_3}(p_3)\notag\\
&\hspace{-3cm}+\sdfrac{2}{(d-2)(d-1)}[R]^{\m_2\n_2}(p_2)[R]^{\m_1\n_1\m_3\n_3}(p_1,p_3)\Bigg\}+\text{permutations},
\end{align}
and 
\begin{align}
\big[E\big]^{\m_1\n_1\m_2\n_2\m_3\n_3}(p_1,p_2,p_3)&=\Bigg\{[R_{abcd}]^{\m_1\n_1\m_2\n_2}(p_1,p_2)[R^{abcd}]^{\m_3\n_3}(p_3)+[R_{abcd}]^{\m_2\n_2}(p_2)[R^{abcd}]^{\m_1\n_1\m_3\n_3}(p_1,p_3)\notag\\
&\hspace{-3.7cm}-4[R_{ab}]^{\m_1\n_1\m_2\n_2}(p_1,p_2)[R^{ab}]^{\m_3\n_3}(p_3)-4[R_{ab}]^{\m_2\n_2}(p_2)[R^{ab}]^{\m_1\n_1\m_3\n_3}(p_1,p_3)+[R]^{\m_1\n_1\m_2\n_2}(p_1,p_2)[R]^{\m_3\n_3}(p_3)\notag\\
&\hspace{-3cm}+[R]^{\m_2\n_2}(p_2)[R]^{\m_1\n_1\m_3\n_3}(p_1,p_3)\Bigg\}+\text{permutations}.
\end{align}
We also give the fourth-order derivatives of the counterterms. In compact notation they take the form

\begin{align}
\big[\sqrt{-g}\,C^2\big]^{\m_1\n_1\m_2\n_2\m_3\n_3\m_4\n_4}(p_1,p_2,p_3,p_4)&=\Bigg\{[R_{abcd}]^{\m_1\n_1}(p_1)[R^{abcd}]^{\m_2 \n_2 \m_3\n_3\m_4\n_4}(p_2,p_3,p_4)\notag\\&\hspace{-6.0cm}+[R_{abcd}]^{\m_2 \n_2 \m_3\n_3\m_4\n_4}(p_2,p_3,p_4)[R^{abcd}]^{\m_1 \n_1}(p_1)+[R_{abcd}]^{\m_1 \n_1 \m_2\n_2}(p_1,p_2)[R^{abcd}]^{\m_3 \n_3 \m_4 \n_4}(p_3,p_4)\notag\\&\hspace{-6.0cm}-\sdfrac{4}{d-2}[R_{ab}]^{\m_1 \n_1}(p_1)[R^{ab}]^{\m_2 \n_2 \m_3\n_3\m_4\n_4}(p_2,p_3,p_4)-\sdfrac{4}{d-2}[R_{ab}]^{\m_1 \n_1 \m_2\n_2}(p_1,p_2)[R^{ab}]^{\m_3\n_3\m_4\n_4}(p_3,p_4)\notag\\&\hspace{-6.0cm}-\sdfrac{4}{d-2}[R_{ab}]^{ \m_2 \n_2 \m_3 \n_3 \m_4 \n_4}(p_2,p_3,p_4)[R^{ab}]^{\m_1 \n_1 }(p_1)+\sdfrac{2}{(d-2)(d-1)}[R]^{\m_1\n_1}(p_1)[R]^{\m_2\n_2\m_3 \n_3 \m_4 \n_4}(p_2,p_3,p_4)\notag\\&\hspace{-6.0cm}+\sdfrac{2}{(d-2)(d-1)}[R]^{\m_1\n_1\m_2 \n_2 \m_4 \n_4}(p_1,p_2,p_4)[R]^{\m_3\n_3}(p_3)+\sdfrac{2}{(d-2)(d-1)}[R]^{\m_1\n_1\m_2 \n_2 }(p_1,p_2)[R]^{\m_3\n_3 \m_4 \n_4}(p_3,p_4)\Bigg\}\notag\\& \hspace{-6.0cm}+[\sqrt{-g}]^{\m_1\n_1}\big[C^2\big]^{\m_2\n_2\m_3\n_3 \m_4 \n_4}(p_2,p_3,p_4)+[\sqrt{-g}]^{\m_1\n_1 \m_2 \n_2 }\big[C^2\big]^{\m_3\n_3 \m_4 \n_4}(p_3,p_4)+\text{permutations},
\end{align}
and
\begin{align}
\big[\sqrt{-g}\,E\big]^{\m_1\n_1\m_2\n_2\m_3\n_3\m_4\n_4}(p_1,p_2,p_3,p_4)&=\Bigg\{[R_{abcd}]^{\m_1\n_1}(p_1)[R^{abcd}]^{\m_2 \n_2 \m_3\n_3\m_4\n_4}(p_2,p_3,p_4)\notag\\&\hspace{-6.0cm}+[R_{abcd}]^{\m_2 \n_2 \m_3\n_3\m_4\n_4}(p_2,p_3,p_4)[R^{abcd}]^{\m_1 \n_1}(p_1)+[R_{abcd}]^{\m_1 \n_1 \m_2\n_2}(p_1,p_2)[R^{abcd}]^{\m_3 \n_3 \m_4 \n_4}(p_3,p_4)\notag\\&\hspace{-6.0cm}-4[R_{ab}]^{\m_1 \n_1}(p_1)[R^{ab}]^{\m_2 \n_2 \m_3\n_3\m_4\n_4}(p_2,p_3,p_4)-4[R_{ab}]^{\m_1 \n_1 \m_2\n_2}(p_1,p_2)[R^{ab}]^{\m_3\n_3\m_4\n_4}(p_3,p_4)\notag\\&\hspace{-6.0cm}-4[R_{ab}]^{ \m_2 \n_2 \m_3 \n_3 \m_4 \n_4}(p_2,p_3,p_4)[R^{ab}]^{\m_1 \n_1 }(p_1)+[R]^{\m_1\n_1}(p_1)[R]^{\m_2\n_2\m_3 \n_3 \m_4 \n_4}(p_2,p_3,p_4)\notag\\&\hspace{-6.0cm}+[R]^{\m_1\n_1\m_2 \n_2 \m_4 \n_4}(p_1,p_2,p_4)[R]^{\m_3\n_3}(p_3)+[R]^{\m_1\n_1\m_2 \n_2 }(p_1,p_2)[R]^{\m_3\n_3 \m_4 \n_4}(p_3,p_4)\Bigg\}\notag\\&\hspace{-6.0cm}+[\sqrt{-g}]^{\m_1\n_1}\big[E\big]^{\m_2\n_2\m_3\n_3 \m_4 \n_4}(p_2,p_3,p_4)+[\sqrt{-g}]^{\m_1\n_1 \m_2 \n_2 }\big[E\big]^{\m_3\n_3 \m_4 \n_4}(p_3,p_4)+\text{permutations}.
\end{align}

\providecommand{\href}[2]{#2}\begingroup\raggedright\endgroup

\end{document}